%% file: draft_monopoles.tex
\documentclass[12pt]{article}
\pdfoutput=1
\usepackage{putex2}
\usepackage{graphicx}
\usepackage{caption}
\usepackage{subcaption}
\usepackage{epstopdf}
\usepackage{enumerate}
\usepackage{cite}
\usepackage{tensor}
\usepackage{slashed}
\usepackage[aligntableaux=center]{ytableau}
\usepackage{bbm}
\usepackage{placeins}
\usepackage{cancel}
\usepackage{diagbox}
\usepackage[export]{adjustbox}
\usepackage{multirow}

\numberwithin{equation}{section}

\newcommand{\abs}[1]{\left\lvert #1 \right\rvert}

\newcommand {\be} {\begin {equation}}
\newcommand {\ee} {\end {equation}}

\newcommand {\bes} {\begin {equation*}}
\newcommand {\ees} {\end {equation*}}

\newcommand{\es}[2] {\begin{equation} \label{#1} \begin{split} #2 \end{split} \end{equation}}

\newcommand{\CP}{\mathbb{CP}}
\newcommand{\Z}{\mathbb{Z}}

\newcommand{\R}{\mathbb{R}}

\def\Tr{\mop{Tr}}

\newcommand{\beq}{\begin{equation}}
\newcommand{\eeq}{\end{equation}}

\newcommand{\ov}{\over}
\newcommand{\Det}{{\rm Det}}
\def\le{\left}
\def\ri{\right}

\definecolor{orange}{rgb}{1,0.5,0}
\newcommand{\orange}{\color{orange}}

\def\SSP#1{{}}
\def\MM#1{{}}
\def\ESD#1{{}}

\def\<{\langle}
\def\>{\rangle}
\newcommand\ket[1]{\ensuremath{\lvert{#1}\rangle}}
\newcommand\bra[1]{\ensuremath{\langle{#1}\rvert}}

\begin{document}

\preprint{MIT-CTP-4495}

\institution{MIT}{Center for Theoretical Physics, Massachusetts Institute of Technology, Cambridge, MA 02139}

\title{Monopole Taxonomy in Three-Dimensional Conformal Field Theories
}

\authors{Ethan Dyer, M\'ark Mezei, and Silviu S.~Pufu}

\abstract{
We study monopole operators at the infrared fixed points of Abelian and non-Abelian gauge theories with $N_f$ fermion flavors in three dimensions.  At large $N_f$, independent monopole operators can be defined via the state-operator correspondence only for stable monopole backgrounds.  In Abelian theories, every monopole background is stable. In the non-Abelian case, we find that many (but not all) backgrounds are stable in each topological class.  We calculate the infrared scaling dimensions of the corresponding operators through next-to-leading order in $1/N_f$.  In the case of $U(N_c)$ QCD with $N_f$ fundamental fermions (and in particular in the QED case, $N_c =1$), we find that the monopole operators transform as non-trivial irreducible representations of the $SU(N_f)$ flavor symmetry group.}

\date{}

\maketitle

\tableofcontents

\section{Introduction}

In three-dimensional gauge theories, one can define local disorder operators by requiring the gauge field to have a certain singular profile close to the point where the operator is inserted \cite{Borokhov:2002ib,Borokhov:2002cg}.  These operators are commonly referred to as monopole operators, because in Euclidean signature the gauge field singularity looks like that of a Dirac monopole \cite{Dirac:1931kp} or a non-Abelian generalization thereof \cite{tHooft:1974qc, Polyakov:1974ek, Goddard:1976qe}, as will be the case in this paper.  

Monopole operators are of interest for many reasons.\footnote{We restrict ourselves to the study of monopole operators in three-dimensional gauge theory.  In four-dimensional gauge theories one can define line operators by requiring the gauge field to asymptote to that of a monopole close to the line singularity \cite{Kapustin:2005py}.  These operators play an important role in the geometric Langlands program;  see, for instance, \cite{Kapustin:2006pk}.}  As explained by Polyakov in 3d Maxwell theory without matter, the proliferation of monopoles provides a mechanism for confinement \cite{Polyakov:1975rs,Polyakov:1976fu}.  If one adds enough fermionic or bosonic matter, however, the monopole operators become irrelevant in the renormalization group (RG) sense \cite{Murthy:1989ps, Borokhov:2002ib, Borokhov:2002cg, Metlitski:2008dw}, and in the deep infrared one finds a deconfined quantum critical theory.  As stressed in \cite{Hermele} (see also \cite{alicea}), the existence of these deconfined quantum critical theories relies on not having any monopole operators with scaling dimensions smaller than three.  Another reason why monopole operators are of interest comes from certain spin systems whose low energy physics is described by an emergent gauge theory, such as the $\CP^N$ model \cite{SVBSF,SBSVF}.   In these gauge theories, monopole operators can act as order parameters \cite{Read:1990zza,Read:1989zz} for second-order quantum phase transitions that cannot be described within the Landau-Ginzburg-Wilson paradigm \cite{SVBSF,SBSVF}.   The scaling dimensions of these monopole operators constitute interesting critical exponents that can also be computed directly from quantum Monte Carlo simulations of the spin systems \cite{2012PhRvL.108m7201K, 2013arXiv1302.1408D, 2013arXiv1307.0519B}.  

Monopole operators play a prominent role in supersymmetric theories as well.  For instance, in the model introduced by Aharony, Bergman, Jafferis, and Maldacena (ABJM) \cite{Aharony:2008ug} (see also the related model in \cite{Bashkirov:2010kz}), it was shown that when the Chern-Simons level is $k=1$ or $2$, there exist BPS monopole operators that are Lorentz vectors and have scaling dimension precisely equal to two \cite{ Benna:2009xd,Bashkirov:2010kz,Gustavsson:2009pm}.  In other words, these operators are conserved currents.  The existence of these conserved currents is what makes possible an enhancement in the amount of supersymmetry from ${\cal N} = 6$, which is the manifest supersymmetry of the ABJM Lagrangian, to ${\cal N} = 8$, which is the expected amount of supersymmetry that follows from M-theory.  In the same ABJM model, as well as in many other superconformal field theories with gravity duals \cite{Gaiotto:2009tk, Jafferis:2008qz, Jafferis:2009th, Benini:2009qs, Franco:2009sp, Martelli:2009ga, Imamura:2009ur, Benini:2011cma, Gulotta:2011si,Gulotta:2011aa}, monopole operators are also needed to match the spectrum of supergravity fluctuations in the bulk, and indeed certain holographic RG flows are triggered by operators with non-vanishing monopole charge \cite{Benna:2008zy}.   Monopole operators also play important roles in various supersymmetric dualities (see for example \cite{Aharony:1997gp, Aharony:2013dha, Aharony:2013kma}) and mirror symmetry (see for example \cite{Borokhov:2002cg, Borokhov:2003yu}), where the duality transformations map them to more conventional operators.  It is important to know the quantum numbers of these monopole operators if one wishes to check these dualities.

The goal of this paper is to study monopole operators in (non-supersymmetric) three-dimensional QCD with gauge group $G$ (which includes QED in the case $G=U(1)$) and $N_f$ flavors of fermions transforming in some representation of $G$.  We study these operators perturbatively to next-to-leading order in $1/N_f$.   While in the absence of matter fields, 3d gauge theory with any compact gauge group is believed to confine \cite{Polyakov:1975rs,Polyakov:1976fu,DHoker:1980az,Feynman:1981ss,Ambjorn:1982ts,Ambjorn:1984yu}, in the presence of a sufficiently  large number of matter fields the theory flows to an interacting conformal field theory (CFT) in the infrared (IR) \cite{Appelquist:1989tc}.  We are interested in studying monopole operators at this interacting IR fixed point.  We want to answer the questions:  How many independent monopole operators are there, and what are their quantum numbers?

Of course, starting with any monopole operator, we can take its product with various gauge-invariant local operators built out of the fermions, and construct new monopole operators this way.  Throughout this work, however, we will focus only on the ``bare'' monopole operators, namely those that cannot be written as such composites.  

Like in any CFT, one can use the state-operator correspondence to identify the space of local operators that can be inserted at a given point on $\R^3$ with the Hilbert space of states on $S^2 \times \R$.  In general, if one defines a monopole operator by requiring the gauge field to have a fixed behavior close to the insertion point, the resulting operator will not have a well-defined scaling dimension.  It is quite subtle, in general, to associate a certain monopole profile to an operator with well-defined scaling dimension, or, equivalently, to a certain energy eigenstate on $S^2 \times \R$.  We will discuss this subtlety in Section~\ref{OVERVIEW}.  As we will explain, in the large $N_f$ limit that we study, the subtlety is ameliorated by the fact that the gauge field fluctuations are suppressed, and one can indeed say that certain energy eigenstates on $S^2 \times \R$ correspond to monopole operators.  However, not all possible non-Abelian generalizations of the Dirac monopole can be associated with linearly-independent monopole operators of well-defined scaling dimension.

 As we will review in Section~\ref{BACKGROUND}, the non-Abelian generalization of a Dirac monopole involves several discrete parameters referred to as Goddard-Nuyts-Olive (GNO) charges \cite{Goddard:1976qe};  the monopoles are also classified topologically by $\pi_1(G)$ \cite{Lubkin:1963zz}, and unless $G = U(1)$, there are infinitely many GNO monopoles in the same topological class.  One of our main results is that, at least in the limit of large $N_f$, only certain sets of GNO charges yield independent bare monopole operators.  These sets are the ones for which the corresponding monopole background is stable, in the sense that it is a local minimum of the gauge effective action on $S^2 \times \R$.  A surprising result is that we find more than one independent monopole operator per topological class.

The monopole operators must transform as representations of the global symmetry group, which includes the conformal group and the flavor group.  The quantum numbers under the conformal group are the spin and the scaling dimension.  We devote a significant part of our work to computing the scaling dimensions of the monopole operators to second order in $1/N_f$.  For clarity, we first present our computations in the case where the gauge group is $G = U(N_c)$ and where the fermions are two-component complex spinors transforming in the fundamental representation of $U(N_c)$.  We later generalize our computations to other gauge groups and/or other representations of the fermion flavors.  Our work improves on existing results in the literature:  in the QED case, $N_c=1$, the leading large $N_f$ behavior of the monopole operator dimensions was found in \cite{Borokhov:2002ib};  for the monopole with lowest charge, the first subleading correction was computed in \cite{Pufu:2013vpa};  lastly, in $U(N_c)$ QCD with $N_f$ fundamental fermions, the dimensions of the monopole operators at leading order in $N_f$ were found in \cite{Borokhov:2003yu}.  Related computations can be found in \cite{Murthy:1989ps, Metlitski:2008dw, Pufu:2013eda} in non-supersymmetric theories, and in \cite{Borokhov:2002cg, Bashkirov:2010kz, Benna:2009xd, Benini:2011cma} in a supersymmetric context.

We also calculate the representations of the monopole operators under the flavor symmetry group, but only in the case where the gauge group is $G = U(N_c)$.  In this case, the flavor group is $SU(N_f)$.  We find that the monopole operators transform in irreducible representations of $SU(N_f)$ whose Young diagrams are rectangles with a number of rows equal to $N_f/2$ and a number of columns that depends on the GNO charges.  Our results apply, of course, in particular to the QED case, $N_c = 1$, and agree with the results of \cite{Borokhov:2002ib} for the monopole of smallest charge, but disagree with \cite{Borokhov:2002ib} on the monopole with two units of charge.  (Our computation for monopole operators with greater than two units of charge is a new result.)  We also find disagreement with the results of \cite{Borokhov:2003yu} in the case of the simplest GNO monopole in $U(N_c)$ QCD\@.

The rest of this paper is organized as follows.  Section~\ref{OVERVIEW} is a rather non-technical and highly recommended read that includes a definition of monopole operators in 3d gauge theory (in particular in QED$_3$ and QCD$_3$), as well as a discussion of how these operators can be studied via the state-operator correspondence.  In Section~\ref{DIMGEN} we set up the computation of the scaling dimensions of the monopole operators in $U(N_c)$ QCD with $N_f$ fundamental fermion flavors as an expansion in $1/N_f$.  To evaluate these scaling dimensions through order $O(N_f^0)$, we need to compute three functional determinants corresponding to the fluctuations of the fermions, of the ghosts, and of the gauge field.  We study the effective actions of these fields in Section~\ref{DETERMINANTS}.  The gauge field effective action is not positive-definite for all sets of GNO charges, thus making certain GNO monopole backgrounds unstable and the corresponding monopole operators poorly defined.  We discuss this stability issue in Section~\ref{STABILITY}.   For the monopoles that are stable, we collect the results on their scaling dimensions in Section~\ref{DIMENSIONS}.  Our results include the QED case $N_c = 1$ as a particular case.   In Section~\ref{QUANTUMNUMBERS}, we find how the monopole operators transform under the $SU(N_f)$ flavor symmetry group.  In Section~\ref{GENERAL} we generalize the results of Sections~\ref{DIMGEN}--\ref{DIMENSIONS} to other gauge groups and/or representations of the fermions.  Lastly, we end with a discussion of our results in Section~\ref{DISCUSSION}.  The reader interested only in the results can skip Sections~\ref{DIMGEN} and~\ref{DETERMINANTS}.

\section{Monopole operators via the state-operator correspondence}
\label{OVERVIEW}

We now start by addressing some of the preliminaries necessary for studying properties of monopole operators.  In Section~\ref{BACKGROUND}, we introduce classical monopole backgrounds in both Abelian and non-Abelian gauge theories. In Section~\ref{OPERATORS}, we then review the gauge theories of interest for this paper and highlight the role played by large $N_f$ in studying them.  Lastly, in Section~\ref{QMOS}, we introduce carefully the monopole operators that we will study in the rest of the paper, and discuss two ways of defining them that become equivalent in the limit of large $N_f$.

\subsection{Classical Monopole Backgrounds}
\label{BACKGROUND}

To begin discussing monopole operators more explicitly, it is convenient to first think about classical backgrounds.   The simplest and perhaps most familiar such backgrounds can be constructed in Abelian gauge theory as follows.  A monopole of charge $q$ in a $U(1)$ gauge theory in three dimensions is a rotationally-invariant background $\mathcal{A}$ for the gauge field $A$, where the field strength ${\cal F} = d\mathcal{A}$ integrates to $4 \pi q$ over any two-sphere surrounding the center of the monopole.  For a monopole at the origin, we can write the gauge field and its field strength in spherical coordinates as
 \es{AAbelian}{
   {\cal F} = q \sin \theta d\theta \wedge d\phi \qquad \Longrightarrow \qquad
    \begin{cases}
      \mathcal{A}^{(N)} = q (1- \cos \theta) d\phi & \text{if $\theta \neq \pi$} \,, \\
      \mathcal{A}^{(S)} = q (-1-\cos \theta) d\phi  & \text{if $\theta \neq 0$} \,,
    \end{cases}  
 }
where the expressions $\mathcal{A}^{(N)}$ and $\mathcal{A}^{(S)}$ satisfy $d\mathcal{A}^{(N)} = d\mathcal{A}^{(S)} = {\cal F}$ and are defined everywhere away from $\theta = \pi$ (the North chart) and away from $\theta = 0$ (the South chart), respectively.\footnote{The expression for the monopole background is given in the dual coordinate basis $\{dr,d\theta,d\phi\}$.  It is also common to present this background in the frame basis $\{\hat{r}$, $\hat{\theta}$, $\hat{\phi}\}$, where it takes the form $\mathcal{A}^{(N)}=q\frac{1-\cos \theta}{r\sin \theta} \hat{\phi}$ and $\mathcal{A}^{(S)}=-q\frac{1 + \cos \theta}{r \sin \theta} \hat{\phi}$ in the North and South charts, respectively.}  In the overlap region, these two expressions differ by a gauge transformation, $\mathcal{A}^{(N)} - \mathcal{A}^{(S)} = d \Lambda$, with gauge parameter $\Lambda = 2 q \phi$.  The condition that this gauge transformation is well-defined, namely that the same $U(1)$ group element $e^{i \Lambda}$ is associated both with $\phi$ and $\phi + 2 \pi$ (assuming that the $U(1)$ gauge group is a circle of circumference $ 2\pi$), implies the quantization condition $q \in \Z/2$.

For a gauge theory with gauge group $G$, one can define similar monopole backgrounds by simply considering a $U(1)$ subgroup of $G$ for which one can construct a monopole just like \eqref{AAbelian} \cite{Goddard:1976qe}.  For instance, if the gauge group is $G = U(N_c)$, we can write
 \es{NonAb}{
  \mathcal{A} = H (\pm 1 - \cos \theta) d\phi \,,
 }
where $H$ is a constant $N_c \times N_c$ Hermitian matrix, and the two possible signs correspond to the North and South charts, as in \eqref{AAbelian}.  Requiring that on the overlap region between the two charts, the expressions for the gauge field in \eqref{NonAb} should differ by a $U(N_c)$ gauge transformation, implies $e^{4 \pi i H} = 1$.  Making use of the gauge symmetry, we can always rotate $H$ to the diagonal form
 \es{HDiag}{
  H = \diag \{ q_1, q_2, \ldots, q_{N_c} \} \,,
 }
with $q_1 \geq q_2 \geq \cdots \geq q_{N_c}$.  The condition $e^{4 \pi i H} = 1$ implies $q_a \in \Z/2$ for all $a$.  

In general, for a gauge group $G$ there exist monopoles of the form \eqref{NonAb}, with $H$ an element of the Lie algebra of $G$.  Using the gauge symmetry, $H$ can always be rotated into the Cartan of the gauge group \cite{Goddard:1976qe}.  More explicitly, if $h_{a}$ (with $a = 1, \ldots, r$, where $r$ is the rank of the gauge group) is a basis for the Cartan subalgebra, then $H$ can be written as $H=\sum_{a=1}^{r}q_{a}h_{a}$ for some set of numbers $q_a$.  This rotation does not completely exhaust the gauge symmetry, as the Weyl group acts non-trivially on the $h_a$, and consequently on the $q_a$ as well.  One should therefore regard as equivalent any two sets of $q_a$ that are related by a Weyl group transformation.   The collection of numbers $\{ q_{a} \}$ are called GNO charges after the authors of \cite{Goddard:1976qe}.  The GNO charges must satisfy the quantization condition $\exp \left[ {4\pi i \sum_{a=1}^r q_{a}h_{a}} \right]=1$, where ``$\exp$'' is the usual exponential map defined on the Lie algebra and valued in the gauge group.  It is equivalent to say that $\exp \left[ {4\pi i \sum_{a=1}^r q_{a}h_{a}} \right]=1$ in any representation of the gauge group, where the $h_a$ are now matrices, and ``$\exp$'' is the matrix exponential.

It is worth noting that the GNO charges $q_a$ are not all conserved, or equivalently, they do not all provide a topological characterization of the singular gauge configurations \eqref{NonAb}.  Indeed, there exists a much coarser classification of monopoles by elements of the fundamental group $\pi_1(G)$ \cite{Lubkin:1963zz}. As we discussed, in order to make sure that we have a well-defined monopole background \eqref{NonAb}, we need to specify a gauge transformation (i.e.~an element of the gauge group $G$) in the overlap region between the North and South charts.  Since the overlap region has the topology of a circle, these gauge transformations are classified topologically by maps from a circle into the gauge group, or in other words by elements of $\pi_1(G)$.  In the case of $U(1)$ this topological charge is the same as the GNO charge $q$.  When $G = U(N_c)$, the topological charge can be derived from the current
 \es{CurrentNonAb}{
  J^\text{top}_\mu = \frac{1}{4 \pi} \epsilon_{\mu\nu\rho} \tr F^{\nu \rho} \,,
 }
which is conserved provided that the field strength $F_{\mu\nu}$ satisfies the (non-Abelian) Bianchi identity.  It follows that in this case it is only the sum $q_\text{top} \equiv \sum_{a=1}^{N_c} q_a $ that is a conserved topological charge, as opposed to all the individual $q_a$.  In other words, in a non-Abelian gauge theory there are several GNO monopoles (in fact, infinitely many) belonging to the same topological class.

\subsection{Three Dimensional Gauge Theories with Fermions}
\label{OPERATORS}

To make the discussion of monopoles more concrete, let us focus our attention on a specific class of three-dimensional gauge theories.  The class of theories whose monopole operators we want to study is QCD$_3$ with gauge group $G$ and $N_f$ fermion flavors transforming in some representation of $G$.   These theories have a parity anomaly if $N_f$ is odd \cite{Redlich:1983dv, Redlich:1983kn}, so we will restrict the following discussion to the case where $N_f$ is even.  When the gauge group is $G = U(N_c)$, the Lagrangian in Euclidean signature is
 \es{QCDActionPreview}{
  {\cal L} = \frac{1}{4 g_\text{YM}^2} \sum_{a, b = 1}^{N_c} (F_{\mu\nu}^{ab})^2 +\sum_{\alpha = 1}^{N_f} \sum_{a, b =1}^{N_c} \left[\psi_{a, \alpha}^{\dagger} \gamma^\mu (i \delta^{ab} \partial_\mu + A_{\mu}^{ab} ) \psi_{b, \alpha} \right] \,.
 }
Here, the indices $a, b$ are fundamental color indices, and $\alpha$ is a flavor index, while the two-component spinor indices on the fermions and gamma matrices are suppressed.  When $N_f$ is sufficiently large, this theory flows to a CFT in the infrared \cite{Appelquist:1989tc}.  This CFT can be studied by simply erasing the Yang-Mills term from the action, which by dimensional analysis is an irrelevant operator.  We can therefore write the Lagrangian for this CFT as
  \es{CFTLag}{
  {\cal L}_\text{CFT} = \sum_{\alpha = 1}^{N_f} \sum_{a, b =1}^{N_c} \left[\psi_{a, \alpha}^{\dagger} \gamma^\mu (i \delta^{ab} \partial_\mu + A_{\mu}^{ab} ) \psi_{b, \alpha} \right] \,.
 }
In the infrared theory, in the absence of the Yang-Mills term, the only role played by the gauge field is that of a Lagrange multiplier that imposes the constraint that the non-Abelian current vanishes, $j^{ab}_\mu(x) = 0$.\footnote{In the quantum theory, this constraint translates into the condition that $\langle \chi |j^{ab}_{\mu}|\chi \rangle=0$ for all physical states $|\chi\rangle$.} 

Being a conformal field theory, the fixed point \eqref{CFTLag} can be studied on any conformally flat space.  The Lagrangian on such a space differs from \eqref{CFTLag} only in that the partial derivative $\partial_\mu$ should be replaced by the covariant derivative $\nabla_\mu$.  The gauge field components $A_\mu^{ab}$, considered as components of a one-form in the coordinate basis, remain invariant under the Weyl transformation used to map the theory on $\R^3$ to that on a different conformally flat space.

For large numbers of fermion flavors, this theory simplifies. Indeed, integrating out the matter fields, the effective action for the gauge field takes the form
 \es{GaugeEff}{
  S_\text{eff}[A] = -N_f \log \det (i \delta^{ab}  \gamma^\mu \nabla_\mu +  \gamma^\mu A_{\mu}^{ab} ) \,.
 }
The factor of $N_{f}$ in front of the action means that we can perform a semiclassical expansion about any saddle point of \eqref{GaugeEff} that is also a local minimum of the effective action, with $N_{f}$ playing the role of $1/\hbar$. As in any such expansion, the typical size of fluctuations about the saddle is of order $\sqrt{\hbar}$, or $1/\sqrt{N_f}$ in our case, as can be seen from expanding \eqref{GaugeEff} around the saddle point configuration and examining the term quadratic in the fluctuations.\footnote{Note that for a saddle point of the effective action that is not a local minimum, the fluctuations would grow with time and eventually become large.
}

The monopole backgrounds introduced in Section~\ref{BACKGROUND} are rotationally symmetric about their center and invariant under conformal inversions, as one can easily check.  These properties guarantee that they are saddle points of the effective action \eqref{GaugeEff} on any conformally flat space.  It is not guaranteed, however, that they are all local minima of the effective action, which is a fact that will become important in studying monopoles in the quantum theory.

\subsection{Quantum Monopole Operators}
\label{QMOS}

In the previous two subsections we established the existence of monopole saddles in a class of three-dimensional gauge theories.  We now explain how to define local operators with well-defined scaling dimensions that are associated with these saddles in the infrared CFT\@. As we explain below, for small $N_{f}$ there is a tension between defining an operator that corresponds to a classical monopole background, and defining an operator with definite scaling dimension. In particular, the operator most easily identified with a monopole background does not have definite scaling dimension, but rather corresponds to a sum of such operators.  At large $N_{f}$, however, this tension is alleviated, and we can indeed associate an operator of fixed scaling dimension to a monopole background.

Before delving into monopole operators, let us briefly review what we know about local operators in the IR theories of interest. The most familiar local operators are those that can be written as gauge-invariant combinations of the fundamental fields, such as $\mathcal{O}=\sum_{a}\psi^{\dagger}_{a}\psi_{a}$.  These operators are sometimes referred to as order operators \cite{Kapustin:2005py}. In addition to order operators, one can also define local disorder operators, which cannot be written simply in terms of the fundamental fields. Rather than being defined as local products of fields, disorder operators can be thought of as creating singularities for the fundamental fields. In the context of the path integral, we can define a disorder operator inserted at a point $p$ by integrating only over field configurations that asymptotically approach a prescribed singular configuration in a neighborhood of $p$.\footnote{Concretely, this procedure can be realized by cutting out a ball of radius $\epsilon$ about the point $p$, fixing the boundary conditions for fields on the surface of this ball, and only integrating over fluctuations outside the radius $\epsilon$. Away from the insertion point, this regularized disorder operator acts just like a local operator. A similar prescription in four dimensions was used in \cite{Kapustin:2005py}. \label{footnoteReg}}

For the classical monopole backgrounds described in Section \ref{BACKGROUND}, the gauge field has such a localized singularity. We can thus define a local disorder operator associated to a monopole background by requiring that the gauge field asymptotically approach that of the classical background near the insertion point of the operator. In this way we can indeed associate a quantum operator with a classical monopole background. Unfortunately, the disorder operator so defined does not transform nicely under the conformal symmetries present at the infrared fixed point.  In a CFT it is convenient to work in a basis of local operators with definite spin and scaling dimension.  As we will see, the disorder operator defined above does not have a definite scaling dimension, but rather can be written as a sum of operators with definite scaling dimension.

To see that the disorder operator cannot in general have a well-defined scaling dimension, it is convenient to change perspectives from operators on $\mathbb{R}^{3}$ to states on $S^{2}\times\mathbb{R}$ by using the state-operator correspondence.  In a CFT, the state-operator correspondence maps operators inserted at the origin of $\mathbb{R}^{3}$ to normalizable states on $S^{2}\times\mathbb{R}$. The $\R$ coordinate $\tau$ is interpreted as Euclidean time and is related to the radial coordinate $r$ on $\R^3$ through $r = e^\tau$.  The scaling dimension of an operator on $\R^3$ is identified with the energy of the corresponding state on $S^2$. Restricting to disorder operators, the correspondence identifies the disorder operator defined by boundary conditions at a point in $\mathbb{R}^{3}$ to the state on $S^{2}\times\mathbb{R}$ given by a wave-functional on field space with delta-function support on the classical field configuration at $\tau=-\infty$.\footnote{If we regularize the disorder operator by smearing it over a sphere of radius $\epsilon$, as in footnote~\ref{footnoteReg}, the wave-functional would have delta-function support on the classical field configuration at $\tau = \log \epsilon$.}

This state is not an energy eigenstate, but rather a superposition of energy eigenstates. The wave function, which is localized about the classical configuration at early times, spreads out at late times. The corresponding operator on $\mathbb{R}^{3}$ is therefore a sum of operators with definite scaling dimension. In a generic theory, there is no principle that singles out any one operator in this sum, and correspondingly there is a significant distinction between a disorder operator defined by boundary conditions and an operator of definite scaling dimension.

At large $N_{f}$, however, the situation is better. The monopole background is a classical saddle, and for large $N_{f}$ the gauge fluctuations are suppressed. If the saddle is stable, the state corresponding to the disorder operator is close to an energy eigenstate. It is the operator of definite scaling dimension corresponding to this energy eigenstate that we refer to as the monopole operator for the remainder of this paper.\footnote{In the $U(1)$ case each operator in the decomposition of the disorder operator must carry the same topological charge.  As such, even away from large $N_{f}$, there is a natural operator with definite scaling dimension to associate with a monopole background, namely the operator corresponding to the lowest energy state with the given topological charge.  As discussed in Section~\ref{BACKGROUND}, for non-Abelian theories there are many classical backgrounds with the same topological charge, and so one would be able to identify only one monopole operator per topological class this way.  In supersymmetric theories, however, it may be possible to identify BPS monopole operators with certain GNO backgrounds after performing a $Q$-exact deformation of the theory to weak coupling (see, for example, \cite{Bashkirov:2010kz, Benini:2011cma}).}

In the next section we explain how to use the path integral on $S^{2}\times\mathbb{R}$ to calculate the energy of eigenstates associated with stable saddles, and in Section~$\ref{STABILITY}$ we determine which saddles are stable.

\section{Free energy on $S^2 \times \R$}
\label{DIMGEN}

In the previous two sections we discussed at some length the precise definition of monopole operators at the infrared conformal fixed point of QCD$_3$ with many flavors of fermions.  In summary, for the GNO backgrounds \eqref{NonAb} on $S^2 \times \R$ that are local minima of the gauge field effective action, and only for those backgrounds, there exist several degenerate lowest-energy states whose wavefunctions are highly peaked around the saddle \eqref{NonAb};   it is these states that, via the state-operator correspondence, get mapped to the bare monopole operators on $\R^3$ whose properties we want to study.  We will refer to these states on $S^2 \times \R$ as ``ground states'' in the presence of the monopole flux \eqref{NonAb}.  The use of the term ``ground states'' can be justified only in large $N_f$ perturbation theory, where one can define a Fock space of states for every stable GNO configuration, and these Fock spaces mix only non-perturbatively in $1/N_f$.

One aspect of the state-operator correspondence is that the scaling dimensions of operators on $\mathbb{R}^{3}$ are equal to the energies of the corresponding states on $S^{2}\times\mathbb{R}$.  In particular, the scaling dimension $\Delta$ of a bare monopole operator equals the ground state energy in the presence of some constant GNO flux through the two-sphere.  The goal of this section is to exploit this equality in order to calculate the scaling dimensions $\Delta$.  In later sections, we will calculate the other quantum numbers of the bare monopole operators.

The ground state energy in the presence of some constant GNO flux can in turn be calculated by performing the path integral on $S^{2}\times\mathbb{R}$. More explicitly, we have
 \es{DeltaMDef}{
  \Delta =-\log Z[\mathcal{A}] \ \equiv \ F[\mathcal{A}] \,,
 }
where $Z[\mathcal{A}]$ is the Euclidean partition function on $S^{2}\times\mathbb{R}$ in the presence of the background $\mathcal{A}$, and $F[{\cal A}]$ is the corresponding ground state energy (or free energy).\footnote{The expression \eqref{DeltaMDef} should be taken to include only perturbative contributions in the $1/N_f$ expansion.  All non-perturbative contributions in $1/N_f$ should not be taken into account.}  In principle, the quantity $\log Z[{\cal A}]$ should be understood as the limit
 \es{ZLimit}{
   \log Z[{\cal A}]=\lim_{\beta\to\infty}\frac{1}{\beta}\log Z_\beta [\mathcal{A}]  \,,
 }
where $Z_\beta[{\cal A}]$ is a similar partition function on $S^2 \times S^1$ calculated after first compactifying the $\R$ direction into a circle of circumference $\beta$.  In practice, as we will see, it is easy to isolate the leading term in the large $\beta$ expansion of $\log Z_\beta[{\cal A}]$ while working directly on $S^{2}\times\mathbb{R}$.

Our procedure for calculating $F[\mathcal{A}]$ consists of three steps:
 \begin{enumerate}
  \item We perform the path integral over the matter fields.  Integrating out the matter fields generates a gauge-invariant effective action for the gauge fluctuations and leads to a sensible $1/N_f$ expansion.
  \item We fix the gauge by introducing Faddeev-Popov ghosts.
  \item We evaluate the path integral over the gauge fluctuations and the ghosts at next-to-leading order in $1/N_f$.
 \end{enumerate}
We now provide an explanation of this procedure, while the next section is devoted to the details of the calculation.

\subsection{Setup}

As discussed previously, the IR conformally-invariant action for QCD with $U(N_c)$ gauge group and $N_f$ flavors of complex two-component fermions is
 \es{QCDAction}{
  S = \int d^3x \, \sqrt{g} \sum_{\alpha = 1}^{N_f} \sum_{a, b = 1}^{N_c} 
    \left[\psi_{a, \alpha}^{\dagger} \gamma^\mu (i \delta^{ab} \nabla_\mu + {\cal A}_{\mu}^{ab} + a_\mu^{ab} ) \psi_{b, \alpha} \right] \,,
 }
where, in anticipation of having to study this theory in the presence of a background monopole flux, we split the non-Abelian gauge field $A_{\mu}^{ab}$ into a sum between a background ${\cal A}_\mu^{ab}$ and fluctuations $a_\mu^{ab}$.  Here, the indices $a, b$ are color indices, and $\alpha$ is a flavor index.  The spinor indices on the fermions and gamma matrices are suppressed.  The action \eqref{QCDAction} can be more compactly written as
 \es{QCDMBAction}{
  S[\mathcal{A};a,\psi^{\dagger},\psi] 
    &=S_{0}[\mathcal{A};\psi^{\dagger},\psi]+\int d^{3}x\sqrt{g}\sum_{a,b}a_{\mu}^{ab}j^{\mu}_{ba} \,,
 }
where $S_{0}[\mathcal{A};\psi^{\dagger},\psi]$ is the action \eqref{QCDMBAction} with gauge fluctuations set to zero, and
 \es{jmuDef}{
   j^{\mu}_{ba} =\sum_{\alpha}:\psi^{\dagger}_{a,\alpha}\gamma^{\mu}\psi_{b,\alpha}:
 }
is the non-Abelian covariantly conserved current. Double colons stand for normal ordering.  As in any gauge theory, we can write the partition function as
\es{PartitionFunction}{
Z\le[{\cal A}\ri]=\frac 1{ \Vol{(\mathbb{G})}} \int Da\, D\psi^\dagger\,  D\psi \, \exp \biggl[ -S\le[{\cal A}; a,\psi^\dagger,\psi\ri]  \biggr] \,,
}
with $\Vol(\mathbb{G})$ being the volume of the group of gauge transformations, which we need to divide by because we do not want to count gauge-equivalent configurations multiple times.

Let us set up our conventions for this calculation. We write the standard line element on $\R^3$ in spherical coordinates as
\es{R3Metric}{
   ds_{\R^3}^2 &= d\vec{x}^2= e^{2\tau}\,\le[d\tau^2+ d\theta^2 + \sin^2 \theta d \phi^2 \ri] \qquad  \vec{x} \equiv   e^\tau \begin{pmatrix}
     \sin \theta \cos \phi & \sin \theta \sin \phi & \cos \theta 
    \end{pmatrix} \,.
    }
We want to calculate the partition function on $S^2\times \R$. The metric on $S^2\times \R$ is obtained by rescaling the $\R^3$ metric \eqref{R3Metric} by $e^{-2\tau}$:
\es{RTimesS2}{
ds_{S^2\times \R}^2=d\theta^2+\sin^2\theta\, d\phi^2+d\tau^2 \ .
}
Recall that the dynamics of a CFT is insensitive to such a rescaling. We will be doing calculations with spinors on the curved space $S^2\times \R$, hence we need to specify a frame $e^i$. We obtain the frame by the conformal transformation of the standard frame $e^i=dx^i$ on $\R^3$
\es{Frame}{
 e^i = e^{-\tau} dx^i \ .
}
We choose the set of gamma matrices  $\gamma^i = \sigma^i$, where the $\sigma^i$ are the Pauli matrices. All subsequent formulae for spinors are understood to follow these conventions. A point on $S^2\times \R$ will be denoted by  $x=(\tau,\theta,\phi)$. Sometimes we will also use the decomposition $x=(\tau,\hat{n})$, where $\hat{n}$ is a unit vector pointing to a point on $S^2$. The covariant derivative on $S^2\times \R$ will be denoted by $\nabla_\mu$. The gauge covariant derivative for a fundamental fermion $\psi^a$ and current $j_\mu^{ab}$ (which transforms in the adjoint representation of $U(N_c)$) is
\es{GaugeCovariant}{
\le[D_\mu^{(A)} \psi\ri]^a&=\sum_{b=1}^{N_c}\le(\nabla_\mu\, \delta^{ab}- i A_\mu^{ab}\ri)\, \psi^b \,,\\
\le[D_\mu^{(A)} j_\nu\ri]^{ab}&=\nabla_\mu\, j_\nu^{ab}- i \le[A_\mu,\, j_\nu\ri]^{ab}\ ,
}
where $ \le[A_\mu,\, j_\nu\ri]^{ab}=\sum_{c=1}^{N_c}\le(A_\mu^{ac}\, j_\nu^{cb}- j_\nu^{ac}\, A_\mu^{cb}\ri)$ is the matrix commutator.

As explained in the previous sections, for a $U(N_{c})$ gauge group, the most general monopole background can be taken to be
 \es{MonopoleGeneral}{
  {\cal A}^{ab} &=\diag \{q_1, q_2, \ldots q_{N_c} \}  {\cal A}^{U(1)} \,, \qquad
  {\cal A}^{U(1)} \equiv
  \begin{cases} 
    (1 - \cos \theta) d\phi & \text{if $\theta\neq\pi$,}\\
    (-1 - \cos \theta) d\phi & \text{if $\theta\neq0$,} \\
  \end{cases}
 }
with $q_1 \geq q_2 \geq \ldots \geq q_{N_c}$, and $ q_{a}\in\mathbb{Z}/2$.

\subsection{Gauge Field Effective Action} \label{GAUGEEFFECTIVE}

The first step in our general procedure for evaluating the ground state energy $F[{\cal A}]$ is to integrate out the matter fields.  Doing so yields a gauge-invariant effective action for the gauge field fluctuations.  The gauge effective action is defined in such a way that the partition function is simply
 \es{PartFluct}{
  Z[\mathcal{A}] = \frac{1}{\Vol(\mathbb{G})}\int Da\, \exp \Bigl[ -S_\text{eff}[a] \Bigr] \,.
 }  
Comparing with \eqref{PartitionFunction}, using the decomposition of the action in \eqref{QCDMBAction}, and expanding in powers of $a$, one can write
 \es{Seff}{
  S_\text{eff}[a] = -\log Z_0[{\cal A}] + \sum_{n=1}^\infty \frac {(-1)^{n+1}}{n!} \left\langle \left( \int d^3 x\,  \sqrt{g(x)} a_\mu^{ab}(x) j^\mu_{ba}(x) \right)^n \right\rangle_\text{conn} \,,
 }
where the correlators on the right-hand side are evaluated using the action of free fermions in the background ${\cal A}$, namely the action $S_0[{\cal A}; \psi, \psi^\dagger]$ introduced above.  In other words, the effective action $S_\text{eff}[a]$ is the generating functional of connected correlators of the current operator $j^\mu_{ba}$ in this theory of free fermions.  The quantity $Z_0[{\cal A}]$ appearing in \eqref{Seff} is the partition function associated with $S_0[{\cal A}; \psi, \psi^\dagger]$;  it is just a Gaussian integral, which evaluates to
 \es{GotZ0}{
  Z_0[{\cal A}] \equiv \int D\psi^\dagger D\psi \exp \Bigl[-S_0[{\cal A}; \psi, \psi^\dagger] \Bigr]
   = \left(\det (i \slashed{D}^{({\cal A})}) \right)^{N_f} \,.
 }
Here, the subscript $({\cal A})$ denotes a background gauge covariant derivative, as in \eqref{GaugeCovariant} with $A$ replaced by ${\cal A}$.  

For us, the GNO monopole background \eqref{MonopoleGeneral} is static as well as invariant under rotations and time reversal, and therefore the one-point function of the current operator must vanish, $\langle j^\mu_{ba}(x) \rangle = 0$ (see also the last paragraph of Section~\ref{OPERATORS}). Therefore, the term linear in $a$ in \eqref{Seff} vanishes.  In general, the term  quadratic in the $a$ does not vanish, and its coefficient is given by the current-current correlator
 \es{GotK}{
  K^{\mu\nu}_{ab, cd} (x, x') \equiv -\langle j^\mu_{ba}(x) j^\nu_{dc}(x') \rangle_\text{conn}  \,.
 }
This current-current correlator should be thought of as an integration kernel that defines an operator on the space of square-integrable one-forms on $S^2 \times \R$.

The kernel $K^{\mu\nu}_{ab, cd}(x, x')$ can be written more explicitly in terms of a quantity $G_q(x, x')$, which can be identified with the Green's function of a single fermion in an Abelian gauge theory in the presence of a monopole background \eqref{AAbelian}, namely
 \es{GOneFermion}{
  G_{q}(x, x') = \langle \psi(x)  \psi^\dagger (x') \rangle = \left\langle x \middle| \frac{1}{i \slashed{\nabla} + q \slashed{\cal A}^{U(1)}} \middle| x' \right\rangle \,.
 }
(This theory would be described by the action \eqref{QCDAction} with $N_c = N_f = 1$ and $a = 0$.)  Indeed, substituting the normal-ordered expression \eqref{jmuDef} into \eqref{GotK}, and noticing that the contractions between the fermions take the form $\langle \psi_{a, \alpha}(x) \psi_{b, \beta}^\dagger(x') \rangle = \delta_{\alpha \beta} \delta_{ab} G_{q_a}(x, x')$, one obtains
 \es{GotKAgain}{
  K^{\mu\nu}_{ab, cd}(x, x') = N_{f}\delta_{bc}\delta_{ad}\mathcal{K}_{q_{b},q_{a}}^{\mu\nu}(x,x') \,, 
 }
with
 \es{calKDef}{
  {\cal K}^{\mu\nu}_{q_b, q_a}(x, x')  \equiv - \tr \left(\gamma^\mu G_{q_b} (x, x') \gamma^\nu G_{q_a}^\dagger(x, x') \right) \,.
 } 

With the expression \eqref{GotKAgain} in hand, we can write the effective action for the gauge field fluctuations as
 \es{SeffFinal}{
  S_\text{eff}[a] = N_f \left[ \tr \log (i \slashed{D}^{({\cal A})}) + \frac 12 \int d^3x\, d^3x' \sqrt{g(x)} \sqrt{g(x')} a^{ab\dagger}_\mu(x) 
    {\cal K}^{\mu\nu}_{q_a, q_b} (x, x')   a^{ab}_\nu(x')  + \cdots \right] \,.
 }
The ellipses denote terms with higher powers of $a$, which one can easily show are also proportional to $N_f$.  We are now in business.  That the quadratic part of the action is proportional to $N_f$ means that the typical gauge field fluctuations are $a \propto 1/\sqrt{N_f}$, and we can calculate $Z[{\cal A}]$ (see \eqref{PartFluct}) approximately at large $N_f$ using a saddle point approximation.

We can now try to perform the integral over the gauge fluctuations by keeping only the terms up to quadratic order in $a$ in $S_\text{eff}[a]$, and write down the free energy on $S^2 \times \R$ as:
 \es{FreeNaive}{
  F[{\cal A}] = - N_f \tr \log (i \slashed{D}^{({\cal A})}) 
   + \frac 12 ``\tr \log K" + {\cal O}(1/N_f)  \,.
 }
The quotation marks are meant to emphasize that the ${\cal O}(N_f^0)$ term is only rough, because we ignored the issue of gauge invariance when we performed the integral over the gauge fluctuations.  To obtain a more explicit answer, we now proceed to a more careful analysis of gauge invariance.

Just like the original action \eqref{QCDAction}, the effective action $S_\text{eff}[a]$ is invariant under gauge transformations that in the gauge sector act as
 \es{GaugeTransf}{
  ({\cal A}_\mu + a_\mu)  \to  i U \partial_\mu U^\dagger + U ({\cal A}_\mu + a_\mu) U^\dagger \,.
 }
Correspondingly, the integrand in \eqref{PartFluct} has flat directions corresponding to these gauge transformations.  Therefore, one cannot simply identify the functional integral $\int Da \exp[ -a K a]$  with the determinant of the kernel $K$, as this kernel has many eigenvalues that vanish.

It is most convenient to work in background field gauge by imposing the condition:
\es{GaugeCond}{
D_\mu^{({\cal A})} a^\mu=0 \,.
}
This condition distinguishes one gauge configuration in every gauge-equivalence class, so if we restrict our integral over $a$ to configurations that satisfy \eqref{GaugeCond} then the integrand $e^{-S_\text{eff}[a]}$ will no longer have any flat directions.  The condition \eqref{GaugeCond} does not exhaust, however, the group of all possible gauge transformations, because there are residual gauge transformations that leave $a$ completely untouched.  These residual gauge transformations form the isotropy group $\mathbb{H}^{(\cal A)}$.  The Faddeev--Popov trick is to insert
\es{FPDet}{
1=\Det'\le(-D_\mu^{({\cal A})} D^{({\cal A}+a)\mu}\ri)\times {1\ov {\rm Vol}\le(\mathbb{H}^{(\cal A)}\ri)}\int DU\,  \delta\le[D_\mu^{({\cal A})} a^{U,\mu}\ri] \ ,
} 
into the path integral, where $\Det'$ denotes the functional determinant with zero modes omitted, and $a_\mu^U =  i U \partial_\mu U^\dagger + U ({\cal A}_\mu + a_\mu) U^\dagger  - {\cal A}_\mu$ is the gauge transformed $a_\mu$.  Changing variables\footnote{Note that both the measure and the action $S_\text{eff} [a]$ are invariant under this change of variables. } $a\to a^U$ in the partition function \eqref{PartFluct}, inserting \eqref{FPDet}, and then renaming $a^U  \to a$ gives:
\es{PartitionFunction2}{
Z\le[{\cal A}\ri]={1\ov {\rm Vol}\le(\mathbb{H}^{(\cal A)}\ri)} \int Da\, e^{-S_\text{eff} [a] - S_\text{FP}[a]} \,  \left( \delta\le[D_\mu^{({\cal A})} a^{\mu}\ri] \, \sqrt{\Det'\le(-D_\mu^{({\cal A})} D^{({\cal A})\mu}\ri)} \right)\,,
}
with 
 \es{SFPDef}{
  S_\text{FP}[a] &=- \frac 12 \Tr' \log\le(-D_\mu^{({\cal A} + a)} D^{({\cal A}+a)\mu}\ri) \,,
 }
where $\Tr'$ is a trace over the non-zero modes. The factor in the parenthesis in \eqref{PartitionFunction2} multiplying the delta-function is precisely the inverse of the Jacobian factor that one obtains when taking $D_\mu^{({\cal A})}$ outside of the delta-function.  We can therefore perform the path integral \eqref{PartitionFunction2} by integrating only over configurations that satisfy \eqref{GaugeCond} (and thus removing by hand all the flat directions), provided that we supplement the effective action $S_\text{eff}[a]$ by a term $S_\text{FP}[a]$ exhibited in \eqref{SFPDef} that comes from the Faddeev-Popov procedure.  Just like $S_\text{eff}[a]$, $S_\text{FP}[a]$ can be expanded in powers of $a$:
 \es{FPExpansion}{
  S_\text{FP}[a] = - \frac 12 \Tr' \log\le(-D_\mu^{({\cal A})} D^{({\cal A})\mu}\ri)  + {\cal O}(a^2) \,.
 }
The term linear in $a$ in this expansion vanishes by an argument based on the symmetries of the background \eqref{MonopoleGeneral} similar to the one that showed that the linear term in $a$ in \eqref{SeffFinal} vanished.

Before presenting the answer for the free energy, we note that the factor of $1/\Vol(\mathbb{H}^{(\cal A)})$ should be ignored.  This factor would be relevant if we computed the partition function on $S^2 \times S^1$, where the $S^1$ circle has circumference $\beta$, because in this case every generator of $\mathbb{H}^{(\cal A)}$ would contribute a factor proportional to $\beta$ to $\Vol(\mathbb{H}^{(\cal A)})$.  Thus, $\log Z_\beta[{\cal A}]$ would receive a contribution proportional to $\log \beta$ from every generator of $\mathbb{H}^{(\cal A)}$.  However, these contributions disappear when we consider the limit in \eqref{ZLimit}.

Evaluating \eqref{PartitionFunction2} in the saddle point approximation, we have 
 \es{FFinal}{
  F[{\cal A}] &=N_f\, F_0[{\cal A}] +\delta F[{\cal A}] + {\cal O}(1/N_f) \,, \\
 \delta F[{\cal A}] &=F_\text{FP}[{\cal A}]+F_\text{gauge}[{\cal A}] \,,
 }
 where $F_0[{\cal A}]$ is the fermion determinant
 \es{FermionDetDef}{
 F_0[{\cal A}]= -  \Tr \log (i \slashed{D}^{({\cal A})}) \ .
 }
$\delta F[{\cal A}] $ denotes the subleading term in the free energy. It is a sum of two terms, namely $F_\text{FP}[{\cal A}]$, which is the Faddeev--Popov determinant
 \es{FPDetDef}{
 F_\text{FP}[{\cal A}]= - \frac 12 \Tr' \log\le(-D_\mu^{({\cal A})} D^{({\cal A})\mu}\ri) \ ,
 }
 and $F_\text{gauge}[{\cal A}]$, which is the gauge fluctuation determinant
  \es{GaugeDef}{
 F_\text{gauge}[{\cal A}]= \frac 12 \Tr' \log K \ .
 }
The fermion, Faddeev--Popov, and gauge field determinants will be calculated in Sections~\ref{subsec:FermionDet},~\ref{subsec:FPDet}, and~\ref{GAUGEDET}, respectively.

\section{Functional determinants}
\label{DETERMINANTS}

\subsection{The fermion determinant}\label{subsec:FermionDet}

We now start by calculating more explicitly the leading term in \eqref{FFinal}, the fermion determinant $F_0[{\cal A}]$ defined in~\eqref{FermionDetDef}.  This term arises from evaluating the partition function $Z_0[{\cal A}] = e^{- N_f \,F_0[{\cal A}]}$ of non-interacting fermions in the background \eqref{MonopoleGeneral}.  See \eqref{GotZ0}.

Examining the action $S_0[{\cal A}; \psi, \psi^\dagger]$ more closely, we see that because we have taken the monopole background \eqref{MonopoleGeneral} to be diagonal in the color indices, it is not only the fermions of different flavor that decouple from one another, but also those of different color.  Each fermion $\psi_{a, \alpha}$ is only coupled to an Abelian monopole background of charge $q_a$.   Because the fermions are non-interacting, the ground state energy can be written as
 \es{FreeLeading}{
  F_0[{\cal A}] = \sum_{a=1}^{N_c} F_0(q_a) \,,
 }
where $F(q)$ denotes the ground state energy of a single fermion in an Abelian monopole background of charge $q$.   The quantity $F(q) = -\log Z(q)$ can be computed from the partition function corresponding to this free fermion,
 \es{Zq}{
  Z(q) = \int D\psi D\psi^\dagger e^{-S(q)} \,, \qquad
   S(q) = \int d^3x \, \sqrt{g(x)} \psi^{\dagger} \gamma^{\mu}(i\nabla_{\mu}+q \mathcal{A}_{\mu}^{U(1)})\psi \,.
 }
This computation was performed in \cite{Borokhov:2002ib, Pufu:2013vpa} as part of studying the scaling dimensions of monopole operators in $U(1)$ gauge theory with $N_f$ fermion flavors.   We now review this computation briefly, partly because such a review will keep our presentation self-contained, and partly because in doing so we will also introduce some notation that will become useful in the following sections. 

In order to evaluate the integral in \eqref{Zq}, we should first decompose the fermion field $\psi$ into a suitable basis of spinor fields.  Since translations in the Euclidean time direction are a symmetry of the action, it is convenient to consider modes with harmonic time dependence, $\psi \propto e^{-i \omega \tau}$.  Finding a basis for the angular dependence of $\psi$ requires more thought.  If $\psi$ were instead a complex scalar experiencing the same monopole flux $q {\cal A}^{U(1)}$, one could use the basis of monopole harmonics $Y_{q, \ell m_\ell}(\hat n)$, which were defined in \cite{Wu:1976ge,Wu:1977qk} (see also Appendix~\ref{App:Identities}) as eigenfunctions of the gauge-covariant Laplacian on $S^2$.  Here, $\ell \geq \abs{q}$ is the angular momentum, and $m_\ell$ ranges from $-\ell$ through $\ell$.  To find a complete basis for a field of a different spin $s$, we can work in the frame \eqref{Frame} obtained by conformal transformation from $\R^3$ and expand every component of the spin $s$ field in terms of the monopole harmonics $Y_{q, \ell m}(\hat n)$.  More conveniently, we can use the usual angular momentum addition rules to work in a basis where the quantum numbers are $\{j, m, \ell, s\}$, $j$ being the total angular momentum and $m$ the eigenvalue of $J_3$.  For spinor fields where $s=1/2$, we have $j = \ell - 1/2$ or $j = \ell +1/2$, and we can define
 \es{TSDef}{
  T_{q, jm} &= \begin{pmatrix}
   \sqrt{\frac{j + m}{2j}} Y_{q, j-\frac 12, m-\frac 12} \\
   \sqrt{\frac{j - m}{2j}} Y_{q, j-\frac 12, m+\frac 12} 
  \end{pmatrix} \,, \qquad \qquad j = \ell + \frac 12 \geq \abs{q} + \frac 12\,, \\
 S_{q, jm} &= \begin{pmatrix}
  -\sqrt{\frac{j+1-m}{2(j+1)}} Y_{q, j+\frac 12, m-\frac 12 }\\
  \sqrt{\frac{j+1+m}{2(j+1)}} Y_{q,j+\frac 12, m+\frac 12}
 \end{pmatrix} \,, \qquad
    j = \ell - \frac 12 \geq \abs{q} - \frac 12 \,,
 }
where $Y_{q, \ell m}$ are the scalar monopole harmonics.  Note that for total angular momentum $j = \abs{q}-1/2$ we have only the $S_{q, jm}$ harmonics, which in this case have orbital angular momentum $\ell = \abs{q}$, while for larger $j$ we have both $S_{q, jm}$, with orbital angular momentum $\ell = j+1/2$, and $T_{q, jm}$, with orbital angular momentum $\ell = j-1/2$.  For $q = 0$, the $S_{q, jm}$ start at $j = 1/2$.

Expanding the fermion $\psi(x)$ in the basis \eqref{TSDef},
 \es{FermionDecomp}{
  \psi(x)=\int\frac{d\omega}{2\pi} \sum_{j\geq \abs{q}-1/2}\sum_{m=-j}^{j}
    \left(\Psi^{(S)}_{j m}(\omega)S_{q, j m}(\hat n)+\Psi^{(T)}_{j m}(\omega)T_{q, j m}(\hat n)\right)e^{-i\omega \tau} \,,
 }
with anti-commuting coefficients $\Psi^{(S)}_{j m}(\omega)$ and $\Psi^{(T)}_{j m}(\omega)$, one finds that the action $S(q)$ defined above can be written in almost diagonal form, because the gauge-covariant Dirac operator only mixes the modes $S_{q, jm}$ and $T_{q, jm}$ with the same $j$ and $m$ \cite{Borokhov:2002ib,Pufu:2013vpa}:
 \es{zeroact}{
   S(q) = \int\frac{d\omega}{2\pi}\sum_{j,m}
     \begin{pmatrix} \Psi^{(T)}_{j m}(\omega) \\ \Psi^{(S)}_{j m}(\omega)\end{pmatrix}^{\dagger}
     \widetilde {\bf G}_{q, j}(\omega) 
     \begin{pmatrix} \Psi^{(T)}_{j m}(\omega) \\ \Psi^{(S)}_{j m}(\omega)\end{pmatrix} \,.
 }
Here, $\widetilde {\bf G}_{qj}(\omega)$ is a matrix that can be identified with the inverse propagator.  When $j > \abs{q}-1/2$, $\widetilde {\bf G}_{q,j}(\omega)$ is a $2 \times 2$ matrix given by \cite{Borokhov:2002ib,Pufu:2013vpa}
 \es{Gqj}{
   \widetilde {\bf G}_{q, j}(\omega) = \begin{pmatrix}
   -\displaystyle{\frac{q\, \omega}{j + \frac 12}} & \left(i -  \displaystyle{\frac{\omega}{j + \frac 12}} \right) \displaystyle{\sqrt{\left(j+\frac 12 \right)^2 - q^2} }  \\[18pt]
   -\left(i + \displaystyle{\frac{\omega}{j+ \frac 12}} \right)  \displaystyle{\sqrt{\left(j+\frac 12 \right)^2 - q^2} } & \displaystyle{\frac{q\, \omega}{j+\frac 12}}
  \end{pmatrix} \,;
 }
when $j = \abs{q} - 1/2$, $T_{q, jm}$ does not exist, and $\widetilde {\bf G}_{q,j}(\omega)$ should be thought of as a $1 \times 1$ matrix equal to the bottom-right entry of \eqref{Gqj}, namely $\widetilde {\bf G}_{q,j}(\omega) =q\, \omega/\abs{q}$.

The path integral \eqref{Zq} becomes a Gaussian integral over the Grassmannian coefficients $\Psi^{(S)}_{jm}$ and $\Psi^{(T)}_{jm}$.  Performing this integral yields
 \es{Fq}{
  F_0(q) = -\int \frac{d\omega}{2 \pi} \sum_{j=\abs{q}+\frac 12}^{\infty} (2j+1)  \log \left(\omega^2 + \left( j + \frac 12 \right)^2 - q^2 \right)
   - \int \frac{d\omega}{2 \pi} 2\abs{q} \log \omega \,,
 }
where the first term corresponds to $j > \abs{q}-1/2$ while the second term represents the contribution from $j = \abs{q}-1/2$.  As one can easily check, the arguments of the logarithms are nothing but $\abs{\det \widetilde {\bf G}_{q, j}(\omega) }$, with $\widetilde {\bf G}_{q, j}(\omega)$ defined in \eqref{Gqj}.  The pre-factors of the logarithms in \eqref{Fq} come from summing over the allowed values of $m$.

 The integrals over $\omega$ are divergent, but they can be regularized by analytic continuation,
 \begin{eqnarray}
 \int \frac{d\omega}{2\pi} \log\left(\omega^{2}+b^{2}\right)&=& \int \frac{d\omega}{2\pi} \frac{d}{ds}\left(\omega^{2}+b^2\right)^{s}|_{s=0} \ = \ \abs{b}\,.
 \end{eqnarray}
 Using this identity to define the regularized expression, the free energy reduces to
   \begin{eqnarray}
  F_0(q) = -\sum_{j=\abs{q}-\frac{1}{2}}^{\infty}(2j+1)\sqrt{\left(j+\frac{1}{2}\right)^{2}-q^{2}}\,.
\end{eqnarray}
This sum is still divergent and can be regularized by various methods, such as by the Abel-Plana summation formula as in \cite{Borokhov:2002ib} or by zeta-function regularization as in \cite{Pufu:2013eda}.   The result can be written in terms of an absolutely convergent sum and the Hurwitz zeta function $\zeta(s,a)=\sum_{n=0}^{\infty} (n+a)^{-s}$ as
 \es{FqFinal}{
  F_0(q) &= -\sum_{j=\abs{q}-\frac{1}{2}}^{\infty}\left((2j+1)\sqrt{\left(j+\frac{1}{2}\right)^{2}-q^{2}}-\frac{1}{2} (2 j+1)^2+q^2\right)  \\
   &{}-\left[(1/2-q^{2})\zeta(0,q-1/2)+2\zeta(-1,q-1/2)+2\zeta(-2,q-1/2)\right]\,.
 }
In Section~\ref{DIMENSIONS} we will tabulate this sum for a few values of $q$; see Table~\ref{qTable}.  Knowing $F(q)$, one can easily calculate the leading term in the large $N_f$ expansion of the ground state energy in the presence of our GNO background using \eqref{FreeLeading}.

Now that we have a handle on the leading order computation let us move on to the next-to-leading order contribution.

\subsection{The Faddeev--Popov determinant}\label{subsec:FPDet}

The next-to-leading order computation of the free energy has two contributions given by the second and third terms in \eqref{FFinal}.  The second term represents the contribution from the Faddeev-Popov ghosts, while the third term comes from the determinant of the gauge field fluctuations.  Of these two contributions, the Faddeev-Popov one is considerably simpler, because it involves the determinant of a local operator, and we will discuss it first.

The Faddeev-Popov contribution to the ground state energy,
 \es{FPCont}{
  F_\text{FP}[{\cal A}] = -\frac 12 \log \Det' \left(-D_\mu^{({\cal A})} D^{({\cal A}),\mu}\right) \,,
 }
can be written as $-\log Z_\text{FP}[{\cal A}]$, where $Z_\text{FP}[{\cal A}]$ is the partition function for an anti-commuting scalar ghost field $c$ valued in the Lie algebra of the gauge group:
 \es{ZFP}{
  Z_\text{FP}[{\cal A}]  = \int Dc\,  e^{-S_\text{ghost}} \,, \qquad
   S_\text{ghost} = \frac 12 \int d^3x\, \sqrt{g(x)}  \sum_{a, b=1}^{N_c} \abs{\partial_\mu c^{ab} - i [{\cal A}_\mu, c]^{ab} }^2 \,.
 }
Evaluated in the GNO monopole background \eqref{MonopoleGeneral}, the ghost action becomes
 \es{SghostExplicit}{
  S_\text{ghost} = \frac 12 \int d^3x\, \sqrt{g(x)}   \sum_{a, b=1}^{N_c}  \abs{ \left[ \partial_\mu - i (q _a - q_b) {\cal A}_\mu^{U(1)} \right] c^{ab}  }^2 \,.
 }
The interpretation of this formula is that the diagonal components $c^{aa}$ are free real scalar fields, while the off-diagonal components $c^{ab}$, $a \neq b$, whose complex conjugates are $c^{ba} = c^{ab*}$, are free complex scalar fields experiencing an Abelian monopole background $(q_a - q_b) {\cal A}^{U(1)}$.  

To diagonalize the action \eqref{SghostExplicit} we should expand the ghost fields $c^{ab}$ in terms of the monopole harmonics $Y_{Q, JM}$ introduced in the previous section, with $Q = q_a - q_b$.  Explicitly, writing
 \es{GhostExpansion}{
  c^{ab}(x) = \int \frac{d\Omega}{2 \pi} \sum_{J, M}  C^{ab}_{JM}(\Omega) Y_{q_a-q_b, JM}(\hat n) e^{-i \Omega \tau} \,,
 }
and using the fact that the monopole spherical harmonics $Y_{Q, JM}$ have eigenvalue $J(J+1) - Q^2$ under the gauge-covariant Laplacian  \cite{Wu:1976ge,Wu:1977qk} on $S^2$, we can put the ghost action in the form
 \es{SGhostC}{
  S_\text{ghost} = \frac 12 \sum_{a, b=1}^{N_c} \int \frac{d \Omega}{2 \pi} \sum_{J, M}   \abs{C^{ab}_{JM}(\Omega)}^2 
   \left[\Omega^2 + J(J+1) - (q_a - q_b)^2 \right] \,.
 }
Note that here the sum over $J$ runs only from $\abs{q_a - q_b}$ to infinity, and the sum over $M$ runs from $-J$ to $J$, as appropriate for the spin-$J$ representation of $SU(2)$.

The contribution to the free energy can now be computed by integrating over the Grassmannian coefficients $C^{ab}_{JM}$ in \eqref{ZFP}. Because the ghost fields $C^{ab}_{JM}$ do not mix, the result will be a sum
\es{FPexplicit}{
 F_\text{FP}[{\cal A}]&= \sum_{a,b=1}^{N_c} F_\text{FP}(q_a,q_b) \ ,\\
 F_\text{FP}(q,q')&\equiv-\frac12\, \int \frac{d\Omega}{2 \pi} \sum_{J=|Q|}^\infty(2J+1)\, \log\left[J (J + 1) -Q^2 + \Omega^2 \right]\ , \qquad Q\equiv q-q' \ .
}
We will postpone evaluating this expression until after combining it with the contribution coming from the gauge field fluctuations in Subsection~\ref{subsec:Combine}.

\subsection{The gauge fluctuations determinant}
\label{GAUGEDET}

We now turn our attention to the third term in \eqref{FFinal}, which is also the hardest to compute.\footnote{We develop a method slightly different from the calculation for $N_c=1$, $\abs{q}=1/2$ done in~\cite{Pufu:2013vpa}. That approach, although less straightforward, has the advantage of producing simpler formulae than the approach of this paper. However, it seems hard to generalize the method of~\cite{Pufu:2013vpa} to the present case.}  Combining \eqref{GaugeDef} and \eqref{GotKAgain}, we can write this term as
 \es{TrlogK}{
   F_\text{gauge} [{\cal A}] &= \sum_{a, b=1}^{N_c} F_\text{gauge}(q_a, q_b) \ ,\\
   F_\text{gauge}(q, q')& = \frac 12  \Tr' \log {\cal K}_{qq'} \,,
 }
where we recall that the quantity ${\cal K}_{q q'}$ can be written in terms of the Green's function $G_q(x, x')$ of a single fermion in an Abelian monopole background with charge $q$ as 
 \es{Kqqp}{
   \mathcal{K}_{q q'}^{\mu\nu}(x,x')= 
     -\textrm{Tr}(\gamma^{\mu}G_{q}(x,x')\gamma^{\nu}G_{q'}^{\dagger}(x,x')) \,.
 }
Therefore, in order to evaluate \eqref{TrlogK}, we should first write down an explicit expression for $G_q(x, x')$, and then describe how to use it to construct ${\cal K}_{qq'}$ and find its eigenvalues. 

In evaluating \eqref{TrlogK}, we will not be able to find simple analytical formulae such as \eqref{FqFinal} or \eqref{FPexplicit}, and instead we will have to resort to numerics.  In the rest of this section we aim to provide enough details on the steps one has to take in implementing these numerics, and the cross-checks that can be performed.  We will postpone the numerical results until Section~\ref{DIMENSIONS}.

\subsubsection{Green's Functions}
\label{GREEN}

The expression for the Green's function $G_q(x, y)$ of a single fermion in a charge $q$ Abelian monopole background can be read off from Fourier transforming back to position space the inverse ${\bf G}_{q, j}(\omega)  = \widetilde {\bf G}_{q, j}(\omega)^{-1}$ of the expression in \eqref{Gqj}.  It is not hard to check that this inverse is
 \es{GotGG}{
  {\bf G}_{q, j}(\omega) = \frac{\widetilde {\bf G}_{q, j}(\omega)}{\omega^2 + (j+1/2)^2 - q^2} \,,
 }
with $ \widetilde {\bf G}_{q, j}(\omega)$ as in \eqref{Gqj}.  The quantity ${\bf G}_{q, j}(\omega)$ is a $2 \times 2$ matrix if $j > \abs{q} - 1/2$.  When $j = \abs{q}-1/2$, it is a $1 \times 1$ matrix equal to $q/(\abs{q} \omega)$.  The Green's function is then
 \es{GPosition}{
   G_{q}(x,x')  &= \langle\psi(x)\psi^{\dagger}(x^{\prime})\rangle  \\
     &=-\int\frac{d\omega}{2\pi}\sum_{j,m}\left(T_{q,j m}(\hat{n}) \ S_{q,j m}(\hat{n})\right) 
      {\bf G}_{q, j}(\omega) e^{-i\omega(\tau-\tau^{\prime})}
      \left(\begin{array}{c}T_{q,j m}^{\dagger}(\hat{n}^{\prime})\\S_{q,j m}^{\dagger}(\hat{n}^{\prime})\end{array}\right)\,,
 }
where it should be understood that when $j = \abs{q} - 1/2$ and ${\bf G}_{q, j}(\omega)$ is a $1 \times 1$ matrix, we should only consider the $S_{q, jm}$ modes.  

The Green's function \eqref{GPosition} is a $2 \times 2$ matrix whose indices are the spinor indices that we have been consistently suppressing.  By combining \eqref{GPosition} with the explicit expressions for the monopole harmonics \eqref{TSDef}, one can see that each entry in $G_q(x, x')$ can be written as a sum over products of two monopole spherical harmonics:
 \es{Greens}{
   G^{ik}_{q}(x,x^{\prime}) = \sum_{j,m}\sum_{\ell,\ell^{\prime} \in \{j-\frac 12, j+\frac 12\}}g_{q, jm ; \ell,\ell^{\prime}}^{ik}(\tau - \tau')Y_{q, \ell, m-m_{i}}(\hat{n})Y_{q, \ell^{\prime} ,m-m_{k}}^{*}(\hat{n}^{\prime})\,,
 }
with coefficients $g_{q, jm ; \ell,\ell^{\prime}}^{ik}(\tau - \tau')$ that can be easily worked out by integrating with respect to $\omega$ in \eqref{GPosition}.  (In \eqref{Greens}, $m_i = 1/2$ or $-1/2$ if $i=1$ or $i=2$, and similarly for $m_k$.)

The details of expression \eqref{Greens} are crucial for calculating the numerical value of the monopole scaling dimensions, but not so essential if one is only concerned with understanding the general structure of the calculation.  For understanding the structure of the calculation, we can write down \eqref{Greens} schematically as
 \es{GreensSchematics}{
  G_q(x, x') = \sum_{j = \abs{q}-1/2}^\infty \left( \text{$2 \times 2$ matrix $\propto Y_{q, \ldots}(\hat n) Y_{q, \ldots}^*(\hat n')$} \right) \,,
 }
where we emphasized that $G_q$ is a $2 \times 2$ matrix (because it has spin indices).  Each entry can be written as a single infinite sum of products between a monopole harmonic with charge $q$ at $\hat n$ and the conjugate of a monopole harmonic with charge $q$ at $\hat n'$.  There are additional finite sums in \eqref{GreensSchematics} that have not been indicated.

The careful reader may notice that the position space Green's function $G_q(x, x')$ is not unique, because one needs to specify a pole-passing prescription in performing the $\omega$ integral in \eqref{GPosition}.  While usually in Euclidean signature the poles of the propagator are off the real axis and the Fourier transform provides a well-defined position-space Green's function, in our case we have zero-energy modes that generate a pole for the propagator on the real axis.  We choose the prescription for passing around this pole given by principal value integration, whereby
 \es{FTPole}{
  \int \frac{d\omega}{2 \pi} \frac{e^{-i \omega(\tau - \tau')}}{\omega} = -\frac i2 \sgn(\tau - \tau') \,.
 }
 This prescription respects CP invariance of the monopole vacuum, in which the Green's function is an expectation value.

\subsubsection{The kernel ${\cal K}$
}
Now that we have an expression for the Green's function, it is a straightforward matter to write down the kernel $\mathcal{K}_{q q'}^{\mu\nu}(x,y)= -\textrm{Tr}(\gamma^{\mu}G_{q}(x,y)\gamma^{\nu}G_{q'}^{\dagger}(x,y))$ and compute its eigenvalues.  Since the Green's function is a sum over products of two monopole harmonics, the kernel ${\cal K}_{q q'}$ is a sum over products of four spherical harmonics
 \es{posspace}{
    \mathcal{K}_{q q'}(x,x')=\sum_{j = \abs{q}-\frac 12}^\infty
       \sum_{j' = \abs{q'} - \frac 12}^\infty 
        \left( \text{$3 \times 3$ matrix $\propto Y_{q, \ldots}(\hat n)
          Y_{q', \ldots}(\hat n') Y_{q, \ldots}^*(\hat n') Y_{q', \ldots}^*(\hat n)$} \right)  \,,
 }
where we emphasize that this kernel can be written as a $3\times 3$ matrix (since the tangent indices $\mu$ and $\nu$ run over three values), and that each entry of this matrix contains two infinite sums of products involving four spherical harmonics.  It is straight-forward, but tedious, to work out the precise form of ${\cal K}_{q q'}(x, x')$ given the precise form of the $G_q(x, x')$.

The object ${\cal K}_{q q'}^{\mu\nu}(x, x')$ should be thought of as an integration kernel that acts on a space of vector fields $a^\mu(x)$ on $S^2 \times \R$ by
 \es{aEvector}{
   [{\cal K}_{q q'} a]^\mu(x) = \int d^3 x' \sqrt{g(x')} {\cal K}_{q q'}^{\mu\nu}(x, x') a_\nu(x') \,.
 }
Actually, the expansion \eqref{posspace} reveals that the $a^\mu(x)$ must not be regular vector fields on $S^2 \times \R$, but rather sections of a more complicated vector bundle.  Indeed, if we pass from the North chart where ${\cal A}_{(N)}^{U(1)} =  ( 1- \cos \theta) d\phi $ to the South chart where ${\cal A}_{(S)}^{U(1)} =  ( -1- \cos \theta) d\phi$, a scalar monopole harmonic $Y_{q, \ell m}$ picks up a phase,  $Y_{q, \ell m}^{(S)}(\hat n) = Y_{q, \ell m}^{(N)}(\hat n) e^{-2i q \phi }$, as appropriate for how a field with charge $q$ should transform under a gauge transformation ${\cal A}_{(S)}^{U(1)} = {\cal A}_{(N)}^{U(1)} + d\Lambda$, with $\Lambda = -2\phi$.  Consequently, we have
 \es{calKTransf}{
  {\cal K}_{q q'}^{\mu\nu(S)}(x, x') = {\cal K}_{q q'}^{\mu\nu(N)}(x, x')
   e^{-2 i (q - q') (\phi - \phi')} \,.
 }
Imposing the condition that both $a^\mu(x)$ and $[{\cal K}_{q q'} a]^\mu(x)$ transform in the same way when passing from the North to South chart (because otherwise ${\cal K}_{q q'}$ would not be a well-defined operator), we see that \eqref{aEvector} implies
 \es{aTransf}{
  a^{\mu(S)}(x) = a^{\mu(N)}(x) e^{-2 i Q \phi} \,, \qquad Q \equiv q - q' \ .
 }
In other words, $a^\mu(x)$ carries charge $Q$ under ${\cal A}^{U(1)}$.

That ${\cal K}_{q q'}(x, x')$ acts on vector fields carrying charge $Q$ could have been anticipated from the form of the effective action \eqref{SeffFinal}.  Indeed, in \eqref{SeffFinal} we see that ${\cal K}_{q q'}$ (with $q=q_a$ and $q' = q_b$) is multiplied on the right by the $a^{ab}_\mu$ components of the gauge field fluctuations. 
In the background ${\cal A}^{U(1)}$, these components carry precisely charge $Q = q_a - q_b = q-q'$, as can be seen by an argument similar to the one in Section~\ref{subsec:FPDet} that showed that the ghost fields $c^{ab}$ carry charge $Q$ as well. We will call the off-diagonal components $a^{ab}_\mu$ $W$ bosons in the following.

We are interested in finding the eigenvalues of the kernel ${\cal K}_{q q'}$.  To do so, we should make use of translational and rotational symmetry.  Translational symmetry in the Euclidean time direction means that if we expand $a^\mu (x)$ in Fourier modes, the kernel ${\cal K}_{q q'}$ will not mix modes with different frequencies. Similarly, rotational symmetry along the $S^2$ directions implies that ${\cal K}_{q q'}$ only mixes modes that transform in the same representation of $SU(2)_\text{rot}$.   As per \eqref{aTransf} above, a good basis for the angular dependence of $a^\mu(x)$ is given by the vector monopole harmonics with charge $Q$.  We saw in the spinor case that we can construct spinor monopole harmonics from scalar harmonics with any given charge $Q$.  A similar construction can be performed for the vector monopole harmonics \cite{Weinberg:1993sg} (see also Appendix~\ref{App:Identities}).  While in the spinor case we had two independent sets of spinor harmonics, $T_{q, jm}$ and $S_{q, jm}$ with orbital angular momentum $\ell = j+1/2$ and $j-1/2$, respectively, we now have three sets of vector harmonics with total angular momentum quantum numbers $(J, M)$ and orbital angular momentum $L$:
 \begin{align}
  &U_{Q, JM}(\hat n) \,, &  L &= J+1 \,, & J &\geq 
     \begin{cases} \abs{Q}-1\,, & \text{if $\abs{Q}\geq 1$}\,, \\
    \abs{Q}\,, & \text{if $\abs{Q} =0$ or $1/2$}\,,
    \end{cases} \nonumber\\
  &V_{Q, JM}(\hat n) \,,&   L &= J \,, & J &\geq \begin{cases}
   \abs{Q}\,, & \text{if $\abs{Q}>0$}\,, \\
   1 \,, & \text{if $\abs{Q}=0$}\,, 
   \end{cases} \label{UVWDef}\\
  &W_{Q, JM}(\hat n) \,, &  L &= J-1 \,, & J &\geq \abs{Q}+1 \,.\nonumber
 \end{align}
That the orbital angular momentum is $L$ means that in a frame basis one can write down the components of the harmonics \eqref{UVWDef} in terms of scalar monopole harmonics $Y_{Q, L M_L}$, where $M_L \in \{M-1, M, M+1\}$.    Since we must always have $L \geq \abs{Q}$, we obtain the allowed ranges in \eqref{UVWDef}.  Note that the harmonics with $J = \abs{Q}-1$ are defined only when $\abs{Q} \geq 1$;  these harmonics will play an important role shortly.

We can thus expand $a$  in terms of vector monopole harmonics \eqref{UVWDef} and Fourier modes in $\tau$:
 \es{aExpansion}{
   a(x) = \int\frac{d\Omega}{2\pi} \sum_{J, M}\left[
        \mathfrak{a}_{U}^{J M}(\Omega) U_{Q, J M}(\hat{n})
     +\mathfrak{a}_{V}^{J M}(\Omega) V_{Q, J M}(\hat{n})
     +\mathfrak{a}_{W}^{J M}(\Omega) W_{Q, J M}(\hat{n})\right] e^{-i\Omega\tau} \,,
 }
with coefficients $\mathfrak{a}^{JM}_U(\Omega)$, $\mathfrak{a}^{JM}_V(\Omega)$, and $\mathfrak{a}^{JM}_W(\Omega)$.   Then the operator ${\cal K}_{q q'}$ is almost diagonal and mixes together only modes with the same $J$, $M$, and $\Omega$:
 \es{KMatrix}{
   \int d^3x \, d^3 x' \sqrt{g(x)} &\sqrt{g(x')} a_\mu(x)^* {\cal K}_{q q'}^{\mu\nu} (x, x') a_\nu(x') \\
  &= \int \frac{d\Omega}{2 \pi} \sum_{J, M} 
    \begin{pmatrix}
        \mathfrak{a}_{U}^{J M}(\Omega) \\
        \mathfrak{a}_{V}^{J M}(\Omega) \\ 
        \mathfrak{a}_{W}^{J M}(\Omega)
      \end{pmatrix}^\dagger
       {\bf K}^J_{q q'}(\Omega)
      \begin{pmatrix}
        \mathfrak{a}_{U}^{J M}(\Omega) \\
     \mathfrak{a}_{V}^{J M}(\Omega) \\
     \mathfrak{a}_{W}^{J M}(\Omega)
     \end{pmatrix} \,.    
 }
When $J \geq \abs{Q}+1$ there are three such modes for each $M$, ${\cal K}_{q q'}$ acts in this $3$-dimensional space as the $3 \times 3$ matrix ${\bf K}^J_{q q'}(\Omega)$.  When $J = \abs{Q}$, the modes corresponding to $W_{Q, JM}$ are absent, and ${\bf K}^J_{q q'}(\Omega)$ is a $2 \times 2$ matrix.  Lastly, when $J = \abs{Q} - 1$, both $V_{Q, JM}$ and $W_{Q, JM}$ are absent, and ${\bf K}^J_{q q'}(\Omega)$ is a $1 \times 1$ matrix.  Because of rotational invariance, ${\bf K}^J_{q q'}(\Omega)$ is independent of $M$.

From \eqref{KMatrix} it is not hard to extract an inversion formula for the components of the matrix ${\bf K}_{qq'}^J(\Omega)$:
 \es{ExtractK}{
  [{\bf K}_{qq'}^J(\Omega)]_{XY} &2 \pi \delta(\Omega - \Omega') \\
   &= \int d^3 x\, d^3 x' \sqrt{g(x)} \sqrt{g(x')} 
     X^\mu_{Q, JM}(\hat n)^* {\cal K}_{qq' \mu\nu}(x, x')
     Y^\nu_{Q, JM}(\hat n') e^{i \Omega \tau -i \Omega' \tau'} \,,
 }
where $X, Y \in \{U, V, W\}$ denote the indices of ${\bf K}_{qq'}^J(\Omega)$.  This expression is rather unwieldy, especially after plugging in the explicit formula \eqref{posspace} for the kernel ${\cal K}_{qq'}$, which yields, schematically,
 \es{momspace}{
   \textbf{K}^J_{q q'}(\Omega)= \int d\hat{n}d\hat{n}^{\prime}\sum_{j,j^{\prime}} 
      {\bf C}^J_{jj'\ldots}(\Omega) Y_{q, \ldots }(\hat{n})Y_{q', \ldots}^{*}(\hat{n})Y_{Q, \ldots}^{*}(\hat{n})
        Y_{q, \ldots}^{*}(\hat{n}^{\prime})Y_{q', \ldots}(\hat{n}^{\prime})
        Y_{Q, \ldots}(\hat{n}^{\prime}) \,.
 }
This formula for $\textbf{K}^J_{q q'}(\Omega)$ involves two angular integrals over a product of six monopole spherical harmonics, two infinite sums exhibited explicitly in \eqref{momspace}, as well as several finite sums that were omitted.

\begin{figure}[t!]
\begin{center}
   \includegraphics[width=0.7\textwidth]{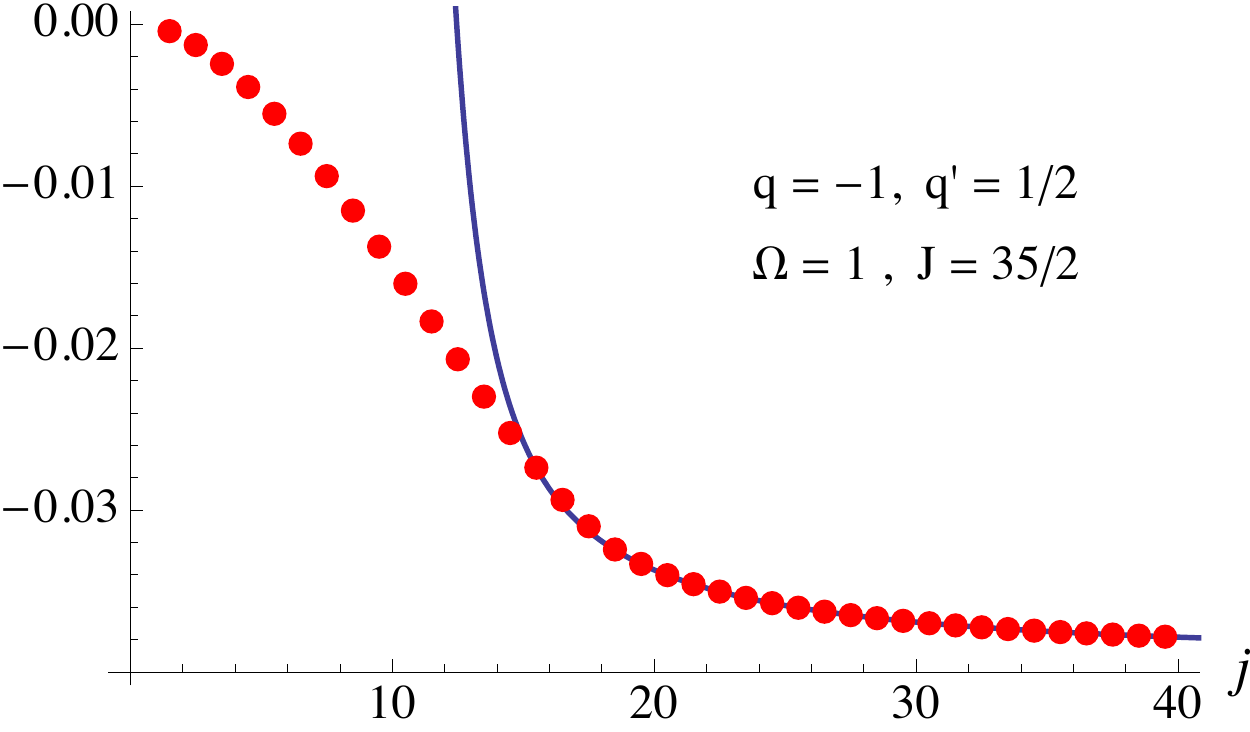}
\caption{We plot the terms in the infinite sum over $j$~\eqref{relnice} that give the matrix element $\le[{\bf K}^J_{q q'}(\Omega)\ri]_{UU}$ for $q=-1,\, q'=1/2,\, \Omega=1,$ and $J=35/2$. We show the stage of the calculation where all the finite sums (over $\delta q,\, l',\, l$, and $j'$) in~\eqref{relnice} have been done and only the infinite sum over $j$ remains.  The dots represent the actual terms in the sum, while the solid line is the asymptotic expansion of the summand to ${\cal O}(1/j^{18})$ that we determined analytically. We perform the infinite sum by zeta-function regularization of the asymptotic form for $j>j_c$, where $j_c$ is the value below which we use the numerical values of the terms in the sum. We check the numerical precision by changing $j_c$ and we reach our goal of $10^{-12}$ precision by choosing $j_c\approx 40$. This precision is needed to get the free energy with $10^{-3}$ precision.}
\label{fig:jasymp}
\end{center}
\end{figure}

We are not discouraged and still determined to evaluate \eqref{momspace} as efficiently as we can.  We can simplify \eqref{momspace} by using rotational invariance, which, as mentioned above, implies that ${\bf K}^J_{q q'}(\Omega)$ is independent of $M$.  So we might as well compute ${\bf K}^J_{q q'}(\Omega)$ after averaging over $M$.  Writing \eqref{ExtractK} as 
 \es{ExtractKBetter}{
  &[({\bf K}_{qq'}^J(\Omega)]_{XY} 2 \pi \delta(\Omega - \Omega') \\
   &=
    \frac{1}{2 J+1} \sum_{M=-J}^J \int d^3 x\, d^3 x' \sqrt{g(x)} \sqrt{g(x')} 
     X^\mu_{Q, JM}(\hat n)^* {\cal K}_{qq' \mu\nu}(x, x')
     Y^\nu_{Q, JM}(\hat n') e^{i \Omega \tau -i \Omega' \tau'} \,,
 }
and plugging in the explicit form of ${\cal K}_{qq'}$ we again obtain an expression of the schematic form in \eqref{momspace}.  This time, however, because we averaged over $M$, this expression is rotationally-invariant and the integrand depends only on the relative angle between $\hat n$ and $\hat n'$.  The integral with respect to $\hat n$ is therefore independent of $\hat n'$, so we can choose $\hat{n}^{\prime}$ to point in the $\hat{z}$ direction and replace the integral with respect to $\hat n'$ by a  factor of $4 \pi$.  Using
\begin{eqnarray}
Y_{q,lm}(\hat{z})&=&\delta_{q,-m}\sqrt{\frac{2l+1}{4\pi}} \,,
\end{eqnarray}
we get rid of three of those pesky monopole harmonics in \eqref{momspace}.   The remaining angular integral over the product of three harmonics can be evaluated using some properties of monopole harmonics \cite{Wu:1977qk}.
 \es{YConj}{
   Y_{q, \ell m} (\hat n)^* = (-1)^{q+m}Y_{-q, \ell,-m} (\hat n) \,,
 }
and
 \es{ThreeJ}{
\int d\hat{n}\,&Y_{q,\ell m}(\hat{n}) Y_{q^{\prime},\ell^{\prime}m^{\prime}}(\hat{n})Y_{q^{\prime\prime},\ell^{\prime\prime}m^{\prime\prime}}(\hat{n}) \\
  &=(-1)^{\ell+\ell^{\prime}+\ell^{\prime\prime}}\sqrt{\frac{(2\ell+1)(2\ell^{\prime}+1)(2\ell^{\prime\prime}+1)}{4\pi}}
  \begin{pmatrix}
   \ell & \ell' & \ell'' \\
   q & q' & q''
  \end{pmatrix}
    \begin{pmatrix}
   \ell & \ell' & \ell'' \\
   m & m' & m''
  \end{pmatrix} \,,
 }
where $\begin{pmatrix}
 j & j' & j'' \\
 m & m' & m''
\end{pmatrix}$ is the Wigner 3$j$ symbol.  After using these identities, ${\bf K}^J_{q q'}(\Omega)$ can be put in the form:
 \es{relnice}{
   {\bf K}^J_{q q'}(\Omega)= \sum_{j,j^{\prime}} \sum_{\ell =j-\frac 12}^{j+\frac 12} \sum_{\ell' =j'-\frac 12}^{j'+\frac 12} \sum_{\delta q = -1}^1 
   \widetilde{\textbf{C}}^J_{j j^{\prime}, \ldots}(\Omega)
     \begin{pmatrix}
      \ell & \ell'  & L \\
      q & -q' & -Q
     \end{pmatrix}
     \begin{pmatrix}
      \ell & \ell'  & L \\
      -q + \delta q & q' - \delta q & Q
     \end{pmatrix} \,,
  }
where $L = J-1$, $J$, or $J+1$ (depending on which component of ${\bf K}^J_{q q'}(\Omega)$ we are computing), and where the coefficients $\widetilde {\bf C}^J_{jj', \ldots}(\Omega)$ have fairly complicated expressions that will not be reproduced here. The lesson to be learned is that one can write ${\bf K}^J_{q q'}(\Omega)$ explicitly in terms of two potentially infinite sums (over $j$ and $j'$) and several finite sums over products of $3j$ symbols.  In fact, for fixed $j$ and $J$ the sum over $j'$ is finite, because the $3j$ symbol vanishes if its arguments are not triangular.  There is therefore only a single infinite sum over $j$ in the expression for $\textbf{K}^J_{q q'}(\Omega)$, which can be evaluated using zeta-function regularization. The terms in the infinite sum are shown in an example in Figure~\ref{fig:jasymp} for one matrix element.

\subsubsection{Properties of ${\bf K}^J_{q q'}(\Omega)$}
\label{PROPERTIES}

On general grounds, the matrix ${\bf K}^J_{q q'}(\Omega)$ should satisfy certain properties that can be used as checks on the explicit formulae \eqref{relnice}.  For instance, $\textbf{K}^J_{q q'}(\Omega)$ is Hermitian, $\textbf{K}^{J\dagger}_{q q'}(\Omega)=\textbf{K}^J_{q q'}(\Omega)$, and due to invariance of the monopole background under CP, one can show that
 \es{KCP}{
   [\textbf{K}^J_{q'q}(\Omega)]_{XY}
    = [\textbf{K}^J_{-q,-q'}(\Omega)]_{XY} =
       (-1)^{X+Y} [\textbf{K}^J_{q q'}(\Omega)]_{XY}   \,.
 }
Here, the range of $X, Y \in \{U, V, W\}$ depends on whether $\textbf{K}^J_{q q'}(\Omega)$ is a $3 \times 3$, $2 \times 2$, or $1 \times 1$ matrix.

It follows from gauge invariance that ${\bf K}^J_{q q'}(\Omega)$ has a zero eigenvalue.\footnote{When $J=\le|Q\ri|-1$ ${\bf K}^J_{q q'}(\Omega)$ is a $1\times 1$ matrix that does not vanish. For $J=\le|Q\ri|$ it is a $2\times 2$  and for $J\geq\le|Q\ri|+1$ it is a $3\times 3$ matrix with one zero eigenvalue. In the following we assume $J\geq\le|Q\ri|+1$.} To leading order in the large $N_f$ expansion the current conservation equation takes the form:
\es{CovConsLead}{
  0 = D^{({\cal A})}_\mu j^\mu_{ba} = \partial_\mu j^\mu_{ba} - i\le[{\cal A}_{\mu}, j^\mu\ri]_{ba} =\le[\partial_\mu-i(q_b-q_a)\, {\cal A}_\mu^{U(1)}\ri] j^\mu_{ba}  \, ,
 }
 where we dropped terms proportional to the gauge fluctuation $a$ and used that the monopole background is diagonal in the gauge indices. The gauge kernel, $K^{\mu\nu}_{ab, cd} $  defined in~\eqref{GotK} is a current two point function. Applying~\eqref{CovConsLead} to the second current in $K^{\mu\nu}_{ab, cd}$ we get 
\es{Ward}{
0= \le[ {\partial\ov \partial x'^\nu} +i(q-q'){\cal A}_\nu^{U(1)}\ri] {\cal K}^{\mu\nu}_{qq'}(x, x')
}
where we used~\eqref{GotKAgain}. Note that in this Ward identity the delta function is absent, as follows from Lorentz covariance and dimensional analysis. From \eqref{Ward} we determine the eigenvector with zero eigenvalue of ${\bf K}^J_{q q'}(\Omega)$. Indeed, inverting~\eqref{ExtractK} we obtain:
\es{ExtractKInverse}{
{\cal K}_{qq'}^{\mu\nu}(x, x')=\int d\Omega \,\sum_{J,M} \sum_{X, Y \in \{U, V, W\}} X^\mu_{Q, JM}(\hat n)\, [{\bf K}_{qq'}^J(\Omega)]_{XY}\,  Y^\nu_{Q, JM}(\hat n')^*\, e^{-i\Omega(\tau-\tau')} \,,
}
and then acting with the derivative in~\eqref{Ward}, we obtain:
\es{Ward2}{
0=\sum_{Y \in \{U, V, W\}}[{\bf K}_{qq'}^J(\Omega)]_{XY}\, \le[ {\partial\ov \partial x'^\nu} +iQ\,{\cal A}_\nu^{U(1)}\ri] Y^\nu_{Q, JM}(\hat n')^*\, e^{i\Omega \tau'} \ .
}
We compute the divergence of vector spherical harmonics in Appendix~\ref{App:Divergence}. Thus, the eigenvector with zero eigenvalue of ${\bf K}^J_{q q'}(\Omega)$ is:
\es{KZeroEigen}{
 {\bf K}^J_{q q'}(\Omega)
   \begin{pmatrix}
      -(L +i\Omega) \displaystyle{\sqrt{{(L + 1)^2 - Q^2\ov(L + 1) (2 L + 1)}}} \\[15pt]
      \displaystyle{\frac{Q (1 -i\Omega)}{\sqrt{L (L + 1)}}} \\[15pt]
      -(L + 1 - i\Omega) \displaystyle{\sqrt{\frac{L^2 - Q^2}{ L (2 L + 1)}}}
   \end{pmatrix}   = 0 \ .
}
This property provides an essential check of our numerical results.

The same result can be understood in a different way. The gauge field effective action should be gauge invariant, hence pure gauge modes should be zero eigenvectors of the real space kernel. We set the gauge fluctuation to be pure gauge by taking $a_\mu= D^{(\cal A)}_\mu Y_{Q,JM}e^{-i\Omega \tau}$ in~\eqref{aEvector} to get
\es{KZeroEigen2}{
0= \le[{\cal K}_{q q'}\le( D^{(\cal A)} Y_{Q,JM}e^{-i\Omega \tau'}\ri)\ri]^\mu(x) = \int d^3 x' \sqrt{g(x')} \,{\cal K}_{q q'}^{\mu\nu}(x, x') \le(D^{(\cal A)}_\nu Y_{Q,JM}(\hat n')e^{-i\Omega \tau'} \ri) \ .
}
Plugging in for ${\cal K}_{q q'}^{\mu\nu}$ the formula~\eqref{ExtractKInverse} we obtain for the zero eigenvector of ${\bf K}^J_{q q'}(\Omega)$ the following expression:
\es{KZeroEigen3}{
\int d^3 x \begin{pmatrix}
     U^\mu_{Q, JM}(\hat n)^*\,e^{i\Omega' \tau} \\[15pt]
      V^\mu_{Q, JM}(\hat n)^*\,e^{i\Omega' \tau}\\[15pt]
     W^\mu_{Q, JM}(\hat n)^*\,e^{i\Omega' \tau}
   \end{pmatrix} &
   \le(D^{(\cal A)}_\mu Y_{Q,JM}(\hat n)e^{-i\Omega \tau} \ri)\\
 & =  2\pi\delta(\Omega-\Omega') \begin{pmatrix}
      -(L +i\Omega) \displaystyle{\sqrt{{(L + 1)^2 - Q^2\ov(L + 1) (2 L + 1)}}} \\[15pt]
      \displaystyle{\frac{Q (1 -i\Omega)}{\sqrt{L (L + 1)}}} \\[15pt]
      -(L + 1 - i\Omega) \displaystyle{\sqrt{\frac{L^2 - Q^2}{ L (2 L + 1)}}}
   \end{pmatrix} \,.
}
This expression agrees with~\eqref{KZeroEigen}. What this formula says is that the zero eigenvector is a pure gauge mode $D^{(\cal A)}_\mu Y_{Q,JM}e^{-i\Omega \tau}$ in the vector monopole harmonic basis. The calculation above is by no means an independent derivation of~\eqref{KZeroEigen}, as current conservation follows from gauge invariance.

\subsubsection{Eigenvalues and determinant of gauge field fluctuations}

Having computed ${\bf K}^J_{q q'}(\Omega)$, it is now easy to find the eigenvalues of the kernel ${\cal K}_{qq'}^{\mu\nu}(x, x')$:  they are simply the eigenvalues of ${\bf K}^J_{qq'}(\Omega)$ for every $J$ and $\Omega$, and they come with multiplicity $2J+1$. To have the terms in~\eqref{TrlogK} we need to compute:
\es{TrlogKResult}{
 \Tr' \log {\cal K}_{q q'}=\int\frac{d\Omega}{2\pi}\sum_{J=|Q|-1}^{\infty} (2J+1)\log \det{}' \, \textbf{K}_{qq'}^J(\Omega)\ ,
}
where $\det'$ indicates that we should only take the product of the non-zero eigenvalues. As shown in~\eqref{KZeroEigen} the presence of pure gauge modes result in one zero eigenvalue for ${\bf K}^J_{q q'}(\Omega)$, and we have to drop the zero eigenvalue as our prescription is to  integrate only over gauge inequivalent configurations. 

\begin{figure}[!h]
\begin{center}
\centering
        \begin{subfigure}[b]{0.8\textwidth}
                \centering
                \includegraphics[width=\textwidth]{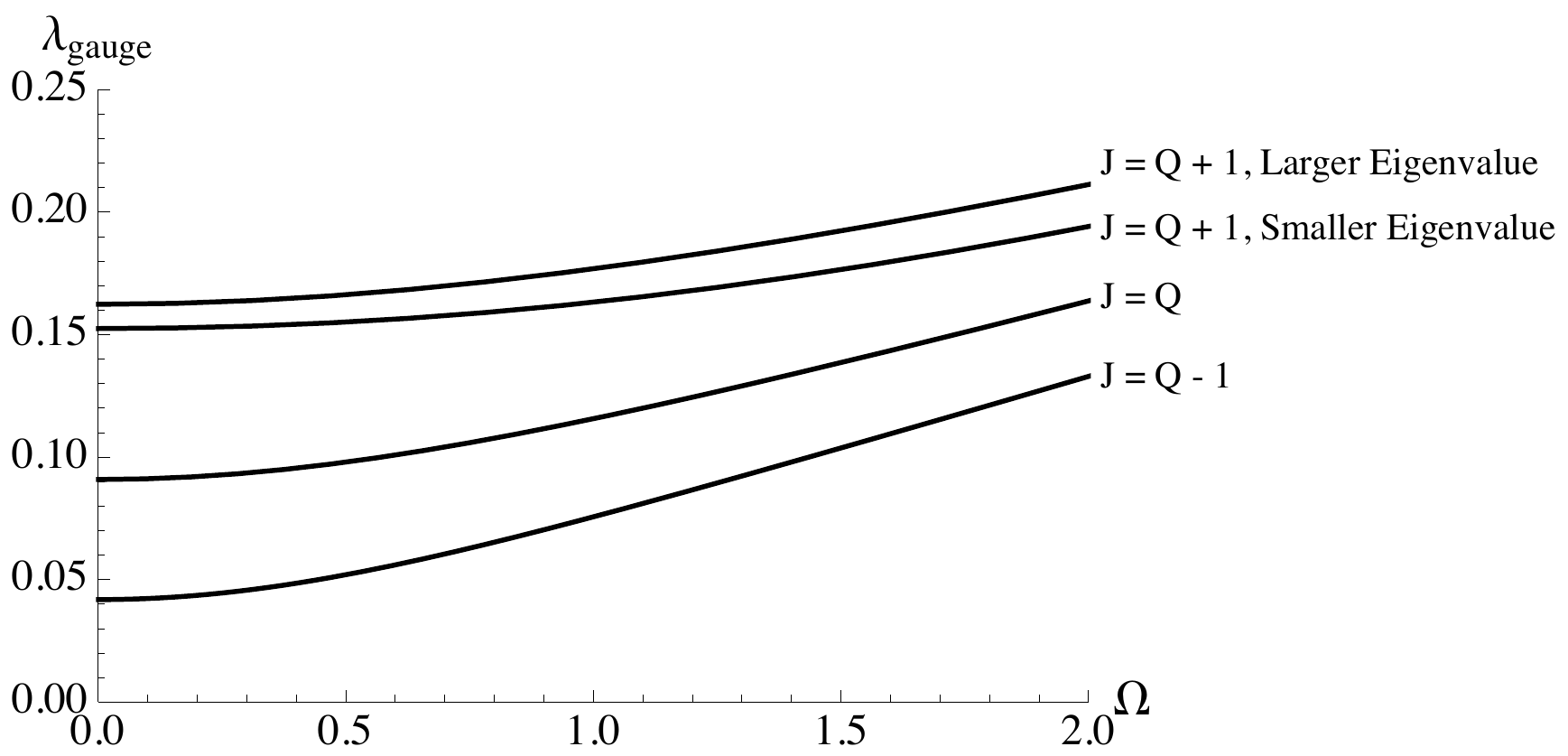}
                \caption{A stable example with $q=-1,\, q'=1/2$.}
                \label{fig:stable}
        \end{subfigure}\\%
        ~ 
   \begin{subfigure}[b]{0.8\textwidth}
                \centering
                \includegraphics[width=\textwidth]{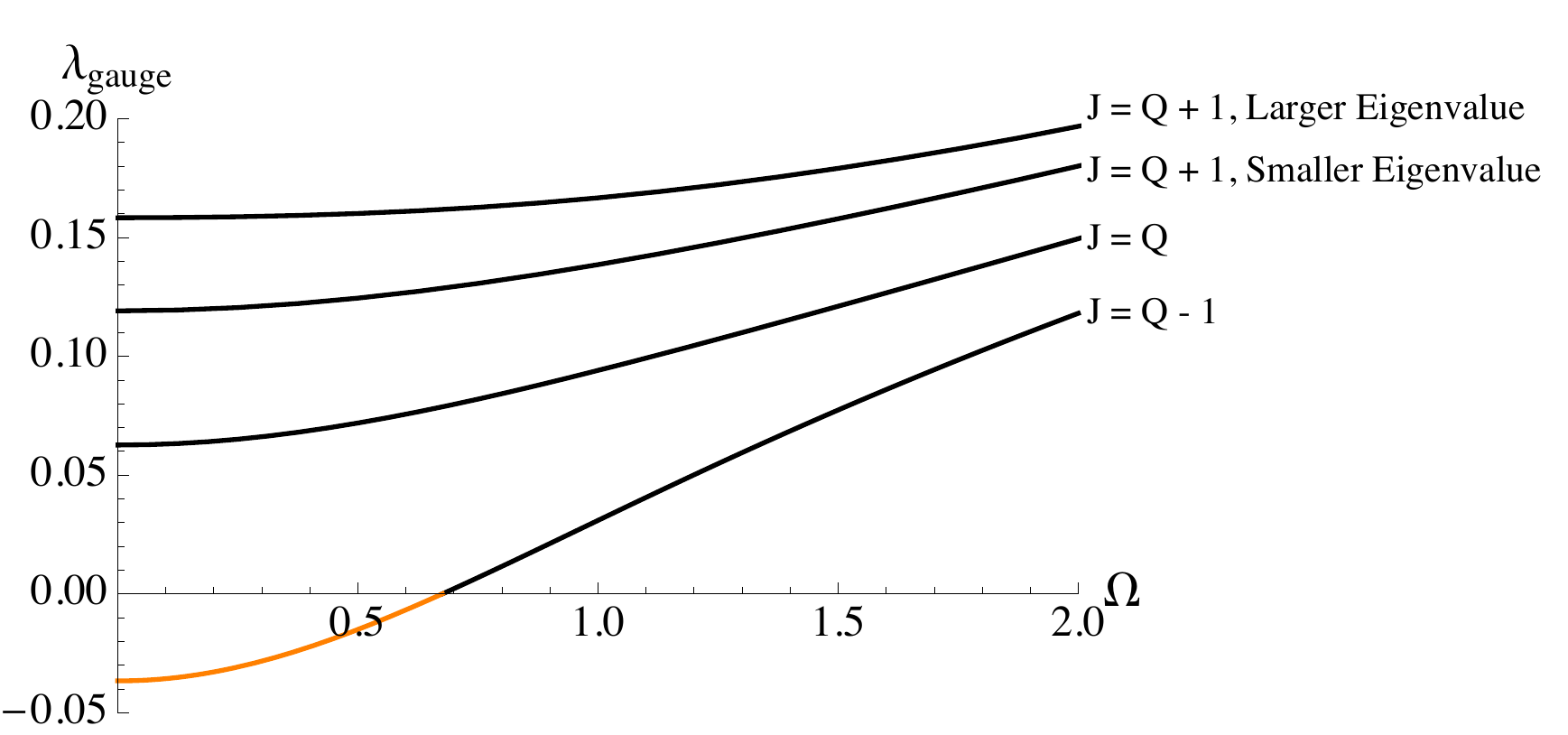}
                \caption{An unstable example with $q=1/2,\, q'=3/2$.  The instability is indicated in orange.}
                \label{fig:unstable}
        \end{subfigure}
        ~ 
                \caption{The eigenvalues of $\textbf{K}_{qq'}^J(\Omega)$ for some example $q,\, q'$ and low $J$  values as a function of $\Omega$. Zero eigenvalues corresponding to pure gauge modes are omitted. Note that the eigenvalues are monotonic in $J$ and $\Omega$, hence it suffices to examine the $\Omega=0$ behavior of the lowest $J$ mode for stability. Also note that in both examples $\le|Q\ri|\geq1$ and the two lowest lying $J$ modes have one non-zero eigenvalue, while higher $J$ modes come with two eigenvalues. (The smaller number of eigenvalues corresponds to the reduced size of the matrix $\textbf{K}_{qq'}^J(\Omega)$.)}
\label{fig:stabilityplots}
\end{center}
\end{figure}

The matrix $\textbf{K}_{qq'}^J(\Omega)$ is not necessarily positive definite. If it has a negative eigenvalue that signals an instability, the corresponding gauge fluctuation gives a wrong sign Gaussian integral in the partition function and makes the free energy complex. In Section~\ref{STABILITY} we discuss the instances when this happens. For illustration, we plot some of the low $J$ eigenvalues in two examples. Figure~\ref{fig:stable} shows a stable monopole background, while Figure~\ref{fig:unstable} an unstable one.

The expression~\eqref{TrlogKResult} is not yet ready to be put on a computer due to various divergences. We find it convenient to combine it with the Faddeev--Popov determinant and introduce a UV cutoff first.

Note that~\eqref{KCP} implies that $ \Tr' \log {\cal K}_{q q'}= \Tr' \log {\cal K}_{q' q}$, which further implies
\es{FCP}{
F_\text{gauge}(q,q')=F_\text{gauge}(q',q) \ .
}
This property is the consequence of CP invariance.  It is also not hard to show that $F_\text{gauge}(q, q') = F_\text{gauge}(-q, -q')$.

\subsection{Combining the subleading terms in the free energy}\label{subsec:Combine}

In the $1/N_f$ expansion of the free energy~\eqref{FFinal} there are two terms at ${\cal O}(N_f^0)$ order, the ghost and the gauge fluctuation contribution. Both contributions involve a sum of $N_c^2$ terms;  see~\eqref{FPexplicit} and~\eqref{TrlogK}. Each term takes the form in~\eqref{FPexplicit} and~\eqref{TrlogKResult}:
 \es{BuildingBlocks}{
 F_\text{FP}(q,q')&\equiv-\frac12\, \int \frac{d\Omega}{2 \pi} \sum_{J=|Q|}^\infty(2J+1)\, \log\left[J (J + 1) -Q^2 + \Omega^2 \right] \,,  \\
 F_\text{gauge} (q,q')&\equiv  \frac 12 \Tr' \log K_{qq'}  = \frac 12 \int\frac{d\Omega}{2\pi}\sum_{J=|Q|-1}^{\infty} (2J+1)\log \det{}' \, \textbf{K}_{qq'}^J(\Omega) \ ,
}
where we used the notation $Q=q-q'$. These expressions only determine a meaningful free energy if $\textbf{K}^J(\Omega)$ only has positive eigenvalues apart from the zero eigenvalue corresponding to pure gauge modes for all $J$ and $\Omega$. If there is a negative eigenvalue, there is an instability that will be discussed in Section~\ref{STABILITY}.

\begin{figure}[t!]
\begin{center}
   \includegraphics[width=0.7\textwidth]{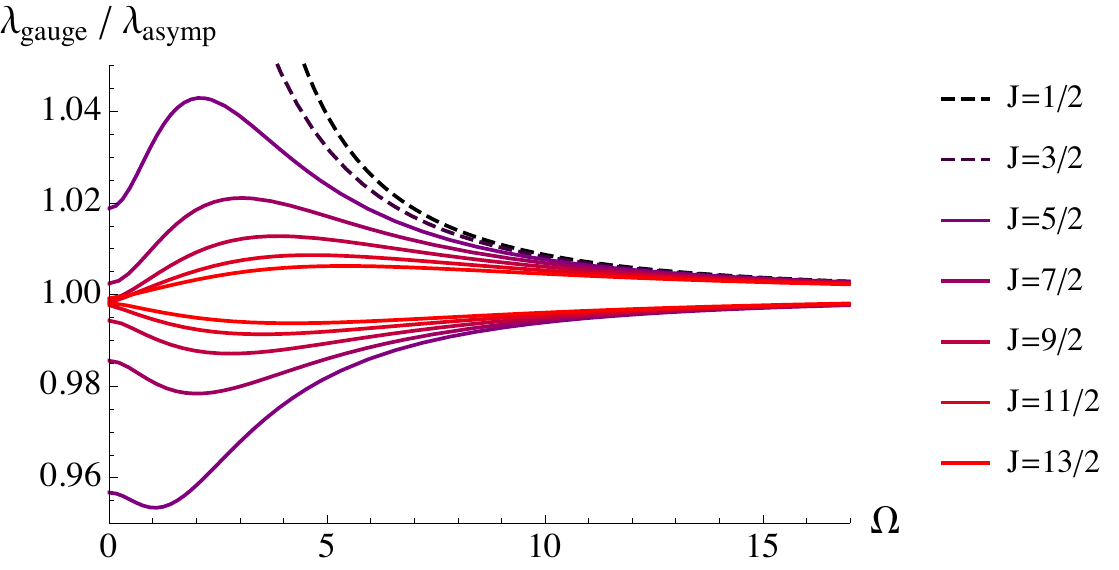}
\caption{We plot the ratio of the non-zero eigenvalues $\lambda_\text{gauge}^J(\Omega)$ of the gauge kernel divided by their asymptotic behavior $\lambda_\text{asymp}^J(\Omega)$. We chose $q=-1,\, q'=1/2$ for this example. Because $|Q|=3/2$ the $J=1/2,\, 3/2$ modes contribute one eigenvalue, while for higher $J$ eigenvalues come in pairs. We used the same colors to plot the pair of eigenvalues for these higher $J$ modes. Because the ghosts give a contribution proportional to $\lambda_\text{asymp}^J(\Omega)$ this plot shows that the low energy modes are the most important in determining the free energy.}
\label{fig:combined}
\end{center}
\end{figure}

Firstly, let us consider the large $J,\,\Omega$ behavior of the eigenvalues $\lambda_{gauge}^J(\Omega)$ of $\textbf{K}_{qq'}^J(\Omega)$, the product of which gives $\det' \textbf{K}_{qq'}^J(\Omega)$. For $J$ and $\Omega$ large we get
\es{UVAsymp}{
\lambda_\text{gauge}^J(\Omega) \sim \lambda_\text{asymp}^J(\Omega)\equiv{\sqrt{J (J + 1) - Q^2 + \Omega^2} \ov 16} \ ,
}
which gives a divergence when integrated over $\Omega$ and summed over $J$. We notice however, the appearance of the Faddeev--Popov determinant~\eqref{FPexplicit2}. Note that the ghost determinant comes with a negative sign and we end up with the ratio inside the logarithm
\es{ResultsCombined}{
 \delta F(q,q')&\equiv F_\text{FP}(q,q')+F_\text{gauge} (q,q')\\
 &=\frac 12 \int\frac{d\Omega}{2\pi}\sum_{J=\le|Q\ri|+1}^{\infty}(2J+1)\log { \det{}' \, \textbf{K}_{qq'}^J(\Omega)\ov J (J + 1) - Q^2 + \Omega^2}+\dots
}
where we introduced the notation $\delta F(q,q')$ for the sum of the gauge and ghost contributions and the dots stand for the low $J$ modes that do not pair up nicely with the ghosts.\footnote{The $J=\le|Q\ri|-1$ and $J=\le|Q\ri|$ cases has to be treated separately with zeta function regularization.} The two beautifully combine to give a well-behaved result for large $J,\,\Omega$.\footnote{Note that according to~\eqref{UVAsymp} the integrand goes to the constant $-\sum_{J=\le|Q\ri|+1}^{\infty}(2J+1)\log 256$. In zeta-function regularization, the $\Omega$ integral of a constant vanishes, hence constants do not give any contribution.} In Figure~\ref{fig:combined} we show that the eigenvalues of the gauge kernel, $\lambda_\text{gauge}^J(\Omega)$ asymptote to~\eqref{UVAsymp}. Because $\le[\lambda_\text{asymp}^J(\Omega)\ri]^2$ asymptotes to that of the ghost contribution we see that the main contribution to the free energy is from the low energy modes. At high energies the ghosts cancel the contribution coming from the gauge fluctuations.

To complete the evaluation of the subleading terms we have to introduce a cutoff that makes the integral definite and the sum finite. Because for large $J$ and $\Omega$ we are probing the UV of the field theory where it should not matter what manifold we are working on, we use a relativistic cutoff
\es{Cutoff}{
J(J+1)-Q^2+\Omega^2\leq \Lambda(\Lambda+1) \ .
}
With this cutoff the sum and the integral in~\eqref{ResultsCombined} are convergent. Evaluating~\eqref{ResultsCombined} for different $\Lambda$ and extrapolating to $\Lambda\to \infty$ we obtain our final result for the subleading term in the free energy. An example is given in Figure~\ref{fig:FreeEnergy}. We give a systematic collection of results in Section~\ref{DIMENSIONS}.

\subsection{Summary and an example} \label{subsec:Example}

In this subsection we summarize the key formulae in the evaluation of the $S^2\times \R$ free energy. We repeat the $1/N_f$ expansion of the free energy~\eqref{FFinal}
\es{FFinal2}{
  F[{\cal A}] = N_f\, F_0[{\cal A}]  + \delta F[{\cal A}]+ {\cal O}(1/N_f)\,.
 }
$F_0[{\cal A}] $ is the fermion determinant in the monopole background given by~\eqref{FreeLeading} and~\eqref{FqFinal}:
 \es{LeadingSummary}{
 F_0[{\cal A}] &= \sum_{a=1}^{N_c} F_0(q_a)\\
 F_0(q) &= -\sum_{j=\abs{q}-\frac{1}{2}}^{\infty}\left((2j+1)\sqrt{\left(j+\frac{1}{2}\right)^{2}-q^{2}}-\frac{1}{2} (2 j+1)^2+q^2\right)  \\
   &{}-\left[(1/2-q^{2})\zeta(0,q-1/2)+2\zeta(-1,q-1/2)+2\zeta(-2,q-1/2)\right] \ .
 }
 $\delta F[{\cal A}]$ is the sum of the gauge and ghost contributions. The Faddeev--Popov determinant is given by~\eqref{FPexplicit}, while the determinant of gauge fluctuation is obtained by~\eqref{TrlogK} and~\eqref{TrlogKResult}:
 \es{FPexplicit2}{
 \delta F[{\cal A}]&= \sum_{a,b=1}^{N_c} \delta F(q_a,q_b)\\
  \delta F(q,q')&\equiv F_\text{FP}(q,q')+F_\text{gauge} (q,q')\\
   F_\text{FP}(q,q')&\equiv-\frac12\, \int \frac{d\Omega}{2 \pi} \sum_{J=|Q|}^\infty(2J+1)\, \log\left[J (J + 1) -Q^2 + \Omega^2 \right]\\
   F_\text{gauge} (q,q')&\equiv  \frac 12 \Tr' \log K_{qq'}  = \frac 12 \int\frac{d\Omega}{2\pi}\sum_{J=|Q|-1}^{\infty} (2J+1)\log \det{}' \, \textbf{K}_{qq'}^J(\Omega)   \ .
}
We combine the subleading terms before evaluating~\eqref{ResultsCombined} numerically.

In this subsection we examine an example in more detail to illustrate some of the steps sketched in the previous subsections. We make the simple choice $G=U(2)$ and $q_1=1/2, \, q_2=-1$. The leading contribution is~\eqref{LeadingSummary}:
\es{LeadingExample}{
F_0[{\cal A}] =F_0(q_1)+F(q_2)=0.265 + 0.673=0.938 \ ,
} 
where we numerically evaluated~\eqref{LeadingSummary}. The list of $F_0(q)$'s will be given in Table~\ref{qTable}.

The subleading term is a sum of four terms 
\es{SubleadingExample}{
 \delta F[{\cal A}]=  \delta F(q_1,q_1)+\delta F(q_1,q_2)+\delta F(q_2,q_1)+\delta F(q_2,q_2)\ .
 }
We pick $\delta F(q_2,q_1)$ from this sum to illustrate the calculation.  The ghost contribution is known analytically~\eqref{FPexplicit2}. To calculate the gauge contribution we need to determine the gauge kernel  $\textbf{K}_{q_2,q_1}^J(\Omega)$ numerically. For every $J$ and $\Omega$ we need to construct this matrix. This construction involves an infinite sum, and the procedure is explained around Figure~\ref{fig:jasymp}.  In that figure we display the matrix element $\le[\textbf{K}_{q_2,q_1}^J(\Omega)\ri]_{UU}$ for a representative choice of $J$ and $\Omega$.  We need to know this kernel to $10^{-12}$ precision.

Gauge invariance determines the eigenvector with zero eigenvalue of the matrix $\textbf{K}_{q_2,q_1}^J(\Omega)$ analytically~\eqref{KZeroEigen}. This eigenvector provides a powerful check of the result and whether we indeed achieved the precision advertised. It turns out that matrices $\textbf{K}_{qq'}^J(\Omega)$ can be reused in the computation for general gauge groups discussed in Section~\ref{GENERALGROUP}.

\begin{figure}[!h]
\begin{center}
   \includegraphics[width=0.7\textwidth]{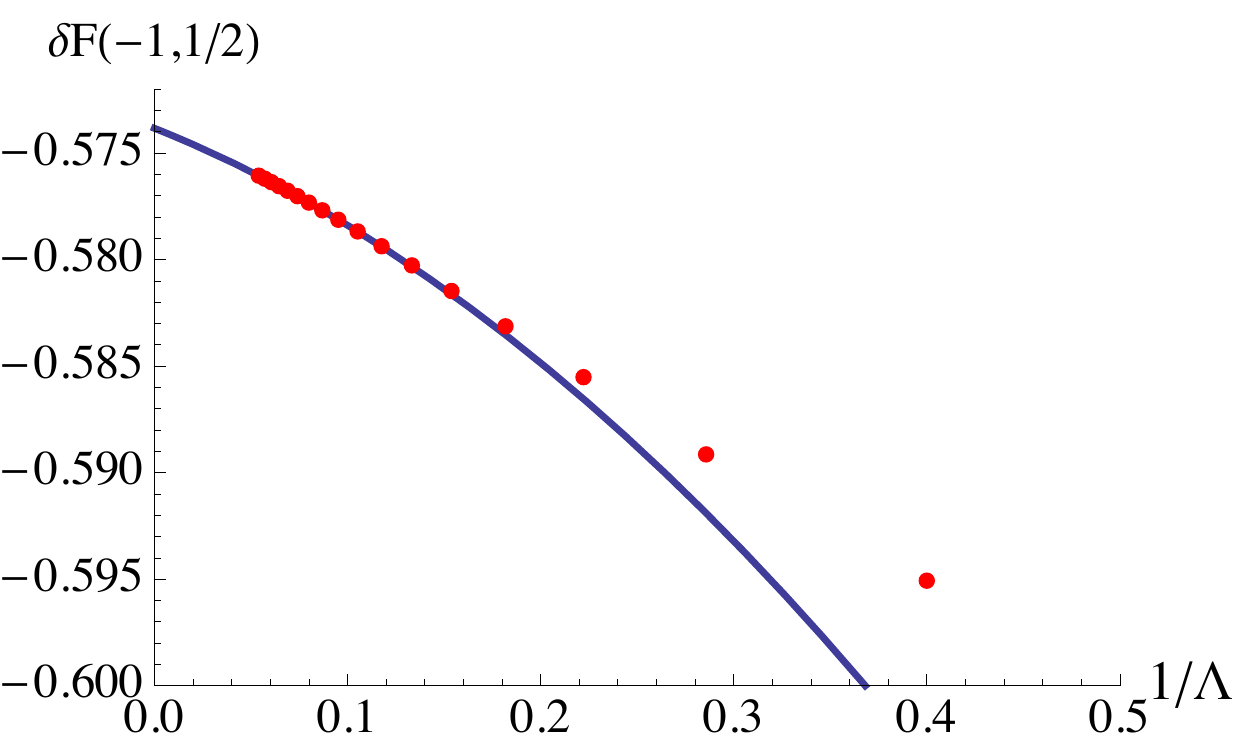}
\caption{We plot the subleading term in the free energy, $  \delta F(q,q')$ for $q=-1,\, q'=1/2$ as a function of the cutoff $\Lambda$. We extrapolate to $1/\Lambda\to 0$ by fitting the data points by a second order polynomial. Our results are reliable to $10^{-3}$ precision.}
\label{fig:FreeEnergy}
\end{center}
\end{figure}

We calculate the eigenvalues of the matrices $\textbf{K}_{qq'}^J(\Omega)$ numerically. Figure~\ref{fig:combined} shows a few eigenvalues for our choice of $q_1,\, q_2$. We drop the zero eigenvalue, and combine the ghost and gauge contributions as explained in Subsection~\ref{subsec:Combine}.  Finally, we calculate the sum over $J$ and the integral over $\Omega$ in~\eqref{FPexplicit2} for different UV cutoffs $\Lambda$ defined in~\eqref{Cutoff}. All the divergences have been regularized in previous steps in zeta-function regularization and the free energy is finite as we take $1/\Lambda\to 0$.  $\delta F(q_2,q_1)$ as a function of $1/\Lambda$ is plotted in Figure~\ref{fig:FreeEnergy}.

The terms that can appear in~\eqref{FPexplicit2} for $|q|,\, |q'|\leq 2$ will be presented in Table~\ref{qqpTable}. We can find the terms needed in~\eqref{SubleadingExample} from that table:
\es{SubleadingExample2}{
 \delta F[{\cal A}]=  -0.0383-0.574-0.574-0.194=-1.38\ .
 }
We conclude that  in $U(2)$ gauge theory the dimension of the GNO monopole operator with charges $q_1=1/2, \, q_2=-1$ is:
\es{FullExample}{
\Delta=0.938 \, N_f -1.38+{\cal O}\le(1/N_f\ri) \ .
}
We discuss the results for monopole operator dimensions more systematically in Section~\ref{DIMENSIONS}.

\section{Stability}
\label{STABILITY}

In the previous section we studied the effective action for the gauge field fluctuations in the presence of a GNO monopole background \eqref{MonopoleGeneral}.  We noticed that the effective action for the $W$ bosons (off-diagonal components of the gauge field) is not always positive-definite (see Figure~\ref{fig:stabilityplots}), which is to say that certain classical monopole backgrounds are unstable.  In this section, we discuss this instability in more detail and characterize which sets of GNO charges yield an unstable background.

The instability of certain GNO backgrounds should come as no surprise, as similar instabilities have been studied in related examples.  Indeed, it is well-known that GNO monopoles in Yang-Mills theory in flat space are generically unstable \cite{Brandt:1979kk, Coleman:1982cx}.  To characterize the unstable configurations, recall that the GNO monopoles organize themselves into classes of topologically-equivalent backgrounds, where each class corresponds to an element of the first fundamental group of the gauge group, $\pi_1(G)$.  In the case $G = U(N_c)$, we have $\pi_1(G) = \Z$, and there is a discrete topological charge that can be identified with the sum of the GNO charges, 
 \es{qtop}{
  q_\text{top} = \sum_{a = 1}^{N_c} q_a \,.
 }
The monopoles that were shown in \cite{Brandt:1979kk, Coleman:1982cx} to be unstable in Yang-Mills theory in flat space were those with $\abs{q_a - q_b} \geq 1$ for at least one pair of GNO charges $(q_a, q_b)$.  It is not hard to convince oneself that each topological class with charge $q_\text{top}$ contains precisely one stable monopole background.\footnote{The stable background has $q_a = \bar q = [q_\text{top} / N_c] +1/2$ for $a \leq (q\, \text{mod}\, N_c)$ and $q_a = \bar q - 1/2$ for the other $a > (q\, \text{mod}\, N_c)$.}  All the rest are unstable.

It is important to note that the flat-space instability of monopoles in Yang-Mills theory discussed in \cite{Brandt:1979kk, Coleman:1982cx}, as well as the instability we noticed in the previous section, occurs only at low frequencies and for $W$ bosons with total angular momentum $J = \abs{q_a - q_b} - 1$.  (This is the lowest value of the total angular momentum provided that $\abs{q_a - q_b} \geq 1$---see \eqref{UVWDef}.)  That the instability is at low frequency and low angular momentum means that it is a property of the infrared dynamics.  Different non-Abelian gauge theories with different IR dynamics can therefore have different sets of stable/unstable GNO configurations.  It just so happened that in the case of Yang-Mills theory in flat space it was all the $W$ bosons with $J = \abs{q_a - q_b} - 1$ that were unstable.  In a different theory, on the other hand, it could be that not all these lowest $J$ modes are unstable.  To assess stability, one has to examine the effective action for the gauge field fluctuations, as well as the fluctuations of other fields, and see whether there are any negative modes.

In the case studied in this paper, namely the IR fixed point of QCD$_{3}$ at large $N_f$, the question of stability is much richer than in pure Yang-Mills theory in flat space.\footnote{One could wonder how many stable monopoles there are in pure Yang-Mills theory on $S^2 \times \R$.  Since Yang-Mills theory in three dimensions is not conformal, one cannot simply borrow the flat space result, so a separate analysis is needed.  We find that if the gauge group $G = U(N_c)$, the quadratic action for the $a_\mu^{ab}$ component of the gauge field fluctuations around the GNO monopole \eqref{MonopoleGeneral} has eigenvalues $\propto \Omega^2 + J(J+1) - (q_a - q_b)^2$ (for physical modes) or $0$ (for pure gauge modes), where $\Omega$ is the frequency and $J$ is the total angular momentum.  There is an instability at low $\Omega$ for $J = \abs{q_a - q_b} - 1$, so the situation is identical to that of Yang-Mills theory in flat space.}  We find that in contrast to the pure Yang-Mills case, in QCD$_3$ there are multiple stable monopoles per topological class.  As we will discuss later, even topologically trivial gauge groups such as $SU(2)$ admit stable monopole backgrounds.

\subsection{A systematic study of monopole stability in QCD$_3$}

In performing a more systematic study of the instability of monopole backgrounds in large $N_f$ QCD with gauge group $U(N_c)$, let us first note that at leading order in $N_f$, where one can treat the gauge field as a background and ignore its fluctuations, there are no instabilities, as this is just a theory of non-interacting fermions.  To decide whether or not a given GNO background is stable, it is important to consider the subleading $1/N_f$ effects described by the effective action for the gauge field fluctuations.

In the previous section we have developed a whole machinery needed to study the eigenvalues and eigenfunctions of the quadratic action for the fluctuations of the gauge field around a GNO monopole background \eqref{MonopoleGeneral}.  In brief, each component $a_\mu^{ab}(x)$ of the gauge field fluctuation can be expanded in terms of Fourier modes in the Euclidean time direction as well as the monopole vector harmonics $U^\mu_{Q, JM}$, $V^\mu_{Q, JM}$, and $W^\mu_{Q, JM}$ defined in \eqref{UVWDef}:
 \es{aExpansionab}{
   a^{\mu, ab}(x) = \int\frac{d\Omega}{2\pi} \sum_{J, M}\biggl[
        \mathfrak{a}_{U}^{ab, J M}(\Omega) U^\mu_{Q_{ab}, J M}(\hat{n})
     +\mathfrak{a}_{V}^{ab, J M}&(\Omega) V^\mu_{Q_{ab}, J M}(\hat{n}) \\
     &{}+\mathfrak{a}_{W}^{ab, J M}(\Omega) W^\mu_{Q_{ab}, J M}(\hat{n})\biggr] e^{-i\Omega\tau} \,.
 }
For the fluctuation $a_\mu^{ab}(x)$, we should take $Q_{ab} = q_a - q_b$.  The quadratic action for the coefficients $\mathfrak{a}_{X}^{ab, JM}$ takes the form
 \es{Quadraticab}{
  N_f \sum_{a, b = 1}^{N_c} \int \frac{d\Omega}{2 \pi} \sum_{J, M} 
    \begin{pmatrix}
        \mathfrak{a}_{U}^{ab, J M}(\Omega) \\
        \mathfrak{a}_{V}^{ab, J M}(\Omega) \\ 
        \mathfrak{a}_{W}^{ab, J M}(\Omega)
      \end{pmatrix}^\dagger
       {\bf K}^J_{q_a q_b}(\Omega)
      \begin{pmatrix}
        \mathfrak{a}_{U}^{ab, J M}(\Omega) \\
     \mathfrak{a}_{V}^{ab, J M}(\Omega) \\
     \mathfrak{a}_{W}^{ab, J M}(\Omega)
     \end{pmatrix} \,.
 }
The matrix ${\bf K}^J_{q_a q_b}(\Omega)$ can be computed tediously by following all the steps presented in the previous section.  As demonstrated in the example presented in Figure~\ref{fig:stabilityplots}, the eigenvalues of ${\bf K}^J_{q_a q_b}(\Omega)$ increase with both $\Omega$ and $J$.  To check whether the action for $a_\mu^{ab}$ is positive-definite, it is therefore sufficient to calculate ${\bf K}^J_{q_a q_b}(\Omega)$ for $\Omega = 0$ and the lowest attainable value of $J = J_{ab}$.  If $\abs{Q_{ab}} <1$, this lowest value is $J_{ab} = \abs{Q_{ab}}$;  if $\abs{Q_{ab}} \geq 1$, it is $J_{ab} = \abs{Q_{ab}} - 1$.

\begin{figure}[t!]
\begin{center}
   \includegraphics[width=0.85\textwidth]{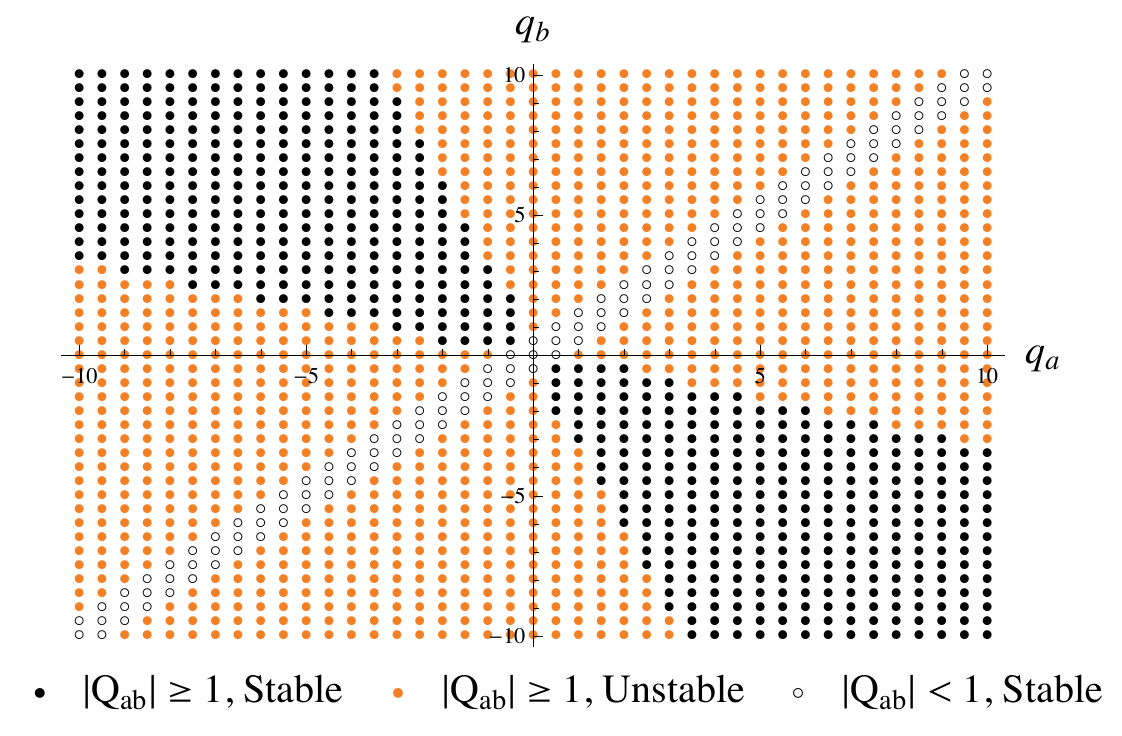}
\caption{A summary plot of the stability of GNO monopoles.  A GNO monopole with charges $\{q_1, \ldots q_{N_c} \}$ is stable provided that all pairs $(q_a, q_b)$ correspond to (open or filled) black circles, and it is unstable otherwise.  We denote $Q_{ab}= q_{a}-q_{b}$, as in the main text. The orange dots correspond to values of $(q_a, q_b)$ for which ${\bf K}_{q_a q_b}^{\abs{Q_{ab}} - 1}(0)<0$, i.e.~the effective action for $a_\mu^{ab}$ has a negative mode with $J = \abs{Q_{ab}} -1$.  The open and filled black circles correspond to values of $(q_a, q_b)$ for which there is no such negative mode.  The difference between the open and filled black circles is that for the filled ones the lowest angular momentum mode has $J = \abs{Q_{ab}} -1$, while for the open ones the lowest value of $J$ is $\abs{Q_{ab}}$. 
}
\label{fig:2dstab}
\end{center}
\end{figure}

We have computed numerically ${\bf K}^{J_{ab}}_{q_a q_b}(0)$ for all possible values of $q_a$ and $q_b$ in the range $-10 \leq q_a, q_b \leq 10$.  From our numerics, we find that it is only the modes with $J_{ab} = \abs{Q_{ab}} - 1$ and $\abs{Q_{ab}} \geq 1$ that are sometimes unstable. 
\begin{figure}[t!]
\begin{center}
   \includegraphics[width=0.6\textwidth]{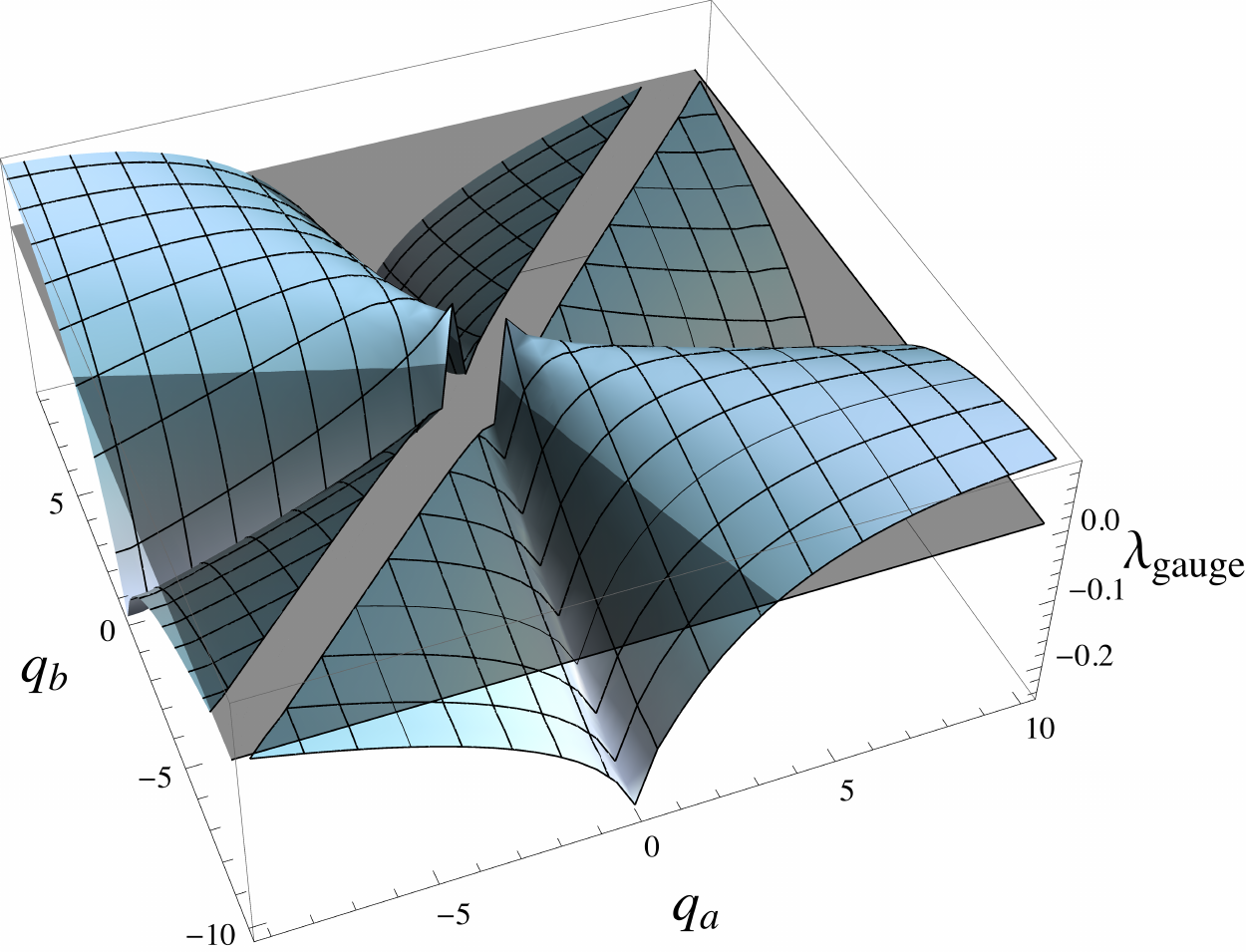}
\caption{The lowest eigenvalue $\lambda = {\bf K}^{\abs{Q_{ab}} - 1}_{q_a q_b}(0)$ of the $a_\mu^{ab}$ component of the gauge field fluctuations around the GNO monopole background \eqref{MonopoleGeneral}.  We have marked explicitly the plane $z = 0$.  The region where this eigenvalue dips below zero corresponds to an instability of $a_\mu^{ab}$.  If this eigenvalue is positive, then the action for $a_\mu^{ab}$ is positive-definite. }
\label{fig:3dstab}
\end{center}
\end{figure}
We have plotted ${\bf K}^{\abs{Q_{ab}} - 1}_{q_a q_b}(0)$ in Figure~\ref{fig:3dstab} as a function of $q_a$ and $q_b$.  In Figure~\ref{fig:2dstab} we have indicated the stable region in black and the unstable region in orange.  A GNO monopole labeled by charges $\{q_{1},q_{2},\ldots,q_{N_{c}}\}$ is stable if every pair of charges lies in the stable region displayed on the plot.

As can be seen from Figure~\ref{fig:2dstab}, we find two stable regions in the $q_a$-$q_b$ plane.  The first such region is where 
 \es{FirstStable}{
   \abs{q_a - q_b} <1 \,,
 }
which, as mentioned above, is the same stability condition as in Yang-Mills theory in flat space.  For these values of the charges there is no instability because the problematic mode with angular momentum $J = \abs{q_a - q_b} - 1$ is simply absent.  The second stable region is new and unexpected.  It occurs where $q_a$ and $q_b$ are comparable in magnitude and of opposite sign.  Asymptotically, at large values of $q_a$ and $q_b$ we can estimate that the second stable region is where
 \es{Stability}{
   -\tan 73^\circ \lesssim \frac{q_b}{q_a} \lesssim - \tan 17^\circ  \,.
 }

The existence of the second stable region implies that there are several stable GNO monopoles per topological class.  Indeed, the first stable region alone implies that each topological class contains at least one stable GNO monopole, and hence the existence of a second stable region means that there can be more than one stable monopoles in the same topological class.  For example, in $U(2)$ gauge theory where the GNO monopoles are indexed by a pair of GNO charges $(q_1, q_2)$ and the topological charge is $q_\text{top} = q_1 + q_2$, there is an infinite number stable GNO monopoles that are topologically trivial, because all monopoles with $q_\text{top} = 0$ lie in the region \eqref{Stability}.  In fact, every topological class in this theory contains an infinite number of stable GNO monopoles, because the monopoles with fixed $q_\text{top}$ lie on a line with slope $-45^\circ$ in the $q_1$-$q_2$ plane that will necessarily lie in the region \eqref{Stability} asymptotically at large $q_1$ and $q_2$.

Having finished the stability analysis we can now complete the discussion of Section~\ref{OVERVIEW}. In the following, we analyze Minkowski time evolution. In the quantum theory all field modes fluctuate about the monopole background. If the background is stable, at large $N_f$ the wave functional is supported on gauge configurations close to the background, as the typical size of the fluctuations is ${\cal O}(1/\sqrt{N_f})$. Conversely, if the background is unstable, the unstable mode grows exponentially. The wave functional that started out as having delta function support on the background spreads, and ends up with broad support. In this case, it is more useful to decompose the wave functional in terms of energy eigenstates. The spread in energy will be wide, comparable to the potential energy difference to a nearby local minimum.\footnote{For very late times tunneling has to be taken into account. We neglect tunneling effects in this discussion.} The Euclidean path integral will be dominated by the lowest energy eigenstate at the bottom of a nearby local minimum, as a result we get the free energy of another monopole. Note that there is no topological obstruction to this scenario, as every topological sector has at least one stable monopole background in it.

One could still consider the disorder operators corresponding to unstable monopole backgrounds. The decomposition of the corresponding state into energy eigenstates translates into this disorder operator being a sum of operators that have a big range of scaling dimensions. In correlation functions, at long distances such a disorder operator would behave as the operator with the lowest scaling dimension from this sum, i.e.~another monopole operator in the same topological sector (or the identity).

\section{Monopole operator dimensions}
\label{DIMENSIONS}

In this section we collect the results for monopole operator dimensions in $U(N_c)$ gauge theories.   We first exhibit the QED case $N_c = 1$ explicitly in Subsection~\ref{QEDDIMENSIONS}, and then we present the results for $N_c \geq 2$ in Subsection~\ref{QCDDIMENSIONS}.  For details on how to obtain the results collected in this section we refer the reader to the example in section~\ref{subsec:Example}.

\subsection{Monopole operator dimensions in QED}
\label{QEDDIMENSIONS}

In $U(1)$ gauge theory all monopole backgrounds are stable because the monopole  charge is a topological quantum number. To obtain the scaling dimensions of the corresponding monopole operators, one should simply set $N_c=1$ in the formulae summarized in Section~\ref{subsec:Example}.   See Table~\ref{qTable} for the scaling dimensions $\Delta_q$ of the monopole operators with $\abs{q}  \leq 5/2$.
\begin{table}[!h]
\begin{center}
\begin{tabular}{c||c}
 $\abs{q}$ & $\Delta_q$ \\
 \hline \hline
 $0$ & $0$ \\
 \hline
 $1/2$ & $0.265 \,N_f - 0.0383 + O(1/N_f)$ \\
  \hline
 $1$ & $0.673  \,N_f  - 0.194 + O(1/N_f)$ \\
  \hline
 $3/2$ & $1.186  \,N_f  - 0.422 +O(1/N_f) $ \\
  \hline
 $2$ & $1.786  \,N_f - 0.706 + O(1/N_f) $ \\
  \hline
 $5/2$ & $2.462  \,N_f - 1.04 + O(1/N_f) $
\end{tabular}
\caption{Monopole operator dimension $\Delta_q$ for monopole charge $q$ in $U(1)$ gauge theory. \label{qTable}}
\end{center}
\end{table}%

Part of these results are not new:  the ${\cal O}(N_f)$ contributions to the scaling dimensions given in Table~\ref{qTable} were first obtained in~\cite{Borokhov:2002ib}, while the subleading correction to the dimension of the monopole operators with $\abs{q} = 1/2$ was also obtained in \cite{Pufu:2013vpa}.

\subsection{Monopole operator dimensions in $U(N_c)$ QCD}
\label{QCDDIMENSIONS}

As mentioned before, in $U(N_c)$ gauge theory not all GNO backgrounds ${\cal A}$ specified by the charges $\{q_{1},q_{2},\ldots,q_{N_{c}}\}$ are stable. Stability is a dynamical question, and we presented the criterion for stability in Section~\ref{STABILITY}---see Figure~\ref{fig:2dstab}.  For the stable backgrounds, we can compute the scaling dimension $\Delta = F[{\cal A}]$ of the corresponding operators using the formulae~\eqref{FFinal2},~\eqref{LeadingSummary}, and~\eqref{FPexplicit2}, which we repeat here for the reader's convenience:
\es{FFinal3}{
  F[{\cal A}] &= N_f\, F_0[{\cal A}]  + \delta F[{\cal A}]+ {\cal O}(1/N_f)\,,
 }
 where
  \es{LeadingSummary2}{
  F_0[{\cal A}] &= \sum_{a=1}^{N_c} F_0(q_a) \,, \qquad
  \delta F[{\cal A}]= \sum_{a,b=1}^{N_c} \delta F(q_a,q_b) \,.
}
The numerical values of $ F_0(q)$ are the same as the coefficients of $N_f$ in the expressions for $\Delta_q$ given in Table~\ref{qTable}.  For $\delta F(q,q')$, see Table~\ref{qqpTable}.  Note that not all the entries in Table~\ref{qqpTable} are numerical;  some of them are instead orange dots, which indicate an instability.   According to the recipe of Section~\ref{STABILITY}, if such a term features in the second sum in~\eqref{LeadingSummary2}, the corresponding monopole background is unstable and does not correspond to a monopole operator with well-defined scaling dimension. 
 
 Using the values listed in Tables~\ref{qTable} and~\ref{qqpTable} one can determine the dimension of any monopole operator with GNO charges obeying $|q_{a}|\leq2$. For higher GNO charges one has to construct larger tables.
 
Note that the subleading terms in Table~\ref{qTable} are equal to the diagonal entries in Table~\ref{qqpTable}.  Note also that Table~\ref{qqpTable} has a reflection symmetry about the diagonal
 \es{CPConsequence}{
 \delta F(q,q')=\delta F(q',q) \,,
 }
as a consequence of CP symmetry~\eqref{FCP}, as well as a reflection symmetry about origin, $\delta F(q, q') = \delta F(-q, -q')$.
\newcommand{\unst}{\orange{$\bullet$}}
\begin{table}[htd!]
\begin{center}
\begin{tabular}{c||c|c|c|c|c|c|c|c|c}

\diagbox{$q$}{$q'$} & $-2$ & $-3/2$ & $-1$ & $-1/2$ & $0$ & $1/2$ & $1$ & $3/2$ & $2$ \
\\ 
 \hline \hline 
$2$ & $-1.90$ & $-1.63$ & $-1.52$ & $-2.16$ & \unst & \unst & \unst & 
$-0.857$ & $-0.706$ \\
 \hline
$3/2$ & $-1.63$ & $-1.26$ & $-1.04$ & $-1.05$ & \unst & \unst & 
$-0.592$ & $-0.422$ & $-0.857$ \\
 \hline 
$1$ & $-1.52$ & $-1.04$ & $-0.730$ & $-0.574$ & \unst & $-0.386$ & 
$-0.194$ & $-0.592$ & \unst \\
 \hline 
$1/2$ & $-2.16$ & $-1.05$ & $-0.574$ & $-0.338$ & $-0.258$ & 
$-0.0383$ & $-0.386$ & \unst & \unst \\
 \hline 
$0$ & \unst & \unst & \unst & $-0.258$ & $0$ & $-0.258$ & \unst & 
\unst & \unst \\
 \hline 
$-1/2$ & \unst & \unst & $-0.386$ & $-0.0383$ & $-0.258$ & $-0.338$ & 
$-0.574$ & $-1.05$ & $-2.16$ \\
 \hline 
$-1$ & \unst & $-0.592$ & $-0.194$ & $-0.386$ & \unst & $-0.574$ & 
$-0.730$ & $-1.04$ & $-1.52$ \\
 \hline 
$-3/2$ & $-0.857$ & $-0.422$ & $-0.592$ & \unst & \unst & $-1.05$ & 
$-1.04$ & $-1.26$ & $-1.63$ \\
 \hline 
$-2$ & $-0.706$ & $-0.857$ & \unst & \unst & \unst & $-2.16$ & 
$-1.52$ & $-1.63$ & $-1.90$ \\
 
\end{tabular}
\caption{$\delta F(q, q')$ for various values of $q$ and $q'$.  The orange dots mean that the corresponding $W$ boson is unstable. \label{qqpTable}}
\end{center}
\end{table}%

Let us consider a few examples:
 \begin{itemize}
\item If we take $\{q_{1},q_{2},\ldots,q_{N_{c}}\}=\{1/2,0,\ldots,0\}$, we have
\es{MonopoleEx1}{
F_0[{\cal A}] &= F_0(1/2)=0.265 \,, \\
 \delta F[{\cal A}]&=  \delta F(1/2,1/2)+2(N_c-1)\,\delta F(1/2,0) \,,\\
\Delta&=0.265\, N_f -0.0383-(N_c-1)\, 0.516+ {\cal O}(1/N_f) \,.
}
This monopole operator has the smallest dimension among all.  

\item If we instead took $\{q_{1},q_{2},\ldots,q_{N_{c}}\}=\{1,0,\ldots,0\}$, we would find that there is no monopole operator with this GNO charge and well-defined scaling dimension, because $\delta F(1, 0)=\text{\unst}$.

\item Finally, if we consider $\{q_{1},q_{2},\ldots,q_{N_{c}}\}=\{1/2,0,\ldots,0, -1/2\}$, the corresponding dimension is
\es{MonopoleEx3}{
F_0[{\cal A}] &= 2F_0(1/2)=0.530 \,, \\
 \delta F[{\cal A}]&=  2\delta F(1/2,1/2)+2\delta F(1/2,-1/2)+4(N_c-2)\,\delta F(1/2,0) \,, \\
\Delta&=0.530\, N_f -0.753-(N_c-2)\, 1.06+ {\cal O}(1/N_f) \,.
}

\end{itemize}

\section{Other quantum numbers of monopole operators}
\label{QUANTUMNUMBERS}

In the previous sections we calculated the energy of the ground state\footnote{As explained before, the use of the term ``ground state'' is not necessarily appropriate.  We are talking about the lowest energy states whose wavefunction at large $N_f$ is highly peaked around the saddle \eqref{NonAb}.} on $S^2 \times \R$ localized around the GNO saddle \eqref{NonAb}.  This computation used the equivalence between the ground state energy and the thermal free energy at zero temperature, and as such does not tell us much about the properties of the ground state, or equivalently, about the quantum numbers of the operator corresponding to it.  In this section we fill this gap.  Of course, the results presented here will only be valid for the GNO saddles that do not have any unstable directions.

The states on $S^2$ must transform in representations of the conformal group and of the flavor symmetry group.  The flavor symmetry group of a theory of $N_f$ fermions and gauge group $U(N_c)$ is $SU(N_f)$.  The conformal group on $S^2 \times \R$ is $SO(4, 1)$, regardless of whether the $\R$ coordinate is Lorentzian or Euclidean time.  We choose to work in Euclidean signature, even though time evolution is a unitary transformation on the Hilbert space of states only in Lorentzian signature.   We expect the bare monopole operators that we studied in the previous sections to be conformal primaries.  We will now determine their spin and $SU(N_f)$ quantum numbers.

\subsection{Quantum numbers of monopole operators in QED}

Before studying the quantum numbers of the GNO monopoles in QCD, it is instructive to study the same question in the QED case, $N_c = 1$, where the monopole operators are labeled by the charge $q \in \Z/2$ and heuristically create the background \eqref{AAbelian}.  The quantum numbers of the monopole operators in QED were calculated in \cite{Borokhov:2002ib} for $\abs{q}= 1/2$.  In this section we present the quantum numbers for arbitrary $q$.\footnote{Our work corrects a slight error in the analysis of  \cite{Borokhov:2002ib} for general $q$.} The result we will find is that the monopole operator of charge $q$ transforms as a Lorentz scalar, and as an irreducible representation of $SU(N_{f})$ given by the rectangular Young diagram with $N_{f}/2$ rows and $2|q|$ columns:
\begin{eqnarray}
{\tiny N_f/2} \Bigl\{   \underbrace{ {\tiny \ydiagram{3, 3}} }_{2 \abs{q}} \,.
\end{eqnarray}
We now explain the derivation of this result.

Because at large $N_f$ the fluctuations of the gauge field around the background \eqref{AAbelian} are suppressed, we should start by canonically quantizing the theory of free fermions in this background, and worry later about the effects of having a dynamical gauge field.  The fermionic modes can be found by solving the Dirac equation
 \es{Dirac}{
  (i \slashed{D} + \slashed{\cal A}) \psi^\alpha = 0 \,.
 }
To solve this equation, one can begin by expanding $\psi$ in terms of the spinor harmonics $S_{q, jm}(\hat n)$ and $T_{q, jm}(\hat n)$, and Fourier modes in time, as in Section~\ref{subsec:FermionDet}:
 \es{FermionExpansion}{
  \psi^\alpha (\hat{n}, t) &=\sum_{j = \abs{q} - \frac 12}^\infty \sum_{m = -j}^j \left[c^{(S)\alpha}_{jm} S_{q, jm}(\hat n) + c^{(T)\alpha}_{jm}T_{q, jm}(\hat n) \right]  e^{-i \omega \tau}   
 } 
 with arbitrary $\omega$ to be determined by solving the Dirac equation.  When $j \geq \abs{q} + 1/2$, the Dirac equation has two solutions (for every $j$, $m$ and flavor $\alpha$) with energy $i \omega_j = \pm E_{q, j}$, where
 \es{Energy}{
  E_{q, j} = \sqrt{(j+1/2)^2 - q^2} \,.
 }
We can denote by $c^{(\pm)\alpha}_{jm}$ the linear combinations of $c^{(S)\alpha}_{jm}$ and $c^{(T)\alpha}_{jm}$ corresponding to energy $\pm E_{q, j}$.  In the quantum theory, the $c^{(\pm)\alpha}_{jm}$ become anti-commuting annihilation operators for the corresponding modes.  Generically, there are $2j+1$ such operators for both choices of sign and every $\alpha$ and $j$.

The case $j =  \abs{q} - 1/2$ is special because the spinor monopole harmonics $T_{q, jm}$ are absent, and the Dirac equation implies that the $S_{q, jm}$ modes have energy $i \omega_j = 0$.  In the quantum theory, the coefficients $c^{(S)\alpha}_{jm}$, which in this case we denote by $c^{\alpha}_{jm}$ for brevity, become annihilation operators for these zero-energy modes.  For each flavor, there are $2j+1=2\abs{q}$ zero-energy modes transforming in the spin $\abs{q} - 1/2$ representation of the $SU(2)_\text{rot}$ rotation group.  There are a total of $2 \abs{q} N_f$ zero-energy modes when we consider all of the flavors.

If there had been no zero-energy modes, the situation would have been quite simple.  The theory would have had a unique rotationally-invariant vacuum $|\Omega\rangle$, corresponding to a Dirac sea filled with particles with negative energy and containing no particles with positive energy.  In other words, this vacuum should be annihilated by all annihilation operators for positive energy modes and by all creation operators for all negative energy modes:
 \es{NaiveVacuum}{
  c^{(+)\alpha}_{jm} |\Omega \rangle = c^{(-)\alpha\dagger}_{jm} |\Omega \rangle = 0 \,, \qquad
   j > \abs{q} - \frac 12 \,.
 }
One could build up the states in the Hilbert space by adding particles with positive energy or removing particles with negative energy.   

The existence of $2 \abs{q} N_f$ zero-energy modes, however, means that the theory of free fermions in the monopole background \eqref{AAbelian} has, in fact, not just one, but $2^{2 \abs{q} N_f}$ degenerate ground states that satisfy the condition \eqref{NaiveVacuum}.  Let us call the Hilbert space spanned by these ground states ${\cal G}$.  One of the states in ${\cal G}$ is the Fock vacuum $|\Omega \rangle$, defined by \eqref{NaiveVacuum} together with the requirement that it should be annihilated by the annihilation operators $c_{jm}^{\alpha}$ with $j = \abs{q} - \frac 12$:
 \es{AdditionalVacuum}{
  c^\alpha_{jm} |\Omega \rangle = 0 \,, \qquad
   j = \abs{q} - \frac 12  \,.
 }
The other linearly independent states in ${\cal G}$ can be obtained by acting with any number of creation operators $c_{jm}^{\alpha\dagger}$ (with $j = \abs{q} - \frac 12$), on the rotationally-invariant Fock vacuum $|\Omega \rangle$.  The full Hilbert space of the theory ${\cal H}$ is obtained by acting on the states of ${\cal G}$ with any number of creation operators $c^{(+)\alpha\dagger}_{jm}$ for positive-energy modes and annihilation operators $c^{(-)\alpha}_{jm}$ for negative energy modes ($j > \abs{q} - \frac 12$). 

This description of the Hilbert space ${\cal H}$ is correct assuming the gauge field is a background field.  For us, however, the gauge field is dynamical and its effect is to remove some of the states in ${\cal H}$ (and consequently some of the ground states in ${\cal G}$).  In the path integral language the gauge field appears in the action only as a Lagrange multiplier that imposes the constraint $j^\mu(x) = 0$. What we mean by this constraint is that all correlation functions of the current should vanish in the full theory
\es{CurrentVanish}{
0=\< j^{\mu_1}(x_1)\, j^{\mu_2}(x_2)\, \dots\,  j^{\mu_n}(x_n)  \>_\text{ full theory} \ .
}
This equation looks perplexing at first sight, as we spent most of the paper determining the gauge kernel $K^{\mu\nu}(x,y)=-\<j^\mu(x) j^\nu(y)\>_\text{conn}$. The resolution of this puzzle is that $K^{\mu\nu}$ is determined by the current-current correlator in the free fermion theory, where the gauge field is treated as a background. 

In canonical quantization language we have a constrained system; the canonical momenta conjugate to $a^\mu(x) $ vanish identically, and we should not define any oscillator modes in the gauge sector. Instead, in analogy with the Gupta--Bleuler prescription, we should require that the positive and zero energy part, $j^{(+)}_\mu$ of the current operator $j_\mu(x) = :\psi^{\alpha\dagger}(x) \gamma_\mu \psi^\alpha(x) :$ annihilates all physical states $|\chi \rangle$:
 \es{CurrentAnnihilates}{
  j^{(+)}_\mu(x) |\chi \rangle = 0 \,.
 } 
This requirement reduces the Hilbert space ${\cal H}$ introduced above to a smaller one ${\cal H}_\text{phys}$, and the $2^{2 \abs{q} N_f}$-dimensional space of ground states ${\cal G}$ to ${\cal G}_\text{phys}$.  Understaning the quantum numbers of the monopole operators means understanding what ${\cal G}_\text{phys}$ is, and how the $SU(N_f)$ flavor symmetry and the rotation group $SU(2)_\text{rot}$ acts on it. The~\eqref{CurrentAnnihilates} condition guarantees that~\eqref{CurrentVanish} is satisfied. Of course, the expectation value has to be taken between states in ${\cal H}_\text{phys}$.\footnote{It is easy to check that imposing only the strictly positive energy part of $j^\mu$ to annihilate physical states is not sufficient to ensure~\eqref{CurrentVanish}. We have to require the stronger condition~\eqref{CurrentAnnihilates}.}

Using \eqref{FermionExpansion} and the definition of $j^\mu(x)$, one can obtain an explicit expression for the current operator in terms of oscillators:
 \es{CurrentExpansion}{
  j^{(+)}_\mu(x) = \sum_{m, m'} \left( c_{jm}^{\alpha\dagger} c_{jm'}^\alpha - {\cal C} \delta_{m m'} \right)  S_{q, jm}^\dagger(\hat n)  \gamma_\mu S_{q, jm'}(\hat n) + \text{(non-zero modes)} \,,
 }
where we wrote down explicitly only the contributions from the oscillators corresponding to the zero-energy modes with $j = \abs{q} - \frac 12$.  (From here on, it should be understood that $j = \abs{q} - 1/2$ and that $m$ runs over $-j$ through $j$ unless otherwise specified.)  The quantity ${\cal C}$ appearing in \eqref{CurrentExpansion} is a $c$-number corresponding to a possible normal-ordering ambiguity when taking the product of $c_{jm}^{\alpha\dagger}$ with $c_{j'm'}^{\alpha'}$.  Such a normal ordering ambiguity is present only when $j=j'$, $m = m'$, and $\alpha = \alpha'$, because any given $c_{jm}^\alpha$ anti-commutes with all the other fermionic creation and annihilation operators except for $c_{jm}^{\alpha\dagger}$.

The normal ordering constant is determined by CP-invariance to be ${\cal C} = N_f /2$.  Indeed, creating a zero-energy mode is related by CP to destroying a zero-energy mode, and if we want to quantize the theory in a CP-invariant way, we better treat the creation and annihilation operators for zero-energy modes on equal footing.  Doing so means that instead of the expression in the parenthesis in \eqref{CurrentExpansion} we should have written
 \es{parenthesis}{
  \frac 12 \left( c_{jm}^{\alpha\dagger} c_{jm'}^\alpha - c_{jm'}^\alpha c_{jm}^{\alpha\dagger} \right) \,.
 } 
Anti-commuting the two factors in the second term, summing over $\alpha$, and comparing with \eqref{CurrentExpansion} yields ${\cal C} = N_f / 2$.

Using the explicit expressions for the spinor monopole harmonics, it is straightforward to find the explicit position dependence of the zero-mode contribution to the current operator \eqref{CurrentExpansion}.  For instance, when $q=1/2$, the zero modes have spin $j=0$.  In the North chart, the expression for the only spinor harmonic with $j=0$ is
 \es{Spinor0}{
  S_{\frac 12, 00}(\hat n) = -\frac{1}{\sqrt{4 \pi}} \begin{pmatrix}
   \cos \frac \theta 2 \\
   e^{i \phi} \sin \frac \theta 2
  \end{pmatrix} \,.
 } 
After plugging this expression into \eqref{CurrentExpansion}, a little algebra shows that the zero modes do not contribute to $j^\theta(x)$ and $j^\phi(x)$, while the charge density $\rho(x) \equiv j^\tau(x)$ is
 \es{rho0}{
  q = \frac 12: \qquad \rho(x) = \frac{1}{4 \pi} \left( \hat {\cal N} - N_f/2 \right)  + \text{(non-zero modes)}  \,,
 }
where $\hat {\cal N} \equiv c_{00}^{\alpha\dagger} c_{00}^\alpha$ is the operator that counts the total number of excited zero-energy modes.  More generally, the operator that counts the total number of fermions in the zero-mode sector is
 \es{NDef}{
  \hat {\cal N} \equiv \sum_m c_{jm}^{\alpha\dagger} c_{jm}^\alpha \,.
 } 
 That the charge density $\rho(x)$ annihilates the states means that ${\cal G}_\text{phys}$ consists only of the states in ${\cal G}$ for which $\hat {\cal N} = N_f/2$.  

Similarly, when $q = 1$, the zero modes have spin $j=1/2$.  In the North chart, the spinor harmonics are
 \es{SpinorHalf}{
  S_{1, \frac 12 \frac 12} (\hat n) = \frac{1}{\sqrt{8 \pi}}
   \begin{pmatrix}
    e^{i \phi} \sin \theta \\
    e^{2 i \phi} \left( 1- \cos \theta \right)
   \end{pmatrix} \,, \qquad
    S_{1, \frac 12,- \frac {1}2} (\hat n) = 
     \begin{pmatrix}
      1 + \cos \theta \\
      e^{i \phi} \sin \theta 
     \end{pmatrix} \,.
 } 
Again, using these expressions one can show that the zero modes do not give any contributions to $j^\theta(x)$ and $j^\phi(x)$, while the charge density is
 \es{rhoHalfAgain}{
    q = 1: \qquad \rho(x) &= \frac{1}{\sqrt{4 \pi}} Y_0^0(\hat n)^* (\hat {\cal N} - N_f) -\frac{1}{\sqrt{6 \pi}} \sum_m Y_1^{-m}(\hat n)^* \, S_{m}
     + \text{(non-zero modes)} \,.
 }
Here, $\hat {\cal N}$ is the fermion number operator in the zero-mode sector defined in \eqref{NDef} and $S_m$ is the total spin of the zero-energy modes organized as states in the spin-$1$ angular momentum basis:
 \es{TotalSpin}{
  S_{1} &= -c_{\frac 12 \frac 12}^{\alpha\dagger} c_{\frac 12, -\frac 12}^\alpha = -S_x - i S_y \,, \\
  S_0 &= \frac{1}{\sqrt{2}} \left[ c_{\frac 12 \frac 12}^{\alpha\dagger} c_{\frac 12 \frac 12}  - c_{\frac 12, -\frac 12}^{\alpha\dagger} c_{\frac 12, -\frac 12}^\alpha \right]
   = \sqrt{2} S_z \,, \\
  S_{-1} &= c_{\frac 12, -\frac 12}^{\alpha\dagger} c_{\frac 12 \frac 12}^\alpha = S_x- i S_y \,.
 }
The requirement that the charge density should vanish for any physical states implies that ${\cal G}_\text{phys}$ consists only of the states in ${\cal G}$ that satisfy $\hat {\cal N} = N_f$ and $\vec{S} = 0$.

The expressions \eqref{rho0}--\eqref{rhoHalfAgain}, as well as the characterization of ${\cal G}_\text{phys}$ as a subspace of ${\cal G}$, generalize to arbitrary $q$ in the following way.  From the creation and annihilation operators in the zero-energy mode sector one can construct $SU(N_f)$-singlet operators of the form
 \es{Bilinear}{
  {\hat {\cal O}} = \sum_{m, m'} {\cal O}^{mm'}  c^{\alpha \dagger}_{j m} c^{\alpha}_{jm'} \,,
 }
where ${\cal O}^{mm'}$ is a $2 \abs{q} \times 2 \abs{q}$ Hermitian matrix.  There are $\left(2 \abs{q}\right)^2$ linearly independent such matrices, and hence $\left(2 \abs{q}\right)^2$ linearly independent operators $\hat {\cal O}$, which organize themselves according to irreducible representations of the rotation group $SU(2)_\text{rot}$.  The representations that appear are precisely those in the product of two spin-$j$ irreps of $SU(2)$, namely all the ones whose spin is between $\ell = 0$ and $\ell = 2j$.  If we denote the spin-$\ell$ operators by $\hat {\cal O}_{\ell m_\ell}$, then $\hat {\cal O}_{00}$ is proportional to the total fermion number $\hat {\cal N}$, $\hat {\cal O}_{1m}$ is proportional to the total spin $S_m$, and so on.  

The expression for the charge density operator in \eqref{rho0} and \eqref{rhoHalfAgain} then generalizes to 
 \es{rhoGeneral}{
  \rho(x) = \frac{1}{4 \pi} \Bigl( \hat {\cal N} - \abs{q} N_f \Bigr) + \sum_{\ell=1}^{2\abs{q} - 1}  \sum_{m_\ell} Y_\ell^{-m_\ell} (\hat n) \hat {\cal O}_{\ell m_\ell} + \text{(non-zero modes)}\,,
 }
where $Y_\ell^{m_\ell}(\hat n)$ are the usual spherical harmonics, and the operators $\hat {\cal O}_{\ell m_\ell}$ come with specific normalizations.  The precise normalization of $\hat {\cal O}_{\ell m_\ell}$ is not essential for the argument we are about to make.  That the charge density annihilates all the states means that out of the $2^{2 \abs{q} N_f}$ degenerate ground states in ${\cal G}$ we should only consider the ones where 
 \es{Requirements}{
  \hat {\cal N}  |\chi \rangle = \abs{q} N_f |\chi \rangle  \qquad \text{and} \qquad
   \hat {\cal O}_{\ell m_\ell} |\chi \rangle = 0 \,, \qquad \text{for $\ell\geq 1$ and all $m_\ell$.}
 }

The first requirement in \eqref{Requirements} means that ${\cal G}_\text{phys}$ contains only states of the form
 \es{States}{
  \prod_{i = 1}^{\abs{q} N_f} c^{\alpha_i \dagger}_{j m_i} |\Omega \rangle \,.
 }
where we act with precisely $\abs{q} N_f$ fermion creation operators (out of the total of $2 \abs{q} N_f$) on the Fock vacuum $|\Omega \rangle$ defined in \eqref{NaiveVacuum}--\eqref{AdditionalVacuum}.  The second requirement in \eqref{Requirements} requires more thought.  It can be understood most simply by enlarging the $SU(2)_{\textrm{rot}}$ symmetry to an $SU(2\abs{q})$ symmetry, where the $(2|q|)^{2}-1$ operators $\hat{\mathcal{O}}_{lm_{l}}$ generate $SU(2\abs{q})$ and the states $c_{jm}^{\alpha\dagger}|\Omega\rangle$ transform in the fundamental representation. The second condition in \eqref{Requirements} means that all operators \eqref{Bilinear} where ${\cal O}^{mm'}$, is a traceless Hermitian matrix should annihilate the physical states $|\chi\rangle$.\footnote{Note that the non-traceless part of $j$ is included in the operator $\hat{\mathcal{N}}$.} This is just the requirement that, infinitesimally, $|\chi\rangle$ should be invariant under $SU(2 \abs{q})$ transformations.  In other words, ${\cal G}_\text{phys}$ consists of the states of ${\cal G}$ that are of the form \eqref{States} and, in addition, are also $SU(2 \abs{q})$ singlets.

Each fermionic creation operator transforms as a fundamental of $SU(N_{f})$, so we are looking for singlets under $SU(2\abs{q})$ in the product of $\abs{q} N_{f}$ fundamentals of $SU(N_{f})$. There is a further wrinkle, however. As we are considering anti-commuting creation operators, the states must be totally antisymmetric.

To count how many such states there are and see how they transform under $SU(N_f)$, it is convenient to introduce a bigger group that contains both $SU(N_f)$ and $SU(2 \abs{q})$:  if we make a list of all the zero-energy mode creation operators $c^{\alpha\dagger}_{jm}$, we can consider $SU(2 \abs{q} N_f)$ transformations under which $c^{\alpha\dagger}_{jm}$ form a fundamental vector.  Similarly, the annihilation operators $c^\alpha_{jm}$ transform in the anti-fundamental representation of the same $SU(2 \abs{q} N_f)$ group.  

The benefit of considering this larger group is that constructing totally antisymmetric states is simple. The states of ${\cal G}_\text{phys}$ are constructed by decomposing the anti-symmetric products of $\abs{q} N_f$ fundamentals of $SU(2 \abs{q} N_f)$ under $SU(2\abs{q})\times SU(N_{f})$ and selecting those which are singlets under the $SU(2\abs{q})$ factor.  We therefore need to identify all the $SU(2\abs{q})$ singlets in the decomposition of the rank-$\abs{q} N_f$ totally antisymmetric representation of $SU(2 \abs{q} N_f)$, 
 \es{YoungAntisymm}{
  \abs{q} N_f \left\{ {\tiny \ydiagram{1, 1, 1, 1, 1, 1} }\right.
 }
under 
 \es{Map}{
  SU(2 \abs{q} N_f) \supset SU(2\abs{q}) \times SU(N_f) \,.
 }
Such a group theory exercise is common in atomic physics where one needs to construct a totally anti-symmetric wavefunction for several identical particles with given angular momentum and spin.  In general, the rank $r$ anti-symmetric representation of $SU(NM)$ decomposes under $SU(N) \times SU(M)$ as the sum (see, for example, \cite{GroupTheory})
 \es{SumDiagrams}{
  \bigoplus_\nu (\nu, \tilde \nu)
 }
over all possible irreps with Young diagrams $\nu$ with a total of $r$ boxes (whose conjugates are denoted by $\tilde \nu$), such that $\nu$ has at most $N$ rows and $\tilde \nu$ has at most $M$ rows.  Each ordered pair $(\nu, \tilde \nu)$ appears precisely once in this decomposition.  For our problem, we have
 \es{OurDecomposition}{
  \abs{q} N_f \left\{ {\tiny \ydiagram{1, 1, 1, 1, 1, 1} }\right. 
   \to  \left(  {\tiny 2 \abs{q} }\Biggl\{  \underbrace{ {\tiny \ydiagram{2, 2, 2}  } }_{N_f /2} \  , \  \underbrace{ {\tiny \ydiagram{3, 3}} }_{2 \abs{q}} \Bigr\} {\tiny N_f/2} \right) \oplus \cdots \,,
 }
where on the RHS the first Young diagram of any given pair corresponds to $SU(2 \abs{q})$ and the second to $SU(N_f)$.  Of this infinite sum, we want to pick out the terms for which the first factor is an $SU(2\abs{q})$ singlet. Only the diagram explicitly exhibited in \eqref{OurDecomposition} has this property.   Consequently, the states of ${\cal G}_\text{phys}$ transform as the $SU(N_f)$ irrep whose Young diagram is a rectangle with $N_f/2$ rows and $2 \abs{q}$ columns:
 \es{SUNfRep}{
  {\tiny N_f/2} \Bigl\{   \underbrace{ {\tiny \ydiagram{3, 3}} }_{2 \abs{q}} \,.
 }
 (Recall that we should only consider an even number of flavors in order to avoid a parity anomaly.)  The dimension of this irrep is
 \es{IrrepDimension}{
  \prod_{i=0}^{2 \abs{q}-1} \frac{i! (i + N_f)!}{\left( (i + N_f /2 )! \right)^2} \,.
 }
See Table~\ref{SampleDiagrams} for a few examples.

\begin{table}[t]
\begin{center}
\begin{tabular}{c||c|c|c|c}
  & $N_f = 2$ & $N_f = 4$ & $N_f = 6$ & $\cdots$ \\
 \hline
 \hline
 $q = 1/2$ & ${\tiny \ydiagram{1}}\, ({\bf 2}) $ & $ {\tiny \ydiagram{1, 1}}\, ({\bf 6})$ & $ {\tiny \ydiagram{1, 1, 1}}\, ({\bf 20})$ & $\cdots$ \\[3ex]
 \hline
 $q = 1$ & ${\tiny \ydiagram{2}}\, ({\bf 3}) $ & $ {\tiny \ydiagram{2, 2}}\, ({\bf 20})$ & ${\tiny \ydiagram{2, 2, 2}}\, ({\bf 175})$ & $\cdots$ \\[3ex]
  \hline
 $q = 3/2$ & ${\tiny \ydiagram{3}}\, ({\bf 4}) $ & $ {\tiny \ydiagram{3, 3}}\, ({\bf 50})$ & ${\tiny \ydiagram{3, 3, 3}}\, ({\bf 980})$ & $\cdots$ \\[3ex]
 \hline
 $\vdots$ & $\vdots$ & $\vdots$ & $\vdots$ & $\ddots$
\end{tabular}
\end{center}
\caption{The transformation properties of the first few (bare) monopole operators under the flavor $SU(N_f)$ global symmetry of QED$_3$ with $N_f$ flavors.  The dimensions of the irreps were calculated using \eqref{IrrepDimension}.  All these monopole operators are singlets under spatial rotations.}
\label{SampleDiagrams}
\end{table}%

This discussion also shows that the physical ground states ${\cal G}_\text{phys}$ are singlets under the rotation group $SU(2)_\text{rot}$.  Indeed, $SU(2)_\text{rot}$ can be embedded as a subgroup of $SU(2 \abs{q})$, and the states that are $SU(2 \abs{q})$ singlets must also be $SU(2)_\text{rot}$ singlets.

We have thus found the quantum numbers of the (bare) monopole operators of charge $q$ in QED with $N_f$ flavors.  Their topological charge is $q$, and their conformal dimensions were computed in the previous section at large $N_f$ (see Table~\ref{qTable}).  In this section we determined that these operators transform as the irrep \eqref{SUNfRep} (see also Table~\ref{SampleDiagrams}) under the flavor $SU(N_f)$ symmetry and as singlets under the $SU(2)_\text{rot}$ group of spatial rotations.

\subsection{Quantum numbers of monopole operators in $U(N_c)$ QCD}

The careful analysis of the previous section can be generalized to the more complicated GNO monopole operators in $U(N_c)$ QCD with $N_f$ flavors in the fundamental representation.  As in the QED case, when $N_f$ is large we can start by quantizing the theory of free fermions in the GNO background \eqref{NonAb}, and then we can take into account the effects of having a dynamical gauge field.  The result is that the monopole operators now transform in an irrep of the $SU(N_f)$  flavor symmetry corresponding to a Young diagram with $N_f/2$ rows and $2 \sum_{a=1}^{N_c} \abs{q_a}$ boxes in each row,
 \es{SUNfRepNonAb}{
  {\tiny N_f/2} \Bigl\{   \underbrace{ {\tiny \ydiagram{3, 3}} }_{ 2 \sum_a \abs{q_a}} \,,
 }
where $\{q_a\}$ is the set of GNO charges.  In addition, the monopole operators are singlets under the $SU(2)_\text{rot}$ rotation group.  We obtained the same result in the Abelian case, but now $\abs{q}$ is replaced with $\sum_{a}\abs{q_{a}}$. The rest of this section provides the derivation of these quantum numbers in the non-Abelian case.

In the non-Abelian case, the fermions $\psi^{a, \alpha}$ carry a color index $a=1,\dots,N_c$ in addition to the flavor index $\alpha$.  In the GNO background \eqref{NonAb}, the action for $\psi^{a, \alpha}$ is the same as that of a QED fermion in an Abelian monopole background \eqref{AAbelian} with charge $q = q_a$.  We therefore have $2 \abs{q_a} N_f$ zero energy modes for each value of $a$, with some corresponding creation operators $c^{a, \alpha\dagger}_{j_a m_a}$ and annihilation operators $c^{a, \alpha}_{j_a m_a}$ (here, $j_a = \abs{q_a} - 1/2$ and $m_a$ ranges from $-j_a$ through $j_a$).  In addition to the zero energy modes, we also have positive and negative energy modes.  As in the Abelian case, we can define the vacuum in the non-zero mode sector by requiring that all positive-energy annihilation operators and all negative-energy creation operators annihilate this vacuum.  These conditions leave $2^{2 N_f \sum_{a} \abs{q_a}}$ degenerate ground states (as appropriate for having $2 N_f \sum_{a = 1}^{N_c} \abs{q_a}$ fermionic oscillators with zero energy) that span a Hilbert space ${\cal G}$.  This Hilbert space has a Fock vacuum $|\Omega\rangle$, which by definition is annihilated by all $c^{a, \alpha}_{j_a m_a}$.  The other linearly independent states in ${\cal G}$ can be constructed by acting with any number of fermionic creation operators $c^{a, \alpha\dagger}_{j_a m_a}$ on $|\Omega \rangle$. 

The analysis so far did not take into account the dynamical gauge field, which, as in the QED case, acts as a Lagrange multiplier that imposes the constraint that the positive and zero energy part of the current, $j^{(+)}_{ba,\mu}(x)$, should annihilate all physical states.  This constraint reduces ${\cal G}$ to a smaller Hilbert space ${\cal G}_\text{phys}$, whose transformation properties under the flavor group $SU(N_f)$ and the rotation group $SU(2)_\text{rot}$ we need to understand, as each state in ${\cal G}_\text{phys}$ corresponds to a monopole operator.

The creation operators transform as fundamentals under $SU(N_{f})$. In the Abelian case, we saw that it was useful to consider $SU(2\abs{q})$ acting on the creation operators. The condition of vanishing current in that case translated into two constraints that select the states of ${\cal G}_\text{phys}$ from ${\cal G}$.  The first constraint required the number of creation operators be equal to $\abs{q}N_{f}$, and the second required that the states in ${\cal G}_\text{phys}$ be singlets under this $SU(2\abs{q})$. Similarly, in the case of $U(N_{c})$ we can consider the action of $SU(2\sum_{a}\abs{q_{a}})$ on the set of all creation operators of a fixed flavor.  Note that this group mixes fermions of different color and spin.   We will argue that the condition of vanishing current in this case translates into the following two constraints: each state in ${\cal G}_\text{phys}$ is created by acting with $\sum_{a}\abs{q_{a}}N_{f}$ creation operators on the Fock vacuum, and it should transform as a singlet under $SU(2\sum_{a}\abs{q_{a}})$.  The problem of finding physical states is then just the same group theory problem we solved in the Abelian case with $\abs{q}$ replaced by $\sum_{a}\abs{q_{a}}$. 

As in the Abelian case, each diagonal component $j_{aa, \mu}^{(+)}$ imposes the constraint that the number of creation operators of color $a$ equal $\abs{q_{a}}N_{f}$ and that the physical states $\ket{\chi}$ are invariant under $SU(2 \abs{q_a})$. It will be more useful, however, to consider the overall constraint coming from $\sum_{a}j_{aa, \mu}^{(+)}$, which says that the total number of generators of all colors is $\sum_{a}\abs{q_{a}}N_{f}$.  The other constraints coming from $j_{aa, \mu}^{(+)}$ imply that the physical states are invariant under $U(1)^{N_{c}-1}\times\prod_{a}SU(2\abs{q_{a}})$.\footnote{The factors of $U(1)$ come from the separate particle number constraints for each color, with one removed corresponding to the total particle number.}  We will now argue that invariance of the physical states under this latter group enhances to invariance under a full $SU(2\sum_{a}\abs{q_{a}})$ when one also examines the off-diagonal generators $j_{ba, \mu}^{(+)}$ with $b \neq a$.

 For simplicity we start by considering $N_c = 2$, where the discussion above implies that the conditions coming from $j_{11, \mu}^{(+)}$ and $j_{22, \mu}^{(+)}$ require invariance under $U(1)\times SU(2\abs{q_{1}})\times SU(2\abs{q_{2}})$.  Recall that, in general, $U(1)\times SU(M)\times SU(N)$ is a maximal subgroup of $SU(M+N)$. Therefore, if a state is a singlet under $U(1)\times SU(M)\times SU(N)$ and is annihilated by any other generator of $SU(M+N)$, it is automatically a singlet under the whole $SU(M+N)$. For the $N_c =2$ case, the off diagonal current $j_{12, \mu}^{(+)}$ provides at least one additional condition independent from the ones coming from $j_{11, \mu}^{(+)}$ and $j_{22, \mu}^{(+)}$, which required that the physical states be annihilated by the generators of $U(1)\times SU(2\abs{q_{1}})\times SU(2\abs{q_{2}})$.  As such, the physical states must be singlets under the full $SU(2\abs{q_{1}}+2\abs{q_{2}})$.

 For the general case, $U(1)^{N_{c}-1}\times\prod_{a}SU(2\abs{q_{a}})$ is not quite a maximal subgroup of $SU(2\sum_{a}\abs{q_{a}})$.  Instead, for each pair of indices $(a, b)$ with $a\neq b$, the subgroup $U(1)\times SU(2\abs{q_{a}})\times SU(2\abs{q_{b}})\subset SU(2\abs{q_{a}}+2\abs{q_{b}})$ is maximal.\footnote{We only need to consider pairs of indices, $( a, b)$, where $q_{a}\neq0$ and $q_{b}\neq0$. For vanishing charge the $SU(2\abs{q})$ factor is not present.} Repeating the argument above from the $N_c = 2$ case, the off-diagonal current $j_{ba, \mu}^{(+)}$, $b \neq a$, is non-vanishing, so it provides an additional constraint on the physical states beyond the invariance under $U(1)\times SU(2\abs{q_{a}})\times SU(2\abs{q_{b}})$ required by $j_{aa, \mu}^{(+)}$ and $j_{bb, \mu}^{(+)}$.    The physical states $\ket{\chi}$ are therefore singlets under $SU(2\abs{q_{a}}+2\abs{q_{b}})$ for every pair $(a, b)$. Iterating this procedure for all pairs of color indices leads to the singlet condition under the full $SU(2\sum_{a}\abs{q_{a}})$.
 
Putting everything together, the states in ${\cal G}_\text{phys}$ are the $SU(2 \sum_a \abs{q_a})$ singlets in the decomposition of the totally anti-symmetric tensor of $SU(2 N_f \sum_a \abs{q_a})$ with $N_f \sum_a \abs{q_a}$ indices under $SU(N_f) \times SU(2 \sum_a \abs{q_a})$. It follows that the states of ${\cal G}_\text{phys}$ transform under $SU(N_f)$ as the irrep \eqref{SUNfRepNonAb}.  The corresponding monopole operators are singlets of $SU(2)_\text{rot}$ because $SU(2)_\text{rot}$ is embedded in $SU(2 \sum_a \abs{q_a})$ and we selected only the $SU(2 \sum_a \abs{q_a})$ singlets.

A generalization of these results to more complicated groups and/or representations of the fermion flavors is left for future work.

\section{Monopoles in general gauge theories}
\label{GENERAL}

In this section we generalize the computation of the dimension of GNO monopole operators, as well as the stability analysis included in Section~\ref{STABILITY}, to arbitrary gauge groups.  We will see that the computation proceeds analogously to the $U(N_c)$ case, and, moreover, no new ingredients are needed.  In particular, to complete the study of gauge field fluctuations around a monopole background in QCD$_3$ with gauge group $G$, all that is needed are the properties of the kernel ${\cal K}_{qq'}(x, x')$ analyzed in the $U(N_c)$ case.

In gauge theory with gauge group $G$, the most general monopole background centered at the origin is 
 \es{calAAgain}{
  {\cal A} = H (\pm 1 - \cos \theta) d\phi \,,
 }
where $H$ is an element of the Lie algebra of $G$, and the two signs correspond to the North and South charts.  As explained in \cite{Goddard:1976qe}, each such configuration is gauge equivalent to one where $H$ points along the Cartan, namely
\es{HCartan}{
H=\sum_{i=1}^r q_i\, h_i\equiv q\cdot h \,,
}
where $r$ is the rank of the gauge group and $h_i$ are the Cartan generators.  There is still some remaining gauge redundancy in \eqref{HCartan}, as the Weyl group acts non-trivially on the $q_i$, and we should hence consider configurations of the form \eqref{HCartan} only as equivalence classes under the action of the Weyl group.

The Dirac quantization condition is \cite{Goddard:1976qe}
 \es{DiracCond}{
  \ket{w}_R=e^{4\pi  i\, H}\ket{w}_R=e^{4\pi i\, q\cdot w}\ket{w}_R
 }
for any state $\ket{w}_R$ in any representation $R$ of $G$.  Here, $w$ is the weight vector corresponding to $\ket{w}_R$ (such that $h_i \ket{w}_R = w_i \ket{w}_R$), and so it belongs to the weight lattice of $G$.  The quantization condition \eqref{DiracCond} implies that $q\cdot w\in\mathbb{Z}/2$ for any $w$.  The set of all $q$ with this property form themselves a lattice that can be identified with a rescaled version of the weight lattice of a dual group $\widetilde G$.  The group $\widetilde G$ is referred to as the GNO dual (or Langlands dual) of $G$.

\subsection{Anomalous dimensions for general groups}
\label{GENERALGROUP}

In general, the fermions transform in some representation $R$ of $G$.  Let us denote the states of this representation by $\ket{w}$, suppressing from now on the index $R$ that we introduced above.  In terms of these states, the fermions can be decomposed as 
\es{FermionRep}{
\psi(x)=\sum_{w \in R} \psi_w(x)\, \ket{w} \ ,
}
with $\psi_w(x)$ being anti-commuting spinor coefficients.  To avoid clutter, the flavor and spinor indices are suppressed.  Like in the $U(N_c)$ case, having $N_f$ flavors of fermions has the only effect of multiplying the gauge field effective action by a factor of $N_f$.   

Similarly, the gauge field background and fluctuations can be decomposed in terms of the states in the adjoint representation of $G$.  Some of the components point along the Cartan generators $h_i$, and some along the root directions $E_\alpha$:
\es{GaugeDirections}{
\mathcal{A}=q\cdot h\,  {\cal A}^{U(1)} \,, \qquad
 a = \sum_{i=1}^r a_i\, h_i+\sum_{\alpha \in \text{roots}} a_{\alpha}\, E_{\alpha}  \,.
}
The $h_i$ and $E_\alpha$ are defined such that they satisfy the standard commutation relations
\es{RootsComm}{
\le[h_i,\, h_j\ri]=0 \qquad \le[h_i,\, E_\alpha\ri]=\alpha_i\, E_\alpha \,.
}

As in the $U(N_c)$ case, the large $N_f$ expansion is equivalent to an expansion in the gauge field fluctuations $a_\mu$.  To leading order in $N_f$ we can thus treat the gauge field as a background and write the action for the fermions in the background ${\cal A}$ as
\es{S0General}{
S_0\le[{\cal A};\psi^\dagger,\psi\ri]=\int d^3x\, \sqrt{g}\, \sum_w \psi_w^\dagger \le(i \slashed{\nabla} + q\cdot w \, \slashed{\cal A}^{U(1)}\ri) \psi_w \ .
}
Since this action does not mix fermions with different weights $w$, the Green's function takes the form 
 \es{psiwGreen}{
 \langle \psi_w(x) \psi^\dagger_{w'}(x') \rangle = \delta_{ww'}G_{ q\cdot w}(x, x') \,,
 }
where $G_{ q\cdot w}(x, x')$ is as defined in~\eqref{GOneFermion}. 

The corrections to \eqref{S0General} come from the coupling between the fermions and the gauge fluctuations, which is
\es{CurrentFermion}{
S_\text{int}\le[a,\psi^\dagger,\psi\ri]&=\int d^3x\, \sqrt{g}\, \sum_{w,w'} \bra{w'} a^\mu \ket{w}\, j_{w w', \mu} \ , \qquad
j_{w w'}^\mu \equiv \,\,  :\psi^\dagger_{w'} \gamma^\mu \psi_w:\ .
}
In complete analogy with the discussion of Section~\ref{GAUGEEFFECTIVE}, we can obtain the effective action for the gauge field fluctuations by integrating out the fermions. As in~\eqref{GotK}, it will be useful to define the kernel
 \es{GotKGeneral}{
  K^{\mu\nu}_{uv, ws} (x, y) \equiv -\langle j^\mu_{vu}(x) j^\nu_{sw}(y) \rangle_\text{conn}  \ ,
 }
and rewrite this kernel in terms of the single fermion Green's function using~\eqref{psiwGreen}.  We have
 \es{GotKAgainGeneral}{
  K^{\mu\nu}_{uv, ws} (x, y) = N_{f}\delta_{vw}\delta_{us}\, \mathcal{K}_{q\cdot v,q\cdot u}^{\mu\nu}(x,y) \,, 
 }
with $ {\cal K}^{\mu\nu}_{q_b, q_a}(x, y)$ defined in~\eqref{calKDef}. Finally, using~\eqref{CurrentFermion}, we can write the effective action for the gauge fluctuations as 
\es{SeffFinalGeneral}{
  S_\text{eff}[a] &= N_f  \tr \log (i \slashed{D}^{({\cal A})})  + S^{(2)}_\text{eff}[a] + \cdots \\
  S^{(2)}_\text{eff}[a] &\equiv \frac {N_f}2 
     \int d^3x\, d^3y \sqrt{g(x)} \sqrt{g(y)} \,\sum_{w,w'} \bra{w'} a_\mu(x) \ket{w} \, 
    {\cal K}^{\mu\nu}_{q\cdot w,q\cdot w'} (x, y)\, \bra{w} a_\nu(y) \ket{w'}  \,.
 }
This expression is the analog of~\eqref{SeffFinal} from the $U(N_c)$ case. We can be more explicit and decompose the gauge fluctuations in terms of internal directions as in~\eqref{GaugeDirections}.  Using the fact that the states $\ket{w}$ are orthonormal, we can write
 \es{aDecomp}{
  \bra{w'} a^\mu(x) \ket{w} = a^\mu\cdot w\, \delta_{w'w}+\sum_{\alpha \in \text{roots}} a_{\alpha}^\mu\, \bra{w'} E_{\alpha}\ket{w} \,,
 }
where the short-hand notation $a\cdot w=\sum_{i=1}^r a_i\, w_i$ involves only a sum over the Cartan components.  Combining \eqref{aDecomp} with \eqref{SeffFinalGeneral}, we see that the cross terms between the fluctuations along the Cartan and root directions are proportional to $\bra{w} E_{\alpha}\ket{w}$, which vanishes, so these two sets of fluctuations decouple from each other.  Furthermore, the gauge field fluctuations in different root directions do not mix either, because
\es{RootProperty}{
\bra{w'} E_{\alpha}\ket{w}\bra{w} E_{\beta}\ket{w'}=\begin{cases}
&0 \qquad \qquad\qquad \qquad\quad\text{if $\beta\neq-\alpha$ or  $w'\neq w-\alpha$\,,}\\
&\le|\bra{w-\alpha} E_{\alpha}\ket{w}\ri|^2 \qquad \;\text{if $\beta=-\alpha$ and $w'=w-\alpha$\,.}\\
\end{cases}
}
We finally obtain:
\es{SeffFinalGeneral2}{
  S^{(2)}_\text{eff}[a] &=   \frac {N_f }2 \int d^3x\, d^3y \sqrt{g(x)} \sqrt{g(y)} \,\le[\sum_{i,j=1}^r   a_{i,\mu}(x)  \, 
    \le(\sum_w w_iw_j\,{\cal K}^{\mu\nu}_{q\cdot w,q\cdot w} (x, y)\ri)\, a_{j,\nu}(y) \ri. \\
    &\le.+   \sum_{\alpha \in \text{roots}}  a_{\alpha,\mu}(x) \, 
   \le( \sum_{w}\le|\bra{w-\alpha} E_{\alpha}\ket{w}\ri|^2\,  {\cal K}^{\mu\nu}_{q\cdot w,q\cdot (w-\alpha)} (x, y)\ri)\,  a_{-\alpha,\mu}(y)\ri]\,.
 }
Note that this expression can be computed using only the kernel ${\cal K}^{\mu\nu}_{q q'}(x, y)$ analyzed in Section~\ref{DETERMINANTS}.

Having found the effective action for the gauge field, we can now compute the free energy on $S^2 \times \R$ by evaluating the path integral on this space in the saddle point approximation.   Let us first examine the fermion determinant term in \eqref{SeffFinalGeneral}, as this term gives the leading contribution to the free energy. As can be seen from~\eqref{S0General}, $S_0[{\cal A}; \psi, \psi^\dagger]$ decomposes into a sum where each fermion $\psi_w$ is only coupled to an Abelian monopole background of charge $q \cdot w$.  In analogy with~\eqref{FreeLeading}, we obtain 
\es{FreeLeadingGeneral}{
F_0[{\cal A}] =\sum_{w} F_0(q \cdot w) \ , 
}
where $F_0(q)$ is the same quantity as defined in~\eqref{FqFinal} that is equal to the ground state energy of a single fermion in an Abelian monopole background of charge $q$.  To leading order in $N_f$, the ground state energy on $S^2 \times \R$, or equivalently the scaling dimension of the corresponding GNO monopole operator, equals
 \es{DeltaLeading}{
  \Delta = N_f F_0[{\cal A}] + {\cal O}(N_f^0) \,.
 }

As in the $U(N_c)$ case, the ${\cal O}(N_f^0)$ contribution to the scaling dimension \eqref{DeltaLeading} receives contributions both from the Faddeev-Popov ghosts and from the gauge field fluctuations. Let us start by examining the Faddeev-Popov contribution, which can be computed from a generalization of the path integral in~\eqref{ZFP}:
 \es{ZFPGeneral}{
  Z_\text{FP}[{\cal A}]  = \int Dc\,  e^{-S_\text{ghost}} \,, \quad
   S_\text{ghost} = \frac 12 \int d^3x\, \sqrt{g}  \, 
     \Bigl\langle \partial^\mu c - i [{\cal A^\mu}, c] \, \Big\vert \, 
      \partial_\mu c - i [{\cal A_\mu}, c]  \Bigr\rangle\,,
 }
where $c$ is an anti-commuting scalar ghost valued in the Lie algebra.   In \eqref{ZFPGeneral}, $\langle \cdot \vert \cdot \rangle$ is the standard inner product on the Lie algebra, defined such that (see for example \cite{Georgi:1999wka})
\es{Traces}{
 \langle h_i \vert  h_j \rangle=\delta_{ij}\,,
    \qquad 
  \langle h_i \vert E_\alpha \rangle = 0 \,, \qquad 
  \langle E_\alpha \vert E_\beta \rangle =\delta_{\alpha\beta}  \,.
}
Transforming in the adjoint representation of the gauge group, the ghosts $c$ can be decomposed just like the gauge field fluctuations in~\eqref{GaugeDirections} into components along the Cartan and components along the root directions:
 \es{GhostDecomposition}{
c=\sum_{i=1}^r c_i\, h_i+\sum_{\alpha \in \text{roots}} c_\alpha\, E_{\alpha}  \ .
}
Using the commutators \eqref{RootsComm} and the normalization conditions \eqref{Traces}, we obtain the analog of~\eqref{SghostExplicit}:
 \es{SghostExplicitGeneral}{
   S_\text{ghost} = {1\ov2} \int d^3x\, \sqrt{g(x)}  \, \le[\sum_{i=1}^r \abs{\partial_\mu c_i }^2+  \sum_{\alpha  \in \text{roots}} \abs{\partial_\mu c_\alpha - i \,q\cdot\alpha\,  {\cal A}^{U(1)}_\mu\, c_\alpha }^2\ri] \,.
 }
Following the same steps as in Section~\ref{subsec:FPDet}, we find that the ghost contribution to $F$ is:
\es{FPexplicitGeneral}{
 F_\text{FP}[{\cal A}]= -\frac 12 \int \frac{d\Omega}{2 \pi}&\le[r \sum_{J=0}^\infty(2J+1)\, \log\left[J (J + 1) + \Omega^2 \right]  \ri.\\
 &\le.+\sum_{\alpha  \in \text{roots}} \sum_{J=|q\cdot \alpha|}^\infty(2J+1)\, \log\left[J (J + 1) - (q\cdot\alpha)^2 + \Omega^2 \right]  \ri] \,.
}
As can be seen from the first line of this expression, the $r$ ghosts in the Cartan directions give equal contributions.

To obtain the full ${\cal O}(N_f^0)$ correction to \eqref{DeltaLeading}, we should also include the contribution coming from the gauge field fluctuations.  We can split this contribution as
\es{GaugeResultGeneral}{
F_\text{gauge} [{\cal A}] &=F_\text{Cartan} +\sum_{\alpha  \in \text{roots}} F_\text{root} (\alpha)\ ,
}
where $F_\text{Cartan}$ is obtained by integrating out the Cartan components of the gauge field, and each $F_\text{root} (\alpha)$ comes from the component along the root $\alpha$. 

In general, all the Cartan contributions mix with one another.  Performing the same decomposition in vector spherical harmonics as in Section~\ref{GAUGEDET}, we obtain
\es{GaugeResultCartan}{
F_\text{Cartan} &\equiv  \frac 12 \Tr' \log \le(\sum_w w_iw_j\,{\cal K}_{q\cdot w,q\cdot w}\ri)  \\
&= \frac 12 \int\frac{d\Omega}{2\pi}\sum_{J=1}^{\infty} (2J+1)\log \det{}' \, \le(\sum_w w_iw_j\,\textbf{K}_{q\cdot w,q\cdot w}^J(\Omega)\ri) \,,
}
where $\textbf{K}$ is the same object as the one appearing in the $U(N_c)$ result~\eqref{BuildingBlocks}.  For each $J$ and $\Omega$ we now have to calculate the determinant of the matrix $\sum_w w_iw_j\,\textbf{K}_{q\cdot w,q\cdot w}^J(\Omega)$.  The dimension of this matrix is $3r \times 3r$ because there are $3$ spatial directions (or equivalently, there are $3$ vector harmonics $U_{Q, JM}$, $V_{Q, JM}$, and $W_{Q, JM}$), and $r$ Cartan elements.  As a sanity check, we note that all fluctuations in the Cartan directions are regular vector fields on $S^2\times \R$, and therefore they can be decomposed in terms of the $Q = 0$ vector spherical harmonics. The total angular momentum $J$ hence takes integer values. We also note that $\sum_w w_iw_j\,\textbf{K}_{q\cdot w,q\cdot w}^J(\Omega)$ has $r$ vanishing eigenvalues, as required by gauge invariance, and that the contribution of the $r$ uncharged ghosts exhibited explicitly in the first line of~\eqref{FPexplicitGeneral} cancels the integrand in \eqref{GaugeResultCartan} asymptotically at  large $J$ and $\Omega$, just as in the $U(N_c)$ case.

As mentioned before, the fluctuations in the root directions decouple and can be examined individually.  Each such fluctuation gives a contribution equal to
\es{GaugeResultRoots}{
F_\text{root}(\alpha) &\equiv  \frac 12 \Tr' \log \le( \sum_{w}\le|\bra{w-\alpha} E_{\alpha}\ket{w}\ri|^2\,  {\cal K}_{q\cdot w,q\cdot (w-\alpha)} \ri)\\
&= \frac 12 \int\frac{d\Omega}{2\pi}\sum_{J=\le|q\cdot\alpha\ri|-1}^{\infty} (2J+1)\log \det{}' \, \le( \sum_{w}\le|\bra{w-\alpha} E_{\alpha}\ket{w}\ri|^2\,  \textbf{K}_{q\cdot w,q\cdot (w-\alpha)}^J(\Omega) \ri)\ .
}
Unlike in the Cartan case, here we need to add the matrices $\textbf{K}$ instead of taking their tensor product, so in evaluating \eqref{GaugeResultRoots} we only need to calculate the eigenvalues of a $3\times3$ matrix. A sanity check in this case is that the vector spherical harmonics that appear in the decomposition of ${\cal K}_{q\cdot w,q\cdot (w-\alpha)}$ have the same $Q=q\cdot w-q\cdot (w-\alpha)=q\cdot\alpha$ for all the terms of the sum over $w$ in \eqref{GaugeResultRoots}. For every root there is a ghost that cancels the integrand in \eqref{GaugeResultRoots} asymptotically at large $J$ and $\Omega$.

We should emphasize that in preparing the results presented in Section~\ref{DIMENSIONS} in the $U(N_c)$ case, we calculated the matrices $\textbf{K}_{qq'}^J(\Omega)$ for all $|q| \leq 2$ and $|q'|\leq2$.  It is not hard to use the same matrices combined with the needed group theory data in \eqref{GaugeResultCartan}--\eqref{GaugeResultRoots} in order to calculate the scaling dimension of any monopole operator that has $|q\cdot w|\leq2$ for all weights $w$ of the matter representation.  

We can also see easily which GNO backgrounds are unstable.  The instability only arises for the gauge fluctuations in the root directions for which  $\le|q\cdot\alpha\ri|\geq1$. For the lowest $J=\le|q\cdot\alpha\ri|-1$ mode, $\textbf{K}_{q\cdot w,q\cdot (w-\alpha)}^{\le|q\cdot\alpha\ri|-1}(\Omega)$ is simply a number, as the $V_{Q, JM}$ and $W_{Q, JM}$ modes in \eqref{UVWDef} are absent.  Hence the condition of stability is
\es{StabilityGeneral}{
0< \sum_{w}\le|\bra{w-\alpha} E_{\alpha}\ket{w}\ri|^2\,  \textbf{K}_{q\cdot w,q\cdot (w-\alpha)}^{\le|q\cdot\alpha\ri|-1}(0) \ .
 }
Here, $\textbf{K}_{q\cdot w,q\cdot (w-\alpha)}^{\le|q\cdot\alpha\ri|-1}$ is evaluated at $\Omega = 0$ because, as one can check, it is a monotonically increasing function of $\Omega$;  it is thus sufficient to check its sign for $\Omega = 0$.  The expression on the right-hand side of this equation can be easily evaluated in particular cases.   We now provide a few examples.

\subsection{Examples} \label{subsec:ResultsGeneral}

In this subsection we use the formulae derived in the previous subsection to obtain some of the monopole dimensions for various gauge groups.  We start with $G=U(N_c)$ and demonstrate the equivalence of the results obtained in the previous subsection with those in the previous parts of the paper. We then move on to discuss several gauge groups with rank $r=1$ and $2$.

\subsubsection{Another perspective on $U(N_c)$ QCD with $N_f$ fundamental fermions}

As a first example, let us see how the $G = U(N_c)$ results presented in Sections~\ref{STABILITY} and \ref{DIMENSIONS} fit within the general group framework of this section.  The $N_f$ fermions transform in the fundamental representation of $U(N_c)$, whose weight vectors $w$ are
 \es{wFundUN}{
  w^a = e^a \,,
 }
where $a = 1, \ldots, N_c$, and the $e^a$ form the standard unit frame in $\R^{N_c}$.  In components, $e^a_i = \delta^a_i$.  The adjoint representation has $N_c$ Cartan elements $h_i$, as well as $N_c^2 - N_c$ roots
 \es{rootsUN}{
  \alpha^{ab} = e^a - e^b \,, \qquad a \neq b \,.
 }
(We could identify the Cartan elements with $\alpha^{aa} = 0$.)  Since the set of all possible weights is the lattice $\Z^{N_c}$, the Dirac quantization condition implies that $q \in (\Z/2)^{N_c}$.  In other words, the GNO monopoles are indexed by $N_c$ charges $q_a$ that are half-integers.

Using the weights \eqref{wFundUN} and roots \eqref{rootsUN} in \eqref{FreeLeadingGeneral}, \eqref{FPexplicitGeneral}, \eqref{GaugeResultCartan}, and \eqref{GaugeResultCartan}, one can straightforwardly reproduce the $U(N_c)$ formulae in \eqref{LeadingSummary} and \eqref{FPexplicit2}.  In doing so, it is helpful to note that in \eqref{GaugeResultRoots} we have $\le|\bra{w^c-\alpha^{ab}} E_{\alpha^{ab}}\ket{w^c}\ri| = \delta^{bc}$.

\subsubsection{$SU(2)$ QCD with fundamental fermions}

\begin{figure}[h!]
\begin{center}
        \includegraphics[width=0.49\textwidth]{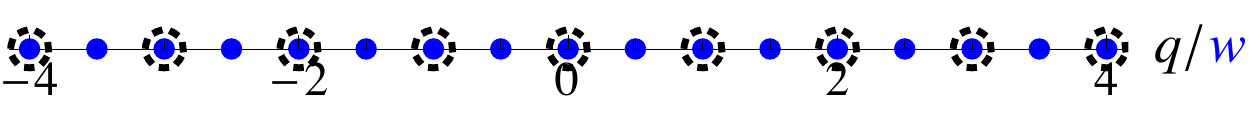}
\caption{The weight lattice of $SU(2)$ (blue dots) as well as the lattice of all possible monopole charges (dotted circles).  The monopole charges are defined modulo the action of the Weyl group, which in this case is $\Z_2$ and acts as reflections about the origin.}
\label{su2FigureLattice}
\end{center}
\end{figure}

Our second example is where $G = SU(2)$ with $N_f$ fermions transforming in the fundamental representation of $SU(2)$.  The group $SU(2)$ has rank $r = 1$, so its roots and weights are simply numbers.  The weights of the fundamental representation are 
 \es{SU2Weights}{
  w^1 = \frac 12 \,, \qquad w^2 = -\frac 12 \,.
 }
The adjoint consists of a Cartan element and two generators with roots 
 \es{SU2Roots}{
  \alpha^{12} = w^1 - w^2 = 1 \,, \qquad \alpha^{21} = w^2 - w^1 = -1 \,.
 }
The weight lattice is generated by the fundamental weights \eqref{SU2Weights} and is therefore $\Z/2$.  The monopole charges that satisfy the condition \eqref{DiracCond} are all $q \in \Z$, modulo the Weyl group---see Figure~\ref{su2FigureLattice}.  The Weyl group consists of $\Z_2$ reflections about the origin, so the monopoles with charge $q$ and $-q$ should be identified. Concretely, we can think of a monopole with charge $q$ as the background where 
 \es{ASU2}{
  {\cal A} = \begin{pmatrix} q/2 & 0 \\
  0 & -q/2 \end{pmatrix} (\pm 1 - \cos \theta) d\phi \,.
 }

\begin{figure}[h!]
 \begin{minipage}[c]{0.49\textwidth}
          \includegraphics[width=\textwidth]{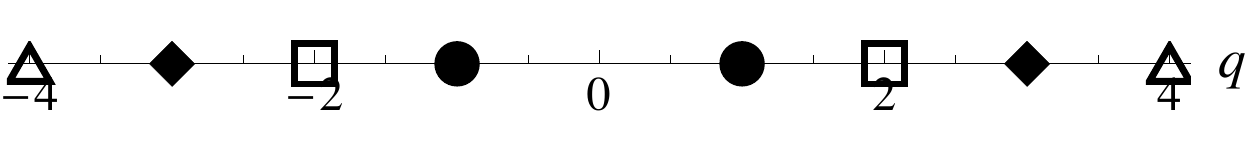}
 \end{minipage}
 \begin{minipage}[c]{0.49\textwidth}        
          \input{su2Table.txt}
 \end{minipage} 
 \begin{center}         
\caption{The $SU(2)$ monopoles appearing as black dotted circles in Figure~\ref{su2FigureLattice}.  In the presence of $N_f$ fundamental fermions these backgrounds are all stable, and we list the scaling dimensions $\Delta$ of the corresponding monopole operators.}
\label{su2Figure2}
\end{center}
\end{figure}

All these monopole backgrounds are topologically trivial because $\pi_1(SU(2))$ is also trivial.  However, they are all stable because the stability condition \eqref{StabilityGeneral} reduces to $\textbf{K}_{q/2,-q/2}^{0}(0)>0$, which can be seen to be true from Figure~\ref{fig:2dstab}.   The scaling dimensions are
 \es{ScalingSU2}{
   \Delta = 2 F_0(q/2) N_f  +  \biggl[ \delta F(q/2, q/2) + 2 \delta F(q/2, -q/2) \biggr]
     + {\cal O}(1/N_f) \,.
 }
The numerical values of $F_0(q/2)$ as well as $\delta F(q/2, \pm q/2)$ can be read out from Tables~\ref{qTable} and~\ref{qqpTable}. See Figure~\ref{su2Figure2} for specific examples.

\subsubsection{$SO(3)$ QCD with fundamental fermions}

Our third example is where the gauge group is $G = SO(3)$.  The Lie algebra of $SO(3)$ is identical to that of its double covering, which is $SU(2)$, but $SO(3)$ has fewer allowed representations than $SU(2)$.  In particular, the spinor representation considered in the previous subsection is not a representation of $SO(3)$.  The smallest irrep of $SO(3)$ is the fundamental, which in this case is the same as the adjoint.  The weights are
 \es{SO3AdjWeights}{
  w^{1} = 1 \,, \qquad w^2 = 0 \,, \qquad w^3 = -1 \,,
 }
where $w^2$ corresponds to the Cartan element, and $w^1$ and $w^3$ to the roots.

\begin{figure}[h!]
\begin{center}
        \includegraphics[width=0.49\textwidth]{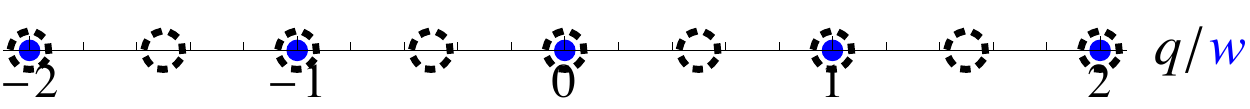}
\caption{The $SO(3)$ weight lattice (blue dots) and its dual lattice (dashed circles). The weight lattice is a sublattice of the $SU(2)$ weight lattice in Figure~\ref{su2FigureLattice}.  The dual lattice contains more monopole charges $q$ than the dual lattice of $SU(2)$.  As in the $SU(2)$ case, the Weyl acts by reflections about the origin, so it provides the identification $q \sim -q$ on the set of monopole charges.}
\label{so3FigureLattice}
\end{center}
\end{figure}

The weight lattice of $SO(3)$ is generated by the fundamental weights \eqref{SO3AdjWeights}, so it can be identified with $\Z$.  It is a subset of the weight lattice of $SU(2)$.  The quantization condition \eqref{DiracCond} implies that in $SO(3)$ the allowed values of $q$ are $q \in \Z/2$, and not just $q \in \Z$ as was the case for $SU(2)$.  See Figure~\ref{so3FigureLattice}.  
As in the case of $SU(2)$, the Weyl group acts by reflections about the origin, so we should identify the monopoles with $\pm q$.  Unlike the case of $SU(2)$, however, we now have a non-trivial fundamental group, as $\pi_1(SO(3)) = \Z_2$.  The topological charge is $(2q) \mod 2$, so the extra values of $q$ that are allowed in $SO(3)$ but not allowed in $SU(2)$ correspond to topologically non-trivial monopole backgrounds.

\begin{figure}[h!]
 \begin{minipage}[c]{0.49\textwidth}
          \includegraphics[width=\textwidth]{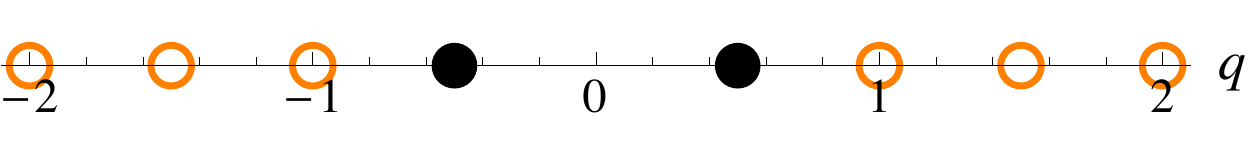}
 \end{minipage}
 \begin{minipage}[c]{0.49\textwidth}        
          \input{so3Table.txt}
 \end{minipage} 
 \begin{center}         
\caption{The $SO(3)$ monopoles appearing as black dotted circles in Figure~\ref{so3FigureLattice}.  Here, we consider these backgrounds in the presence of $N_f$ fermions transforming in the three-dimensional fundamental representation of $SO(3)$.  The orange circles correspond to unstable backgrounds.  For the stable backgrounds (represented in black by various shapes), we list the scaling dimensions $\Delta$ of the corresponding monopole operators.}
\label{so3Figure2}
\end{center}
\end{figure}

The stability condition \eqref{StabilityGeneral} reduces to ${\bf K}^0_{q, 0}(0) > 0$ in this case.  As can be seen from Figure~\ref{fig:2dstab}, the only stable monopole background is that with $\abs{q} =1/2$.  Note that this stable monopole background is also topologically non-trivial.  The scaling dimension of the corresponding monopole operator is 
 \es{ScalingSO3}{
  \Delta = 2 F_0(1/2) N_f + \biggl[\delta F(1/2, 1/2) + 2 \delta F(1/2, 0) \biggr] + {\cal O}(1/N_f) \,.
 }
See Figure~\ref{so3Figure2}.

\subsubsection{$SU(3)$ QCD with fundamental fermions}

Our next example is QCD with gauge group $SU(3)$ and $N_f$ fermions in the fundamental representation.  The rank of $SU(3)$ is $r = 2$, so the roots and weights are points in $\R^2$.

\begin{figure}[h!]
\begin{center}
        \includegraphics[width=0.49\textwidth]{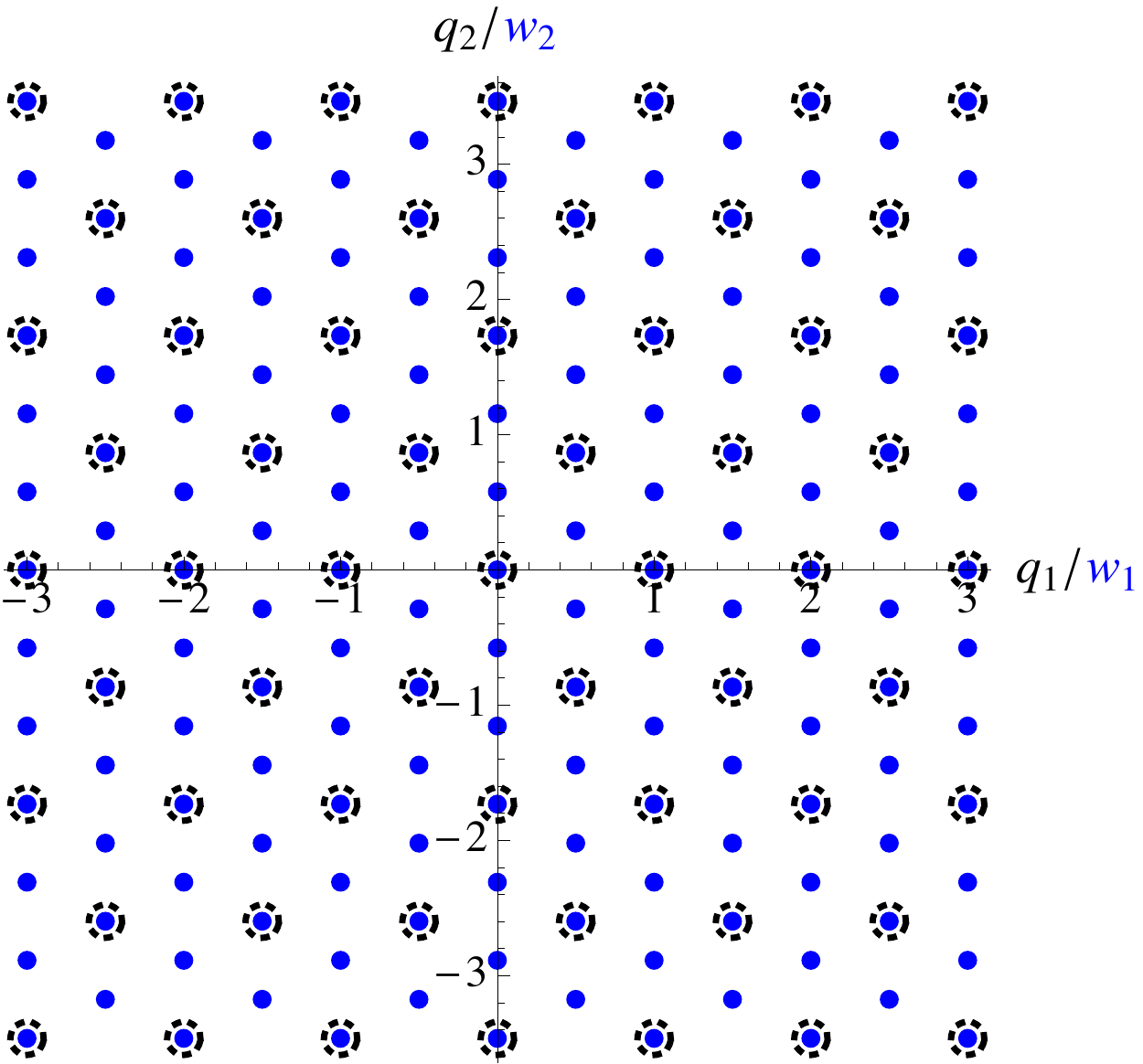}
\caption{The weight lattice of $SU(3)$ (blue dots) as well as the lattice of all possible monopole charges (dotted circles).  The monopole charges are defined modulo the action of the Weyl group, which in this case is $S_3$ and is generated by $120$ degree rotations as well as reflections about the $q_2$ axis.}
\label{su3Figure}
\end{center}
\end{figure}

The weights of the fundamental representation are
 \es{FundWeights}{
  w^1 = \left(\frac 12, \frac{\sqrt{3}}{6} \right) \,, \qquad
   w^2 = \left( -\frac 12, \frac{\sqrt{3}}{6} \right) \,, \qquad
   w^3 = \left( 0 , -\frac{\sqrt{3}}{3} \right) \,.
 }
The adjoint consists of two Cartan generators, as well as six roots given by
 \es{RootsSU3}{
  \pm \left(1, 0 \right) \,, \qquad
   \pm \left(-\frac 12, \frac{\sqrt{3}}{2} \right) \,, 
   \qquad
   \pm \left(\frac 12 , \frac{\sqrt{3}}{2} \right) \,.
 }
The weight lattice is generated by the fundamental weights \eqref{FundWeights}.  Dirac quantization implies that the monopole charges belong to a lattice generated by
 \es{DualSU3}{
  \left(1, 0 \right) \,, \qquad
   \left( -\frac 12, \frac{\sqrt{3}}{2} \right) \,.
 }
This lattice is the weight lattice of the GNO dual group $SU(3) / \Z_3$.  See Figure~\ref{su3Figure}.  

\begin{figure}[h!]
 \begin{minipage}[c]{0.49\textwidth}
          \includegraphics[width=\textwidth]{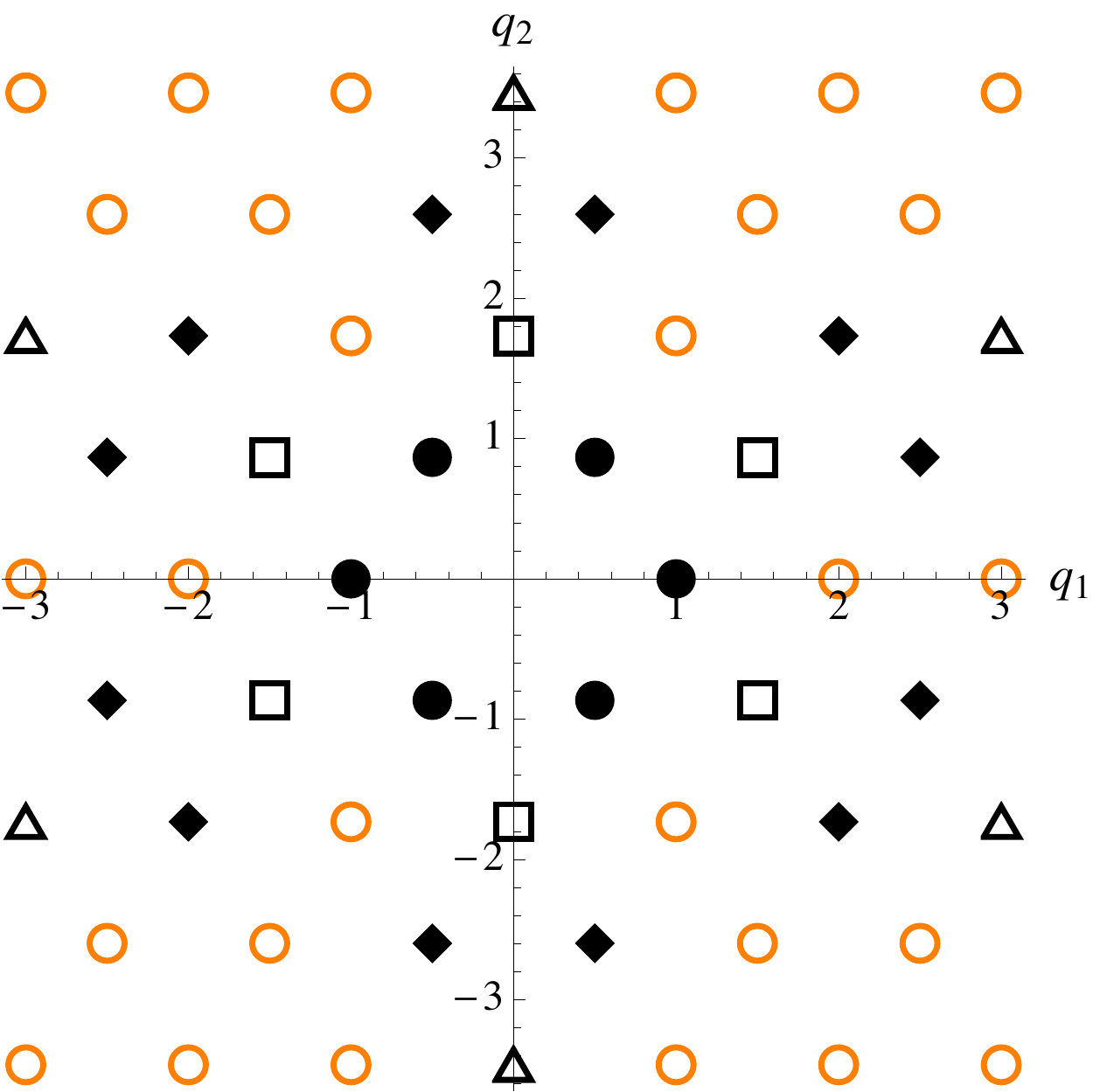}
 \end{minipage}
 \begin{minipage}[c]{0.49\textwidth}        
           \input{su3Table.txt}
 \end{minipage} 
 \begin{center}         
\caption{The $SU(3)$ monopoles appearing as black dotted circles in Figure~\ref{su3Figure}.  Here, we consider these backgrounds in the presence of $N_f$ fermions transforming in the three-dimensional fundamental representation of $SU(3)$.  The orange circles correspond to unstable backgrounds.  For the stable backgrounds (represented in black by various shapes), we list the scaling dimensions $\Delta$ of the corresponding monopole operators.}
\label{su3Figure2}
\end{center}
\end{figure}

In this case, the Weyl group is $S_3$  and is generated by rotations of $2\pi/3$ and reflections about the $q_{2}$ axis.  The independent monopoles are thus given by the points $(q_{1},q_{2})$ in the monopole charge lattice modded out by the action of $S_3$.  See Figure~\ref{su3Figure2} for which of these monopole configurations are stable and for the scaling dimensions of the monopole operators corresponding to the stable backgrounds pictured.  Since $\pi_1(SU(3))$ is trivial, none of these monopoles carry any non-trivial topological quantum numbers. 

\FloatBarrier

\subsubsection{$Sp(4)$ QCD with fundamental fermions}

\begin{figure}[h!]
\begin{center}
        \includegraphics[width=0.49\textwidth]{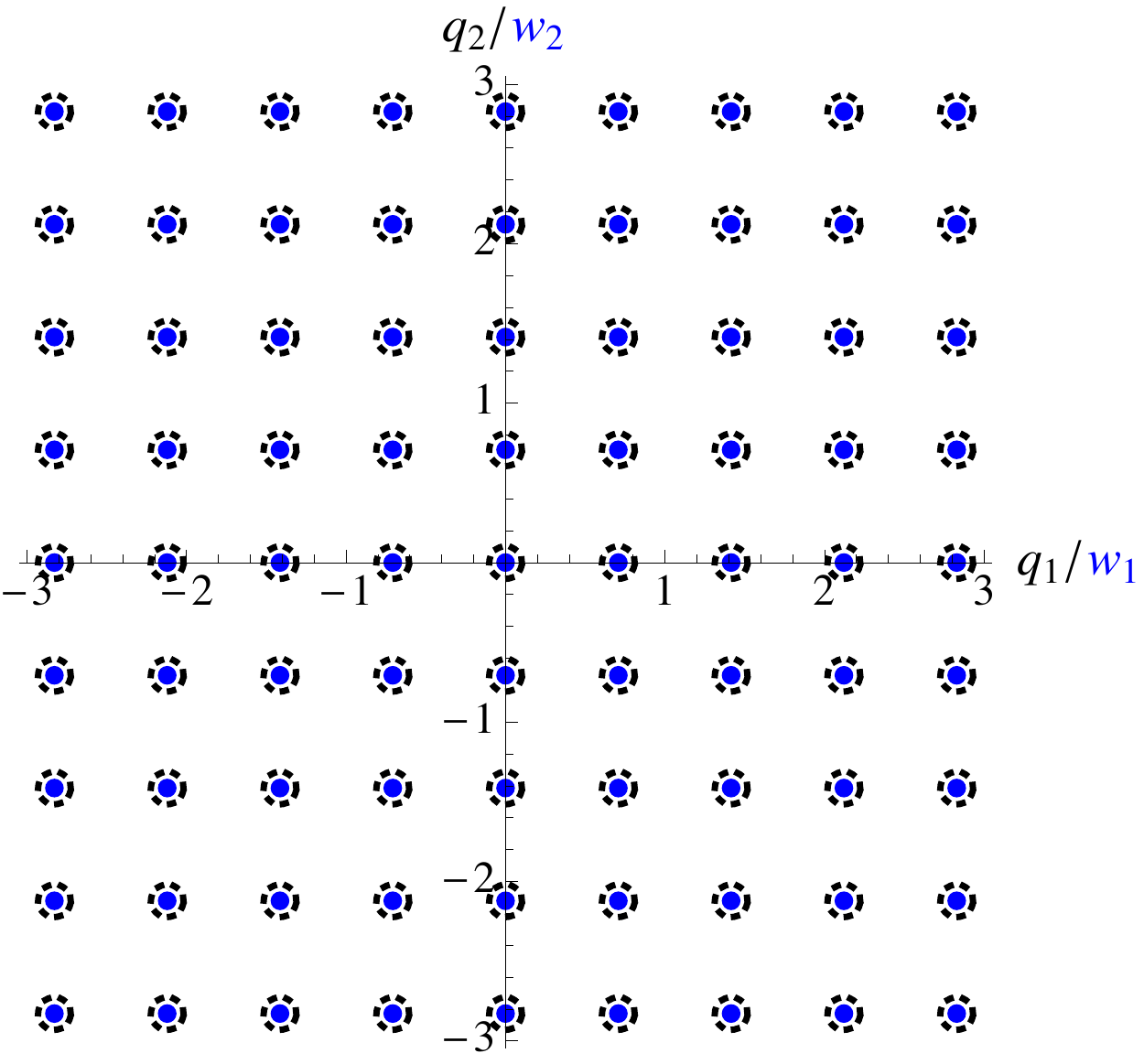}
\caption{The weight lattice of $Sp(4)$ (blue dots) as well as the lattice of all possible monopole charges (dotted circles).  The monopole charges are defined modulo the action of the Weyl group, which in this case is $(Z_2)^3$ and is generated by reflections about the $q_1$ axis, $q_2$ axis, and the line that makes a $45$ degree angle with the $q_1$ axis.}
\label{sp4Figure}
\end{center}
\end{figure}

We can also consider the gauge group $G = Sp(4)$ and $N_f$ fermions transforming in the four-dimensional fundamental representation of this gauge group.  The rank of $Sp(4)$ is $r = 2$, so again the roots and weights are points in $\R^2$.  The weights of the fundamental representation are
 \es{sp4FundWeights}{
  \pm \left(\frac{1}{\sqrt{2}}, 0 \right) \,, \qquad \pm \left(0, \frac{1}{\sqrt{2}} \right) \,.
 }
The adjoint consists of two Cartan elements as well as eight roots:
 \es{Rootssp4}{
  \pm \left(\frac{1}{\sqrt{2}}, \frac{1}{\sqrt{2}} \right) \,,  \qquad
   \pm \left(\frac{1}{\sqrt{2}}, -\frac{1}{\sqrt{2}} \right) \,,  \qquad
   \pm \left(\sqrt{2}, 0 \right) \,, \qquad
   \pm \left(0, \sqrt{2} \right) \,.
 }
The charge lattice in this case is the same as the weight lattice, and is generated by:
 \es{sp4Charges}{
  \left(\frac{1}{\sqrt{2}}, 0 \right) \,, \qquad \left(0, \frac{1}{\sqrt{2}} \right). \,
 }
After scaling, the charge basis vectors also generate the weight lattice of $SO(5)$, which indeed is the GNO dual of $Sp(4)$.  See Figure~\ref{sp4Figure}.

\begin{figure}[h!]
 \begin{minipage}[c]{0.49\textwidth}
          \includegraphics[width=\textwidth]{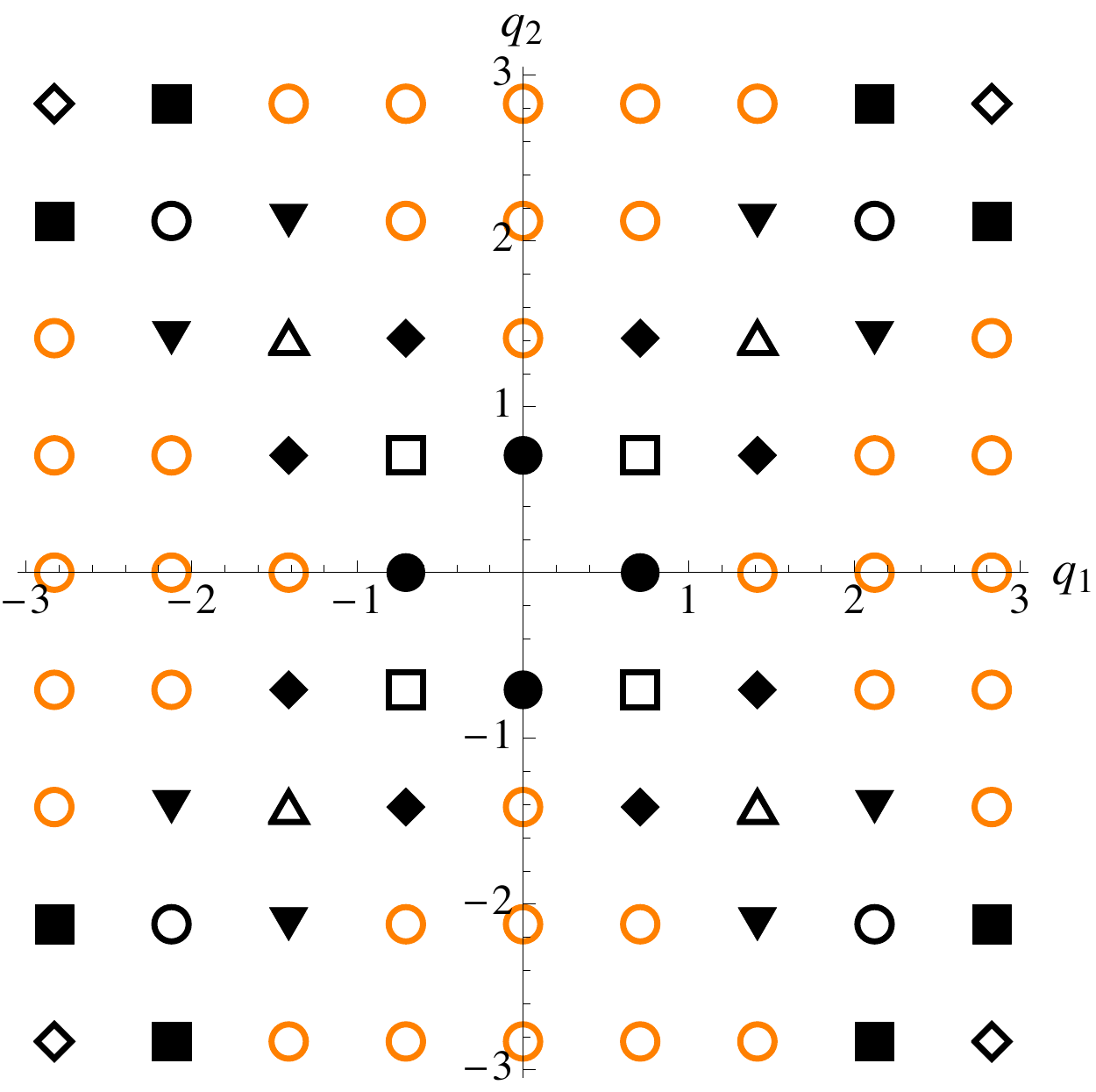}
 \end{minipage}
 \begin{minipage}[c]{0.49\textwidth}        
          \input{sp4Table.txt}
 \end{minipage} 
 \begin{center}         
\caption{The $Sp(4)$ monopoles appearing as black dotted circles in Figure~\ref{sp4Figure}.  Here, we consider these backgrounds in the presence of $N_f$ fermions transforming in the four-dimensional fundamental representation of $Sp(4)$.  The orange circles correspond to unstable backgrounds.  For the stable backgrounds (represented in black by various shapes), we list the scaling dimensions $\Delta$ of the corresponding monopole operators.}
\label{sp4Figure2}
\end{center}
\end{figure}

$Sp(4)$ is simply connected, so all GNO monopoles have trivial topological charge. The stability of various monopoles along with their scaling dimensions are included in Figure \ref{sp4Figure2}.
\FloatBarrier

\subsubsection{$SO(5)$ QCD with fundamental fermions}
\begin{figure}[h!]
\begin{center}
        \includegraphics[width=0.49\textwidth]{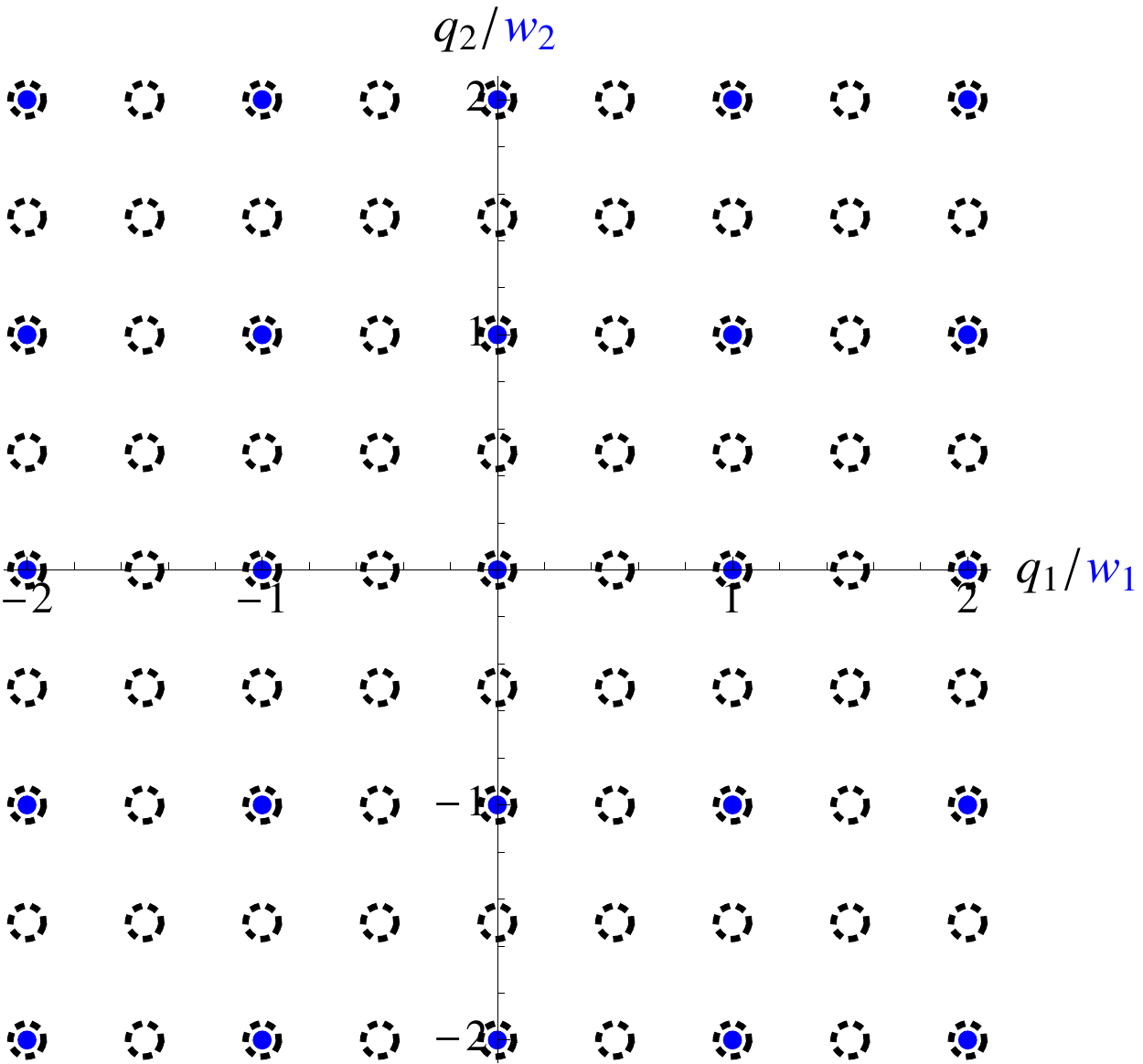}
\caption{The weight lattice of $SO(5)$ (blue dots) as well as the lattice of all possible monopole charges (dotted circles).  The monopole charges are defined modulo the action of the Weyl group, which, as in the $Sp(4)$ case, can be identified with the $(Z_2)^3$ generated by reflections about the $q_1$ axis, $q_2$ axis, and the line that makes a $45$ degree angle with the $q_1$ axis.}
\label{so5Figure}
\end{center}
\end{figure}

Moving onto $G=SO(5)$ with $N_{f}$ fundamental fermions, the weights of the fundamental representation are:
 \es{SO5Weights}{
  \pm \left( 1, 0 \right) \,, \qquad \pm (0, 1) \,, \qquad (0, 0) \,.
 }
The adjoint consists of two generators in the Cartan, as well as generators with roots
 \es{SO5Adjoint}{
  \pm \left(1, 1 \right) \,, \qquad \pm \left(1, -1 \right) \,, \qquad
   \pm \left(1, 0 \right) \,, \qquad \pm \left(0, 1 \right) \,.
 }
The charge lattice is generated by:
 \es{SO5Charges}{
  \left(\frac{1}{2}, 0 \right) \,, \qquad \left(0, \frac{1}{2} \right). \,
 }
See Figure~\ref{so5Figure}. 

\begin{figure}[h!]
 \begin{minipage}[c]{0.49\textwidth}
          \includegraphics[width=\textwidth]{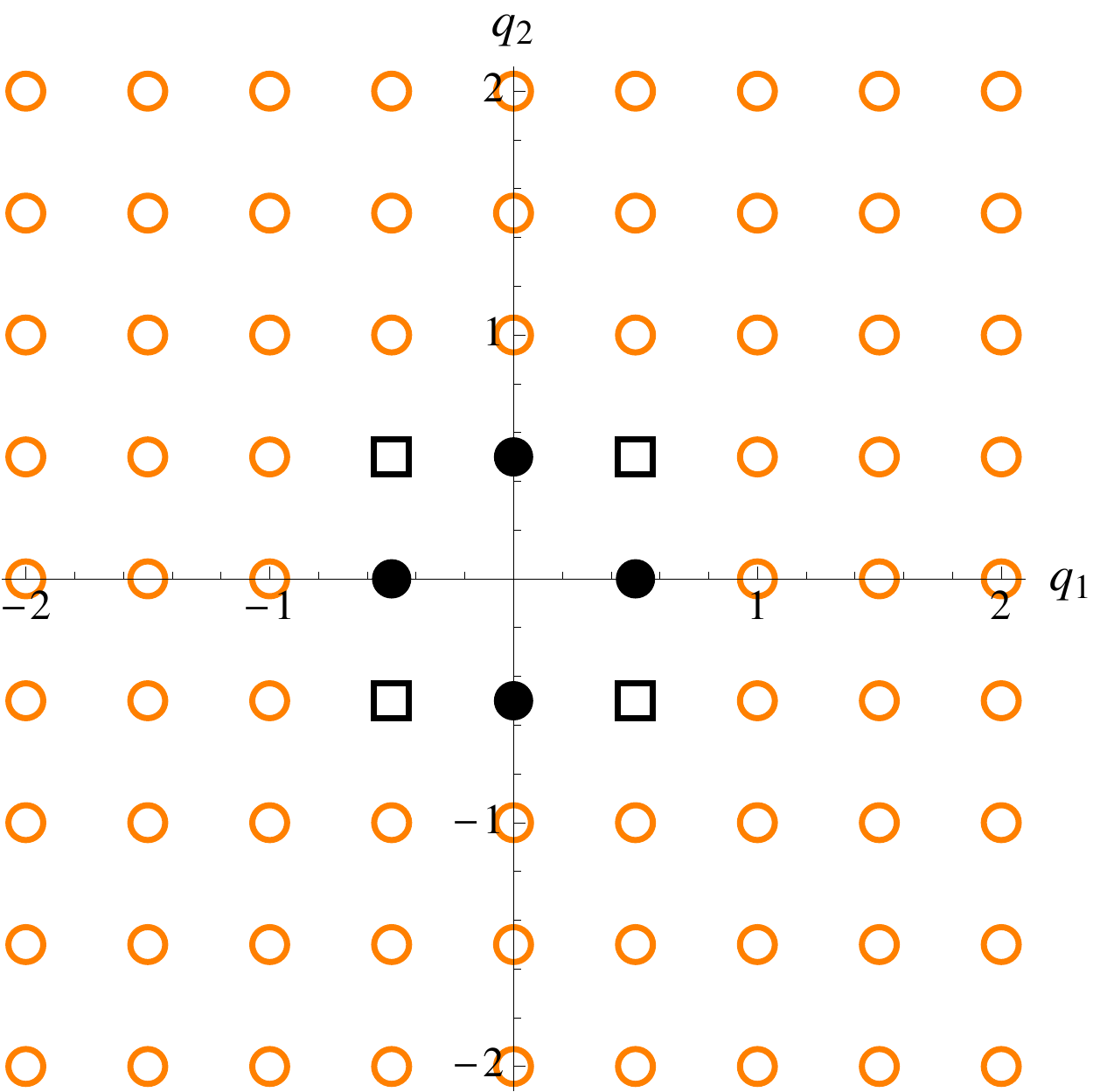}
 \end{minipage}
 \begin{minipage}[c]{0.49\textwidth}        
          \input{so5Table.txt}
 \end{minipage} 
 \begin{center}         
\caption{The $SO(5)$ monopoles appearing as black dotted circles in Figure~\ref{so5Figure}.  Here, we consider these backgrounds in the presence of $N_f$ fermions transforming in the five-dimensional fundamental representation of $SO(5)$.  The orange circles correspond to unstable backgrounds.  For the stable backgrounds (represented in black by various shapes), we list the scaling dimensions $\Delta$ of the corresponding monopole operators.}
\label{so5Figure2}
\end{center}
\end{figure}

In this case the fundament group is non-trivial, $\pi_{1}(SO(5))=\mathbb{Z}_{2}$. The topological charge of a monopole with GNO charges $q_{1}$, $q_{2}$ is $(2 q_1 +2q_2 )\, \text{mod}\, 2$.  For $SO(5)$, there are only two stable monopoles. Monopoles of various charges are plotted in Figure \ref{so5Figure2}.

\subsubsection{$G_2$ QCD with fundamental fermions}

\begin{figure}[h!]
\begin{center}
        \includegraphics[width=0.49\textwidth]{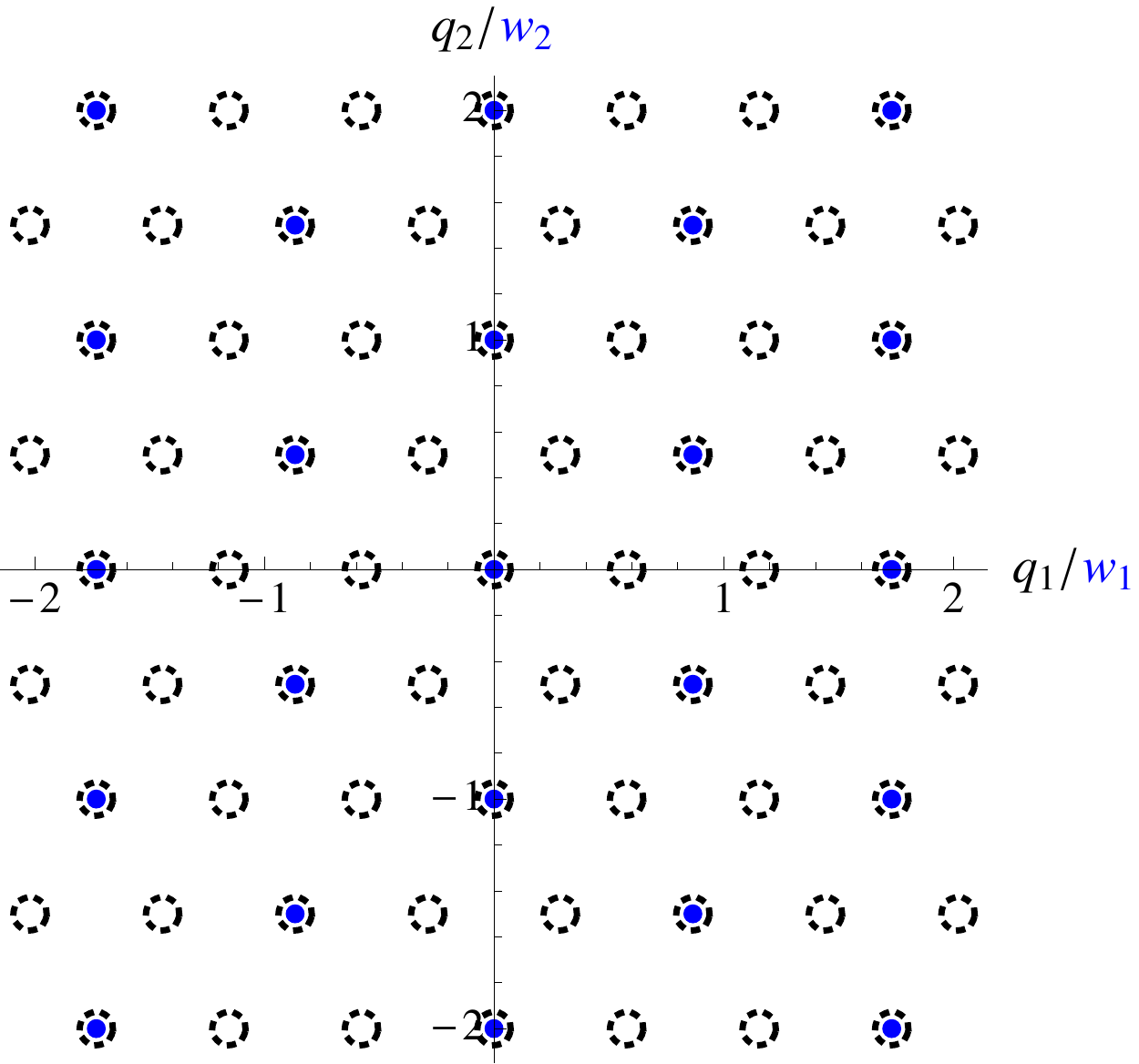}
\caption{The weight lattice of $G_2$ (blue dots) as well as the lattice of all possible monopole charges (dotted circles).  The monopole charges are defined modulo the action of the Weyl group, which in this case is $D_6$ (the dihedral group of order $12$) and is generated by $60$ degree rotations as well as reflections about the line that makes a $45$ degree angle with the $q_1$ axis.}
\label{g2Figure}
\end{center}
\end{figure}

Lastly, we consider $G = G_2$ and $N_f$ fermions transforming in the seven-dimensional fundamental representation of $G_2$.  The weights of the fundamental representation are
 \es{G2WeightsFund}{
  \pm \left(0, 1 \right) \,, \qquad
   \pm \left(\frac{\sqrt{3}}{2}, \frac 12 \right)  \,, \qquad
   \pm \left(\frac{\sqrt{3}}{2}, -\frac 12 \right)  \,, \qquad
   \left(0, 0 \right) \,.
 }
The adjoint representation is fourteen-dimensional and consists of two Cartan elements as well as the roots
 \es{G2Roots}{
  &\pm \left(0, 1 \right) \,, \qquad
   \pm \left(\frac{\sqrt{3}}{2}, \frac 12 \right)  \,, \qquad
   \pm \left(\frac{\sqrt{3}}{2}, -\frac 12 \right)  \,, \qquad
   \pm \left( \frac{\sqrt{3}}{2}, \frac 32 \right) \,, \\
   &\qquad\qquad\qquad\qquad\pm \left( \frac{\sqrt{3}}{2}, -\frac 32 \right) \,, \qquad
   \pm \left( \sqrt{3}, 0 \right) \,. 
 }
The set of all possible monopole charges is generated by the vectors
 \es{G2Charges}{
  \left(\frac{1}{\sqrt{3}}, 0 \right) \,, \qquad \left(\frac{\sqrt{3}}{2}, \frac{1}{2} \right). \,
 }
After scaling and rotating, the charge lattice is identical to the weight lattice, reflecting the fact that $G_{2}$ is its own GNO dual.  See Figure~\ref{g2Figure}. Here, the Weyl group is $D_6$ (the dihedral group with $12$ elements), which is generated by rotations by $60$ degrees as well as reflections about the line that makes $45$ degrees with the $q_1$ axis.

\begin{figure}[h!]
 \begin{minipage}[c]{0.49\textwidth}
          \includegraphics[width=\textwidth]{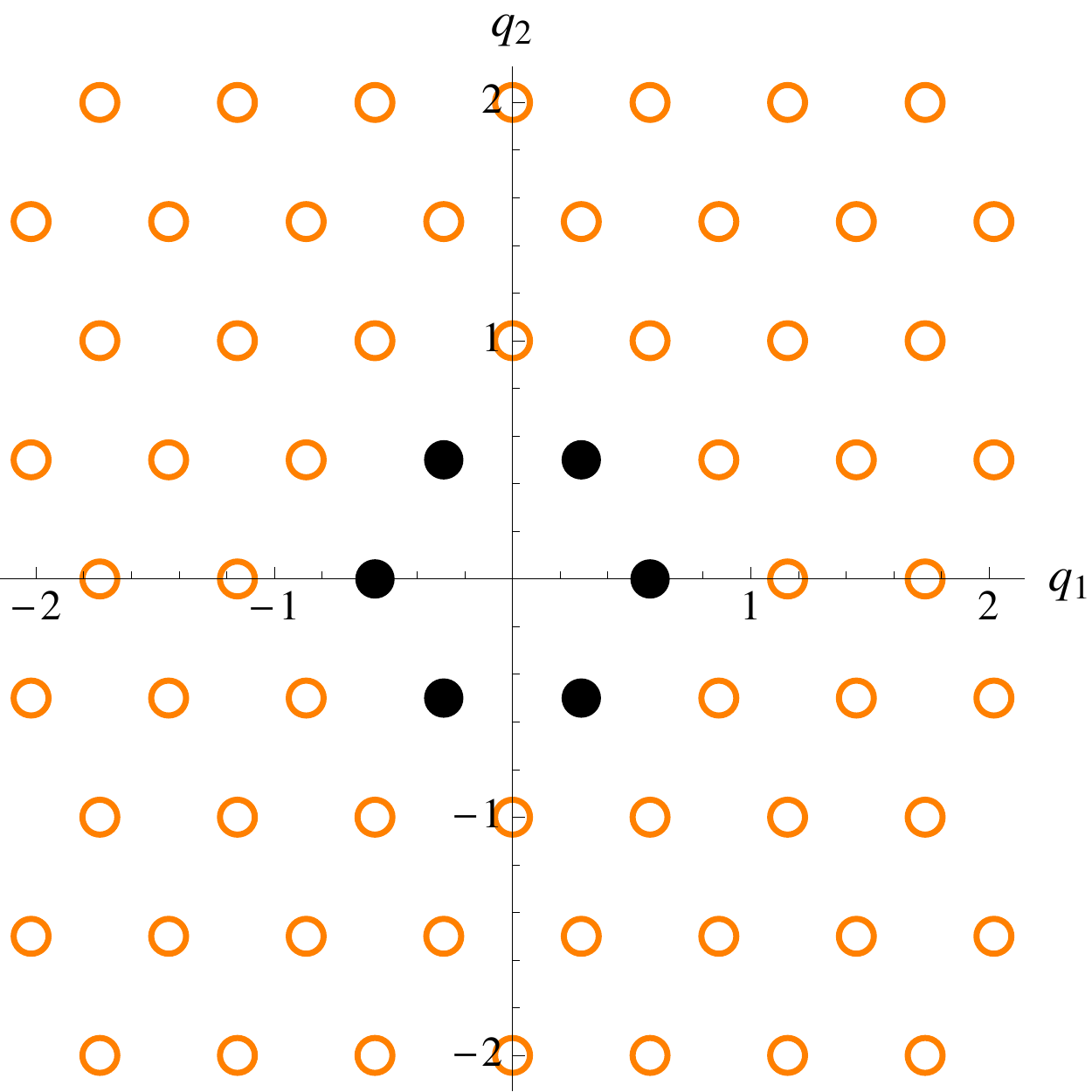}
 \end{minipage}
 \begin{minipage}[c]{0.49\textwidth}        
          \input{g2Table.txt}
 \end{minipage} 
 \begin{center}         
\caption{The $G_2$ monopoles appearing as black dotted circles in Figure~\ref{g2Figure}.  Here, we consider these backgrounds in the presence of $N_f$ fermions transforming in the seven-dimensional fundamental representation of $G_2$.  The orange circles correspond to unstable backgrounds.  For the stable backgrounds (represented in black by various shapes), we list the scaling dimensions $\Delta$ of the corresponding monopole operators.}
\label{g2Figure2}
\end{center}
\end{figure}

$G_{2}$ has a trivial fundamental group, and so there is no topological charge.  The stability of monopoles with different GNO charges as well as the dimensions of the operators corresponding to the stable backgrounds are given in Figure~\ref{g2Figure2}.
\FloatBarrier

\subsubsection{$SU(3)$ QCD with adjoint and fundamental fermions}
While in all of our previous examples, the matter fields were in the fundamental representation of the gauge group, we can also consider matter in other representations.   In this example we consider fermions that transform in the adjoint representation of $SU(3)$. The weights of the adjoint are just the root vectors \eqref{RootsSU3}. The set of possible monopoles is independent of the matter representations, and so the charge lattice is still generated by $\eqref{DualSU3}$ and is shown in Figure~\ref{su3Figure}. 

Unfortunately, for $N_{f}$ copies of the adjoint, there are no stable monopoles.\footnote{The absence of stable monopoles is a common feature of larger representations of any gauge group.} The absence of stable monopoles does not mean that the adjoint representation is uninteresting, however. There is no reason to restrict to matter in an irreducible representation, and we can consider theories with $n_\text{adj} N_f$ adjoint fermions, and $n_\text{fund} N_{f}$ fundamentals. For $n_\text{adj} \ll n_\text{fund}$ this theory should have many stable monopoles, as is the case for $SU(3)$ with only fundamental matter.\footnote{$SU(3)$ with only fundamental matter has infinitely many stable monopole backgrounds.} For $n_\text{adj} \gg n_\text{fund}$ the theory should behave more like the theory with only adjoint matter, and have no stable monopoles. Below we plot the number of stable monopoles as a function of the ratio $n_\text{fund}/n_\text{adj}$. For small values of this ratio, the specific monopoles which become stable are shown.

\begin{figure}[t!]
\begin{center}
   \includegraphics[width=1\textwidth]{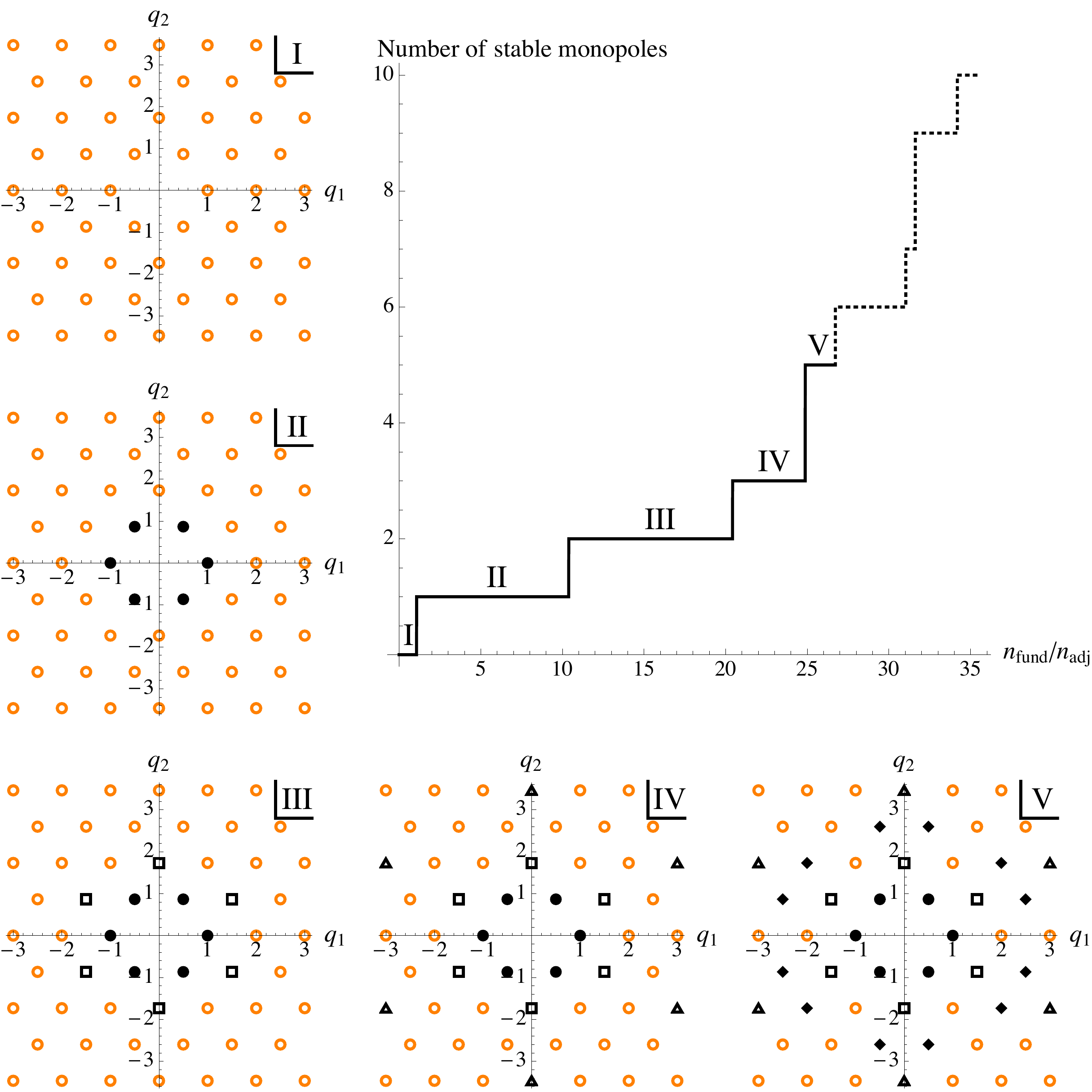}
\caption{In the top right corner we show the number of (inequivalent) stable monopoles for $SU(3)$ gauge theory with $n_\text{fund} N_f$ fundamental fermions and $n_\text{adj} N_f$ adjoint fermions as a function of the ratio $n_\text{fund}/n_\text{adj}$.  The solid line is divided into five regions that correspond to the diagrams on the left and bottom, where we show explicitly which monopoles are stable in each region.  The dashed line is a continuation of the solid line for larger values of $n_\text{fund}/n_\text{adj}$, but in this region we do not show explicitly which monopoles are stable.}
\label{fig:adjstab}
\end{center}
\end{figure}

\FloatBarrier

\clearpage

\section{Discussion}
\label{DISCUSSION}

\subsection{Summary}

In this paper we studied properties of monopole operators in non-supersymmetric QCD$_3$ and QED$_3$ with $N_f$ fermion flavors.  We worked in the limit of large $N_f$, where gauge field fluctuations are suppressed and where the theory flows in the infrared to an interacting CFT\@.  At this infrared fixed point, we used the state-operator correspondence to first define the monopole operators in terms of energy eigenstates on $S^2 \times \R$ and then to study their transformation properties under the conformal and flavor symmetry groups.  As we emphasized in Section~\ref{OVERVIEW}, associating energy eigenstates with certain GNO monopole backgrounds can be done cleanly in the limit of large $N_f$, provided that these GNO backgrounds are stable saddle points of the effective action for the gauge field fluctuations.  

We obtain three main results.   Our first result is that only certain GNO monopole backgrounds are stable saddle points in the CFT\@.   In general, stability is a dynamical issue that can only be decided by studying the effective potential for the gauge field fluctuations.  We provided the criterion for stability in Section~\ref{STABILITY} in the case where the gauge group is $U(N_c)$ (see Figure~\ref{fig:2dstab} for a summary plot).  We later generalized this criterion to theories with other gauge groups in Section~\ref{GENERAL}.   In all these theories, we were thus able to identify precisely for which sets of GNO charges one can define independent monopole operators, at least at large $N_f$.  We found that many, but not all, GNO backgrounds in each topological sector are stable.  For every stable background there is a Fock space of energy eigenstates on $S^2 \times \R$ whose wavefunctions are localized around that background.  Each such energy eigenstate corresponds to an operator on $\R^3$.  We further focused on the lowest energy eigenstate within every Fock space and studied its quantum numbers. We referred to the operator that corresponds to this state as a bare monopole operator.

Our second result is that we computed the scaling dimensions of the bare GNO monopole operators in the $1/N_f$ expansion.  The scaling dimension is a quantity determined by the dynamics. It equals the ground state energy on $S^2 \times \R$ in the GNO monopole background. We obtained the ground state energy by evaluating the path integral on $S^2 \times \R$ to subleading order in the $1/N_f$ expansion.  For large $N_f$ the monopole operators have ${\cal O}(N_f)$ dimension
\es{DimensionSketch}{
\Delta=N_f\, F_0+\delta F+{\cal O}(1/N_f) \ .
}
Explicit results for $\Delta$ for various GNO charges can be found in Section~\ref{DIMENSIONS} for $U(N_c)$ theories, and in Section~\ref{GENERAL} for general gauge groups.  We expect the results we obtained from the large $N_f$ expansion to be reliable down to fairly small values $N_f  \gtrsim \delta F/F_0={\cal O}(1)$. This expectation is supported by the high accuracy of large $N_f$ computations for supersymmetric theories~\cite{Safdi:2012re}, where the answer can be compared to exact results. 

Our third result is that for the case where the gauge group is $G = U(N_c)$, we calculated all the other quantum numbers of the bare monopole operators.  We found that these operators are all spin singlets and that they transform in the irreducible representations of the $SU(N_f)$ flavor symmetry group given by the Young diagrams
 \es{SUNfRepNonAb2}{
  {\tiny N_f/2} \Bigl\{   \underbrace{ {\tiny \ydiagram{3, 3}} }_{ 2 \sum_a \abs{q_a}} \ ,
 }
where the GNO charges are $\{q_1,\, q_2,\, \dots,\, q_{N_c}\}$.  We can therefore completely characterize the quantum numbers of the bare monopole operators in $U(N_c)$ QCD with $N_f$ fundamental flavors:  We know their topological charge, scaling dimension, spin, and representation of the flavor symmetry group.   It would be very interesting to generalize this analysis to other gauge groups and matter representations.

These results are interesting in their own right as they teach us about the operator content of QCD$_3$\@. Using the knowledge that we gained, it is desirable to understand the role that the monopole operators play in the dynamics of the theory.

\subsection{Confinement and chiral symmetry breaking}

From our results, we can learn about the following three theories, one of which describes confinement:

\begin{enumerate}[I.]

 \item \label{YangMills} We can consider Yang-Mills theory coupled to $N_f$ flavors of massless fermions.  This theory is super-renormalizable and asymptotically free, so it is well-defined up to arbitrarily high energies.  If we wish, we could think of it as an effective field theory at large distances that arises from a lattice Hamiltonian that does not allow mass terms for the fermions.
 
 \item \label{Interacting} We can also consider a non-trivial interacting CFT\@.  At large $N_f$, we can define this CFT by erasing the Yang-Mills term from the action of \eqref{YangMills}, as we did throughout this paper.  This description of the CFT should make sense as long as this CFT can be achieved from \eqref{YangMills} without any fine tuning, which should happen for all $N_f$ greater than or equal to some number $N_f^\text{deconf}$ that we will estimate shortly.   Below $N_f^\text{deconf}$, a non-trivial CFT may still exist, but a good description for it may not be readily available.

 \item \label{Confined} A confined or partially confined theory, potentially with some number of Goldstone bosons coming from spontaneous flavor symmetry breaking.  The description of this theory is intentionally vague, as it should be viewed just as an alternative to \eqref{Interacting} for describing the IR physics of \eqref{YangMills}.

\end{enumerate}

Recall that we restrict our discussion to the case where the number of fermions is even, because otherwise we would necessarily be breaking parity \cite{Redlich:1983dv,Redlich:1983kn}.  

The first question we can ask is:  When is the infrared physics of generic RG flows starting from \eqref{YangMills} described by the deconfined CFT \eqref{Interacting}, and when is it not?  In other words, we should estimate $N_f^\text{deconf}$ from the fact that for $N_f \geq N_f^\text{deconf}$, all monopole operators should be irrelevant, i.e. their scaling dimensions should be greater than three.  From \eqref{DimensionSketch} we find:
 \es{Nfdeconf}{
  N_f^{\text{deconf}} \approx \frac{3-\delta F}{F_0} \,.
 }
Here, the values of $F_0$ and $\delta F$ correspond to the monopole operator with the lowest scaling dimension for a given gauge group and matter content.  
 \begin{table}[!h]
 \begin{center}

 \begin{tabular}{c||c} 

Gauge group & $ N_f^{\text{deconf}}$ \\
 \hline \hline 
$U(1)$ & $ 12 $ \\
 \hline 
$U(2)$ & $14 $ \\ 
 \hline 
 $SU(2)$ & $ 8 $ \\ 
 \hline 
 $SO(3)$ & $8 $ \\ 
 \hline 
$SU(3)$ & $ 10$ \\ 
 \hline 
$Sp(4)$ & $10 $ \\ 
 \hline 
 $SO(5)$ & $ 10$ \\ 
 \hline 
 $G_2$ & $6 $ 
 \end{tabular}
\caption{Estimates of the smallest number of fermions, $N_f^{\text{deconf}}$ for which the IR of QCD$_3$ with gauge group $G$ is in a deconfined quantum critical point. Results are listed for various rank one and two gauge groups. \label{tab:Deconf}}
 \end{center}
\end{table}
See Table~\ref{tab:Deconf} for a few particular cases.  As can be seen from this table, $N_f^{\text{deconf}}$ is smaller for groups with fewer monopoles.

We stress that the estimate \eqref{Nfdeconf} as well as the numbers given in Table~\ref{tab:Deconf} are not relying on any assumptions about the physics at $N_f < N_f^\text{deconf}$.  All we can tell for sure is that in this case, Yang-Mills theory with $N_f$ fermions does not generically flow to the deconfined CFT \eqref{Interacting}.

It is possible to obtain an independent estimate of $N_f^\text{deconf}$ if we make some assumptions about what happens for $N_f < N_f^\text{deconf}$.  We will do so only in the cases where the gauge group is $G = U(1)$ and $U(2)$, and leave a more extensive analysis for future work.  As reviewed in \cite{Grover:2012sp}, one expects that below $N_f^\text{deconf}$, the $SU(N_f)$ global flavor symmetry should be broken to $SU(N_f/ 2) \times SU(N_f/2) \times U(1)$.\footnote{In \cite{Vafa:1983tf,Vafa:1984xh} it is shown that if the number of fermions is $N_f \geq 6$ there must be massless particles in the infrared.  It is likely that these particles are Goldstone bosons corresponding to the symmetry breaking pattern mentioned in the main text. This symmetry breaking pattern is usually referred to as chiral symmetry breaking, even though there is no chiral symmetry for fermions in three dimensions. The name ``chiral symmetry'' comes from the fact that if the same theory were realized in 4d by pairing up the $N_f$ Weyl spinors into $N_f/2$ Dirac spinors, then the broken symmetry would be chiral.} 
A simple computation shows that the number of Goldstone bosons is 
\es{GoldstoneNumber1}{
N_G = \frac{N_f^2}{2} \ .
}
As pointed out in \cite{Grover:2012sp}, such a symmetry breaking pattern is constrained by the $F$-theorem \cite{Jafferis:2011zi,Klebanov:2011gs,Myers:2010tj,Casini:2012ei},\footnote{Note that this $F$ stands for the $S^3$ free energy, and should not be confused with the $S^2\times \R$ partition function that was discussed in this paper. }  which states that any three-dimensional Lorentz-invariant RG flow from a UV CFT to an IR CFT should satisfy
\es{Ftheorem}{
F_\text{UV}\geq F_\text{IR} \,,
}
where $F_\text{UV}$ ($F_\text{IR}$) is the $S^3$ free energy of the UV (IR) CFT\@.  

To use the $F$-theorem, we can consider starting with Yang-Mills theory with $N_f = N_f^\text{deconf} - 2$ fermions, which by assumption is the largest value of $N_f$ for which the IR theory consists of $N_f^2/2$ Goldstone bosons.  It is likely that the same IR theory of Goldstone bosons can be obtained by starting with the CFT in \eqref{Interacting} with $N_f^\text{deconf}$ fermions and giving masses to two of them.  These masses should be of opposite sign in order to preserve parity.  The latter flow is the one for which we will use the $F$-theorem.  The $F$-theorem should of course hold for the flow from Yang-Mills theory with $N_f$ fermions to the theory of $N_f^2/2$ Goldstone bosons as well, but in the UV Yang-Mills theory is not conformal and should be assigned $F_\text{UV} = \infty$.

For the flow between the deconfined CFT with $N_f^\text{deconf}$ fermions and the IR theory of $N_f^2/2$ Goldstone bosons, $F_\text{UV}$ can be read off from the results of~\cite{Klebanov:2011td}:
\es{FUV}{
F_\text{UV}=\le({\log 2\ov 4}+{3\zeta(3)\ov 8\pi^2}\ri)\, N_c \, N_f^\text{deconf}+\frac{N_c^2} 2\, \log\le(\pi N_f^\text{deconf}\ov 8\ri)+\log \frac{\Vol (U(N_c))}{\Vol \le(U(1)\ri)^{N_c^2}}+{\cal O}(1/N_f^\text{deconf}) \,,
}
where for $N_c = 1, 2$ we should use $\Vol \le(U(1)\ri)=2\pi$ and $\Vol \le(U(2)\ri)=8\pi^3$.   Because the IR theory is a CFT of $N_f^2/2$ free scalar fields, we have 
 \es{FIR}{
  F_\text{IR}= \frac {N_f^2}2  \, F_\text{scalar} \,,
 }
with $F_\text{scalar} \approx 0.0638$ being the $S^3$ free energy of a single real scalar field~\cite{Klebanov:2011gs}.

Using \eqref{FUV} and \eqref{FIR}, we see that the $F$-theorem inequality \eqref{Ftheorem} holds  for $N_f^\text{deconf} \leq 12$ in the $U(1)$ case and $N_f^\text{deconf} \leq 20$ in the $U(2)$ case.\footnote{In the $U(1)$ case, \cite{Grover:2012sp} obtained $N_f^\text{deconf} \leq 14$ from considering a flow between supersymmetric QED$_3$ and the symmetry broken phase of $N_f^2/2$ Goldstone bosons.  The bound $N_f^\text{deconf} \leq 12$ that we obtain is more constraining than that of \cite{Grover:2012sp}, because we start from a UV theory with fewer degrees of freedom.}  This result is consistent with, but less precise than, the values $N_f^\text{deconf} = 12$ and $N_f^\text{deconf} = 14$ for $G = U(1)$ and $U(2)$, respectively, that we obtained from studying the scaling dimensions of the monopole operators.

\subsection{QED and and algebraic spin liquids}

The analysis of the previous subsection on the minimal value of $N_f$ for which Yang-Mills theory with $N_f$ fermions flows to a deconfined phase considered only ``generic'' such RG trajectories.  This analysis can be refined in the case where $\pi_1(G)$ is non-trivial---and so certain monopole operators carry topological charges---by restricting our attention to RG trajectories that are invariant under a subgroup of the corresponding topological symmetry.  Under this extra assumption, Yang-Mills theory with $N_f$ fermions flows generically to a deconfined CFT provided that the monopole operators that are invariant under the above subgroup are irrelevant;  it does not matter whether the other ones are relevant or not.  Consequently, the values of $N_f^\text{deconf}$ in this case would be smaller than the values obtained in the previous section.

The QED case $N_c=1$ provides a nice example relevant to algebraic spin liquids \cite{Hermele,Rantner:2002zz}.  In this case the topological symmetry is $U(1)_\text{top}$, and the topological charge is $q_\text{top} = q \in \Z/2$.  It was suggested in \cite{HermeleRanLeeWen} that if $U(1)$ QED with $N_f=4$ fermions can be obtained as an effective theory of a spin system on the Kagome lattice, the lattice symmetries are embedded into a $\Z_3$ subgroup of $U(1)_\text{top}$.  So let us restrict our attention to RG trajectories that preserve this $\Z_3$ subgroup as a symmetry.  Under the generator of this $\Z_3$ symmetry, a monopole of charge $q$ is multiplied by a phase equal to $e^{4\pi i q/3}$, so only monopole operators with $q \in 3\Z/2$ are invariant.  If all fermion mass terms are also forbidden by the lattice regularization, it then follows that the IR theory is a deconfined CFT provided that all the monopole operators with $q \in 3 \Z/2$ are irrelevant.  According to Table~\ref{qTable}, these monopole operators have scaling dimension greater than $3$ for $N_f \geq 4$.  This bound is less restrictive than $N_f \geq 12$, which is what we obtained in the previous section by requiring that all monopole operators should be irrelevant.

\section*{Acknowledgments}
We thank A.~Kapustin, Z.~Komargodski, D.~Park, T.~Senthil, J.~Sonner, and E.~Witten for useful discussions.  This work was supported in part by the U.S. Department of Energy under cooperative research agreement Contract Number DE-FG02-05ER41360\@.  SSP was also supported in part by a Pappalardo Fellowship in Physics at MIT\@.

\appendix

\section{Definition and Properties of Monopole Harmonics} \label{App:Identities}
\subsection{Scalar Harmonics}
\subsubsection{Definition}
In this Appendix we review some properties of the monopole harmonics.  We start with the scalar harmonics introduced in \cite{Wu:1976ge,Wu:1977qk}.  The monopole harmonics $Y_{q,\ell m}(\hat{n})$ are eigenfunctions of the angular momentum operator in the presence of a monopole background of charge $q$, \eqref{AAbelian}. 
In this background, the angular momentum operator takes the form:
 \es{AngMom}{
    L_{z}&= -i\partial_{\phi}-q \,, \\
    \vec{L}^{2}&=-\nabla^{2}+\frac{2q}{\sin^{2} \theta}(\cos \theta-1)L_{z} \,. 
 }
The scalar monopole harmonics are defined to satisfy the relations:
 \es{DefRelations}{
  \vec{L}^{2}Y_{q,\ell m}(\hat{n})&=\ell(\ell+1)Y_{q,\ell m}(\hat{n}) \,, \\
   L_{z}Y_{q,\ell m}(\hat{n})&= mY_{q,\ell m}(\hat{n}) \,.
 }
We can write the solutions to these equations explicitly in position space:\footnote{Recall that $\hat{n}$ is a unit vector parameterizing the two-sphere and just shorthand for $\theta,\phi$.}
 \es{ExpForm}{
  Y_{q,\ell m}(\hat{n})&=2^{m-1}\sqrt{\frac{(2\ell+1)(\ell-m)!(\ell+m)!}{\pi(\ell-q)!(\ell+q)!}}\sqrt{\frac{(1+x)^{q-m}}{(1-x)^{q+m}}}P_{\ell+m}^{(-q-m,q-m)}(\cos \theta)e^{(m+q)\phi}  \,.
 }
It is sometimes convenient to write the monopole harmonics in bra-ket notation.
 \es{YPosition}{
  Y_{q, \ell m}(\hat{n})&=\langle\theta, \phi|\ell, m\rangle_{q} \,.
 }

\subsubsection{Identities}
The scalar monopole harmonics have a number of useful properties \cite{Wu:1977qk}. Under charge conjugation the monopole harmonics transform as:
 \es{YCC}{
   Y^{*}_{q,\ell m}(\hat{n})&= (-1)^{q+m}Y_{-q,\ell,-m}(\hat{n}) \,.
 }
When evaluated at the north pole, the scalar harmonics satisfy.
 \es{YNP}{
   Y_{q,l,m}(\hat{z})&=\delta_{q,-m}\sqrt{\frac{2\ell+1}{4\pi}} \,.
 }
Gauge invariant products of monopole harmonics also satisfy integral relations.  The monopole harmonics are normalized such that
 \es{Yint2}{
    \int d\hat{n}\, |Y_{q,\ell m}(\hat{n})|^{2}&= 1 \,.
 }
The integral of a product of three monopole harmonics is given by 
 \es{Yint3}{
   &\int d\hat{n}\, Y_{q,\ell m}(\hat{n})Y_{q^{\prime},\ell^{\prime} m^{\prime}}(\hat{n})Y_{q^{\prime\prime},\ell^{\prime\prime} m^{\prime\prime}}(\hat{n})  \\
&=(-1)^{\ell+\ell^{\prime}+\ell^{\prime\prime}}\sqrt{\frac{(2\ell+1)(2\ell^{\prime}+1)(2\ell^{\prime\prime}+1)}{4\pi}}
 \begin{pmatrix}
  \ell & \ell^{\prime}&\ell^{\prime\prime} \\
  m & m^{\prime}&m^{\prime\prime}
 \end{pmatrix}
 \begin{pmatrix}
   \ell&\ell^{\prime}&\ell^{\prime\prime} \\
q&q^{\prime}&q^{\prime\prime}
 \end{pmatrix} \,.
 }

\subsection{Spin $s$ Harmonics}
Now that we have the scalar monopole harmonics for arbitrary angular momentum, $Y_{q \ell m}(\hat{n})$. It is easy to construct monopole harmonics of arbitrary spin using the Clebsch-Gordon decomposition. Explicitly, we have:
\begin{eqnarray}\label{CSDecomp}
|s \ \ell;j, m\rangle_{q}&=&\sum_{m_{s}=-s}^{s}\sum_{m_{\ell}=-\ell}^{\ell}\langle s \ \ell;m_{s} m_{l}|s\ \ell;j, m\rangle |s, m_{s}\rangle_{0}\otimes|\ell, m_{\ell}\rangle_{q}.
\end{eqnarray}
Here, $\langle s \ \ell;m_{s} m_{\ell}|s\ \ell;j, m\rangle $ is the usual Clebsch-Gordan coefficient, which can also be written in terms of the Wigner 3j symbol.
\begin{eqnarray}
\langle j_{1} \ j_{2};m_{1} m_{2}|j_{1}\ j_{2};j, m\rangle&=&(-1)^{j_{1}-j_{2}+m}\sqrt{2j+1}\left(\begin{array}{ccc}
j_{1} & j_{2} & j\\
m_{1} & m_{2} & -m
\end{array}\right).
\end{eqnarray}

The Clebsch-Gordon coefficient is zero unless $j_{1}$, $j_{2}$, and $j$ satisfy the triangle inequality, $\abs{j_{1}-j_{2}}\leq j \leq \abs{j_{1}+j_{2}}$. In \eqref{CSDecomp} the scalar monopole harmonic $|\ell,m_{\ell}\rangle_{q}$ vanishes unless $\ell\geq\abs{q}$. Together these relations imply that the state $|s \ \ell;j, m\rangle_{q}$ only exists for,
 \es{jrange}{
    j \geq |q|-s \,.
 }
In this paper, we often decompose fields of fixed spin, $s$, and total angular momentum, $j$, in terms of sums over orbital angular momentum, $\ell$. The only terms that contribute have
 \es{jellrange}{
  |j-s|\leq \ell \leq j+s \,,  \ \ \ \textrm{and} \ \ \ell\geq|q| \,.
 }
For large $j$, \eqref{jellrange} gives $2s+1$ states.  For smaller $j$ there are fewer allowed values for $\ell$.
\subsection{Spin 1/2 Harmonics}

It is useful to have explicit expressions for the spin $1/2$ monopole harmonics. The number of independent states depends on the value of $j$.
\\

\noindent$\mathbf{j=\abs{q}-1/2}$\\

From \eqref{CSDecomp} we see that there are two states with $s=1/2$ for each $j$ when $j\geq\abs{q}$, $|1/2 \ j\pm1/2; j, m\rangle$. In position space these take the explicit form:

 \es{S12Harm}{
\langle\theta,\phi|1/2 \ j-1/2; j, m\rangle &\equiv T_{q, \, j m}(\hat{n})  =  
\begin{pmatrix}
\sqrt{\frac{j+m_{j}}{2}}Y_{q,\, j-1/2,m-1/2}(\hat{n})\\
\sqrt{\frac{j-m_{j}}{2}}Y_{q,\, j-1/2,m+1/2}(\hat{n})
 \end{pmatrix} \,, \\
\langle\theta,\phi|1/2 \ j+1/2; j, m\rangle &\equiv S_{q, \, j m}(\hat{n}) = 
\begin{pmatrix}
-\sqrt{\frac{1+j-m_{j}}{2+2j}}Y_{q,\, j-1/2,m-1/2}(\hat{n})\\
\sqrt{\frac{1+j+m_{j}}{2+2j}}Y_{q,\, j-1/2,m+1/2}(\hat{n})
\end{pmatrix} \,.
 }
$\mathbf{j=\abs{q}}-1/2$\\

For $j=\abs{q}$ only the single mode $S_{q, \, j m}(\hat{n})$ exists.

\subsection{Spin 1 Harmonics}
The spin 1 case is similar to the spin 1/2 case, except that now there are two special values of $j$, $j=q-1$ and $j=q$. The vector harmonics take the form:

\noindent$\mathbf{j>\abs{q}}$
 \es{S1Harm}{
\langle\theta,\phi|1 \ j-1; j, m\rangle &\equiv W_{q, \, j m}(\hat{n}) = 
 \begin{pmatrix}
\sqrt{\frac{(j+m-1) (j+m)}{2j (2 j-1)}}Y_{q, j-1,m-1}(\hat{n})\\
\sqrt{\frac{(j-m) (j+m)}{j (2 j-1)}}Y_{q, j-1,m}(\hat{n})\\
\sqrt{\frac{(j-m-1) (j-m)}{2j (2 j-1)}}Y_{q, j-1,m+1}(\hat{n})
 \end{pmatrix}\,, \\
\langle\theta,\phi|1 \ j; j, m\rangle &\equiv V_{q, \, j m}(\hat{n}) = \begin{pmatrix}
-\sqrt{\frac{(j-m+1) (j+m)}{2j (j+1)}}Y_{q, j,m-1}(\hat{n})\\
\frac{m}{\sqrt{j(1+j)}}Y_{q, j m}(\hat{n})\\
\sqrt{\frac{(j-m) (j+m+1)}{2j (j+1)}}Y_{q, j,m+1}(\hat{n})
 \end{pmatrix} \,, \\
\langle\theta,\phi|1 \ j+1; j, m\rangle &\equiv U_{q, \, j m}(\hat{n}) = 
 \begin{pmatrix}
\sqrt{\frac{(j-m+1) (j-m+2)}{(2j+2) (2 j+3)}}Y_{q, j+1,m-1}(\hat{n})\\
-\sqrt{\frac{(j-m+1) (j+m+1)}{(j+1) (2 j+3)}}Y_{q, j+1,m}(\hat{n})\\
\sqrt{\frac{(j+m+1) (j+m+2)}{(2j+2) (2 j+3)}}Y_{q, j+1,m+1}(\hat{n})
 \end{pmatrix} \,.
}

\noindent$\mathbf{j=\abs{q}}$\\

If $j=\abs{q}$ we only have the last two modes, $U_{q, \, j m}(\hat{n})$ and $V_{q \, j m}(\hat{n})$. If $j=q=0$, only the $U_{q,\, j m}(\hat{n})$ mode is non vanishing.
\\

\noindent$\mathbf{j=\abs{q}-1}$\\

In the case $j=\abs{q}-1$ only the mode $U_{q,\, j m}(\hat{n})$ is non-zero. This mode plays a key role for the stability analysis of monopoles.

\label{App:Divergence}
In order to check gauge invariance of the effective action, \eqref{Seff}, it is useful to have an expression for the divergence of the harmonics. 
The gauge covariant divergence of the vector monopole harmonics with charge $q$ is
 \es{Divs}{
  D_\mu \left( e^{-i \omega \tau} U^{\mu}_{q, j m} (\theta, \phi) \right) 
    &= -(j - i \omega) \sqrt{\frac{(j+1)^2 - q^2}{(j+1)(2j+1)} }e^{-i \omega \tau} Y_{q, j m}(\theta, \phi)  \,, \\
  D_\mu \left( e^{-i \omega \tau} V^{\mu}_{q, j m} (\theta, \phi) \right) 
    &= \frac{q ( 1 + i \omega)}{\sqrt{j(j+1)}} e^{-i \omega \tau} Y_{q, j m}(\theta, \phi)  \,, \\
  D_\mu \left( e^{-i \omega \tau} W^{\mu}_{q, j m} (\theta, \phi) \right) 
    &= - (j + 1 +  i \omega ) \sqrt{\frac{j^2 - q^2}{j (2j+1)} } e^{-i \omega \tau} Y_{q, j m}(\theta, \phi)  \,.
 }

\bibliographystyle{ssg}
\bibliography{monopoleQCD}

\end{document}

%% file: su2Table.txt
\begin{tabular}{c||c} 

Symbol & $\Delta$ \\
 \hline \hline 
{\Large $ \bullet $} & $ 0.530 N_f -0.713+{\cal O}(1/N_f) $ \\
 \hline 
$ \square $ & $ 1.35 N_f -1.65+{\cal O}(1/N_f) $ \\
 \hline 
$ \blacklozenge $ & $ 2.37 N_f -2.95+{\cal O}(1/N_f) $ \\
 \hline 
$ \vartriangle $ & $ 3.57 N_f -4.51+{\cal O}(1/N_f) $ \\

 \end{tabular}

%% file: so3Table.txt
\begin{tabular}{c||c} 

Symbol & $\Delta$ \\
 \hline \hline 
{\Large $ \bullet $} & $ 0.530 N_f -0.554+{\cal O}(1/N_f) $ \\ 
 \hline 
 \hspace{0.7mm} {\Large $ {\orange \circ} $ } & Unstable

 \end{tabular}

%% file: su3Table.txt
\begin{tabular}{c||c} 

Symbol & $\Delta$ \\
 \hline \hline 
{\Large $ \bullet $} & $ 0.530 N_f -1.75+{\cal O}(1/N_f) $ \\
 \hline 
$ \square $ & $ 1.20 N_f -2.55+{\cal O}(1/N_f) $ \\
 \hline 
$ \blacklozenge $ & $ 2.12 N_f -5.38+{\cal O}(1/N_f) $ \\
 \hline 
$ \vartriangle $ & $ 3.13 N_f -7.20+{\cal O}(1/N_f) $ \\ 
 \hline 
 \hspace{0.7mm} {\Large $ {\orange \circ} $ } & Unstable

 \end{tabular}

%% file: sp4Table.txt
\begin{tabular}{c||c} 

Symbol & $\Delta$ \\
 \hline \hline 
{\Large $ \bullet $} & $ 0.530 N_f -1.75+{\cal O}(1/N_f) $ \\
 \hline 
$ \square $ & $ 1.06 N_f -2.18+{\cal O}(1/N_f) $ \\
 \hline 
$ \blacklozenge $ & $ 1.88 N_f -4.29+{\cal O}(1/N_f) $ \\
 \hline 
$ \vartriangle $ & $ 2.69 N_f -5.16+{\cal O}(1/N_f) $ \\
 \hline 
$ \blacktriangledown $ & $ 3.72 N_f -7.87+{\cal O}(1/N_f) $ \\
 \hline 
{\Large $ \circ $} & $ 4.75 N_f -9.27+{\cal O}(1/N_f) $ \\
 \hline 
$ \blacksquare $ & $ 5.95 N_f -12.4+{\cal O}(1/N_f) $ \\
 \hline 
$ \lozenge $ & $ 7.15 N_f -14.2+{\cal O}(1/N_f) $ \\ 
 \hline 
 \hspace{0.7mm} {\Large $ {\orange \circ} $ } & Unstable

 \end{tabular}

%% file: so5Table.txt
\begin{tabular}{c||c} 

Symbol & $\Delta$ \\
 \hline \hline 
{\Large $ \bullet $} & $ 0.530 N_f -1.59+{\cal O}(1/N_f) $ \\
 \hline 
$ \square $ & $ 1.06 N_f -1.86+{\cal O}(1/N_f) $ \\ 
 \hline 
 \hspace{0.7mm} {\Large $ {\orange \circ} $ } & Unstable

 \end{tabular}

%% file: g2Table.txt
\begin{tabular}{c||c} 

Symbol & $\Delta$ \\
 \hline \hline 
{\Large $ \bullet $} & $ 1.06 N_f -2.80+{\cal O}(1/N_f) $ \\ 
 \hline 
 \hspace{0.7mm} {\Large $ {\orange \circ} $ } & Unstable

 \end{tabular}